\documentclass[prd,twocolumn,showpacs,showkeys,preprintnumbers,floatfix,nofootinbib,superscriptaddress]{revtex4-1}
\usepackage{amsfonts} 
\usepackage{amssymb} 
\usepackage{amsmath} 
\usepackage{graphicx} 
\usepackage{subfigure} 
\usepackage{array} 
\usepackage{dcolumn} 
\usepackage{bm} 
\usepackage{bbm} 
\usepackage{latexsym} 
\usepackage{longtable} 
\usepackage{hyperref} 
\usepackage{bbold}
\usepackage{color}
\usepackage{multirow}
\usepackage{rotating}
\usepackage[utf8]{inputenc}

\graphicspath{{./figs/}}

\newcommand{\tsep}{\mathop{t_{\rm sep}}\nolimits}
\newcommand{\tsepi}{\mathop{t_{\rm sep} \to \infty}\nolimits}
\newcommand{\tmin}{\mathop{t_{\rm min}}\nolimits}
\newcommand{\tmax}{\mathop{t_{\rm max}}\nolimits}
\newcommand{\tminb}{\mathop{t_{\rm min}^{\rm best}}\nolimits}
\newcommand{\tskip}{\mathop{\tau_{\rm skip}}\nolimits}
\newcommand{\Ntwo}{\mathop{N_{\rm 2pt}}\nolimits}
\newcommand{\Nthree}{\mathop{N_{\rm 3pt}}\nolimits}

\newcommand{\GeV}{\mathop{\rm GeV}\nolimits}

\newcommand{\slashed}{/\!\!\!}

\DeclareMathOperator{\Tr}{Tr}

\definecolor{green}{rgb}{0.1, 0.8, 0.1}

\newcolumntype{.}[1]{D{.}{.}{#1}}

\allowdisplaybreaks
\providecommand{\matrixe}[3]{\langle#1\lvert#2\rvert#3\rangle}

\begin{document}


\title{Isovector charges of the nucleon from 2+1-flavor QCD with clover fermions}
\author{Boram~Yoon}
\email{boram@lanl.gov}
\affiliation{Los Alamos National Laboratory, Theoretical Division T-2, Los Alamos, NM 87545}

\author{Yong-Chull~Jang}
\email{ypj@lanl.gov}
\affiliation{Los Alamos National Laboratory, Theoretical Division T-2, Los Alamos, NM 87545}

\author{Rajan~Gupta}
\email{rajan@lanl.gov}
\affiliation{Los Alamos National Laboratory, Theoretical Division T-2, Los Alamos, NM 87545}

\author{Tanmoy~Bhattacharya}
\affiliation{Los Alamos National Laboratory, Theoretical Division T-2, Los Alamos, NM 87545}



\author{Jeremy~Green}
\thanks{Present address: NIC, Deutsches Elektronen-Synchrotron, 15738 Zeuthen, Germany}
\affiliation{Institut f\"ur Kernphysik, Johannes Gutenberg-Universit\"at Mainz, D-55099 Mainz, Germany}

\author{B\'alint~Jo\'o}
\affiliation{Jefferson Lab, 12000 Jefferson Avenue, Newport News, Virginia 23606, USA}

\author{Huey-Wen~Lin}
\affiliation{Physics Department, University of California, Berkeley, CA 94720}
\author{Kostas~Orginos}
\affiliation{Department of Physics, College of William and Mary, Williamsburg, Virginia 23187-8795, USA and 
 Jefferson Lab, 12000 Jefferson Avenue, Newport News, Virginia 23606, USA}
\author{David~Richards}
\affiliation{Jefferson Lab, 12000 Jefferson Avenue, Newport News, Virginia 23606, USA}

\author{Sergey~Syritsyn}
\affiliation{Jefferson Lab, 12000 Jefferson Avenue, Newport News, Virginia 23606, USA}

\author{Frank~Winter}
\affiliation{Jefferson Lab, 12000 Jefferson Avenue, Newport News, Virginia 23606, USA}
 
\collaboration{Nucleon Matrix Elements (NME) Collaboration}
\preprint{LA-UR-16-20523}
%
\date{\today}
\begin{abstract}
We present high-statistics estimates of the isovector charges of the
nucleon from four 2+1-flavor ensembles generated using Wilson-clover
fermions with stout smearing and tree-level tadpole improved Symanzik
gauge action at lattice spacings $a \approx 0.127$ and $0.09$~fm and with
$M_\pi \approx 280$ and 170~MeV. The truncated solver method with bias
correction and the coherent source sequential propagator construction
are used to cost-effectively achieve $O(10^5)$ measurements on each
ensemble. Using these data, the analysis of two-point correlation
functions is extended to include four states in the fits, and of
three-point functions to three states. Control over excited-state
contamination in the calculation of the nucleon mass, the mass gaps
between excited states, and in the matrix elements is demonstrated by
the consistency of estimates using this multistate analysis of the
spectral decomposition of the correlation functions and from simulations of 
the three-point functions at multiple values of the source-sink
separation.
The results for all three
charges, $g_A$, $g_S$ and $g_T$, are in good agreement with
calculations done using the clover-on-HISQ lattice formulation with
similar values of the lattice parameters.
\end{abstract}
\maketitle
%
%
%
%
\section{Introduction}
\label{sec:into}

This work presents high-statistics estimates of isovector charges of
the nucleon, $g_A^{u-d}$, $g_S^{u-d}$ and $g_T^{u-d}$, on four
ensembles of (2+1)-flavor lattice QCD using clover-Wilson fermions and
a stout smeared tree-level tadpole-improved Symanzik gauge
action~\cite{JLAB:2016}.  With increased precision, we demonstrate
control over excited-state contamination using a multistate analysis
of the spectral decomposition of the correlation functions.

Nucleon charges play an important role in the analysis of standard
model (SM) and beyond the standard model (BSM) physics.  The nucleon axial
charge $g_A^{u-d}$ is an important parameter that encapsulates the
strength of weak interactions of nucleons.  The ratio, $g_A^{u-d}/g_V^{u-d}$, is
best determined from the experimental measurement of neutron beta
decay using polarized ultracold neutrons by the UCNA Collaboration,
$1.2756(30)$~\cite{Mendenhall:2012tz}, and by PERKEO II,
$1.2761{}^{+14}_{-17}$~\cite{Mund:2012fq}. In the SM, $g_V^{u-d}=1$ up to
second order corrections in isospin
breaking~\cite{Ademollo:1964sr,Donoghue:1990ti} as a result of the
conservation of the vector current. Since $g_A^{u-d}$ is so well measured, 
it serves to benchmark lattice QCD calculations and our goal is to 
provide estimates with $1\%$ total uncertainty. 

The isovector charges $g_S^{u-d}$ and $g_T^{u-d}$, combined with the
helicity-flip parameters $b$ and $b_\nu$ extracted from the
measurements of the neutron decay distribution, probe novel scalar and
tensor interactions at the TeV scale~\cite{Bhattacharya:2011qm}.
Assuming that $b$ and $b_\nu$ are measured at the $10^{-3}$ precision
level~\cite{abBA,WilburnUCNB,Pocanic:2008pu}, one requires that
$g_S^{u-d}$ and $g_T^{u-d}$ be calculated with a precision of
10\%--15\%~\cite{Bhattacharya:2011qm}. This precision has recently
been reached using the clover-on-HISQ lattice
formulation~\cite{Bhattacharya:2016zcn} and the current
clover-on-clover analysis is a necessary independent check using a
unitary lattice QCD formulation. The tensor charge is also given by
the zeroth moment of the transversity distributions that are measured
in many experiments including Drell-Yan and semi-inclusive deep
inelastic scattering (SIDIS).  Accurate calculations of the
contributions of the up and down quarks to the tensor charges will
continue to help elucidate the structure of the nucleon in terms of
quarks and gluons and provide a benchmark against which
phenomenological estimates utilizing a new generation of experiments
at Jefferson Lab (JLab) can be compared~\cite{Dudek:2012vr}.  We 
also use the conserved vector current relation to determine the
neutron-proton mass difference in QCD by combining the estimates of
$g_S^{u-d}/g_V^{u-d}$ with the difference of light quarks masses
$(m_d-m_u)^{\rm QCD}=2.67(35)$~MeV obtained from independent lattice
QCD calculations~\cite{Gonzalez-Alonso:2013ura,Bhattacharya:2016zcn}.

Most extensions of the Standard Model designed to explain nature at
the TeV scale have new sources of $CP$ violation, and the neutron
electric dipole moment (EDM) is a very sensitive probe of
these. Planned experiments aim to reduce the current bound on the
neutron EDM of $2.9 \times 10^{-26}\ e$~cm~\cite{Baker:2006ts} to
around $ 10^{-28}\ e$~cm.  To put stringent constraints on many BSM
theories, one requires that matrix elements of novel $CP$-violating
interactions, of which the quark EDM is one, are calculated with the
required precision.  The contributions of the $u,\ d, \ s, \ c$ quark
EDM to the neutron EDM~\cite{Bhattacharya:2015wna,Pospelov:2005pr} are
given by the flavor diagonal tensor charges. Precise results for the
connected contributions to these charges from 2+1+1-flavor
clover-on-HISQ lattice formulation have been reported
in~\cite{Bhattacharya:2016zcn}. Here we present results from a similar
high statistics study using the clover-on-clover formulation. The
needed contributions of disconnected diagrams are being done in a 
separate study~\cite{Gambhir:2016jul}.

The methodology for the lattice QCD calculations of
the matrix elements of quark bilinear operators within the nucleon
state is
well-developed~\cite{Lin:2012ev,Syritsyn:2014saa,Green:2014vxa,Constantinou:2014tga,Bhattacharya:2015wna,Bhattacharya:2016zcn}.
Our goal is to first calculate the charges ${g_\Gamma}(a, M_\pi, M_\pi
L)$ as functions of the lattice spacing $a$, the quark mass
characterized by the pion mass $M_\pi$ and the lattice size $L$
expressed in dimensionless units of $M_\pi L$.  After renormalization
of these lattice estimates, physical results will be obtained by
taking the continuum limit ($a\rightarrow 0$), the physical pion mass
limit (set by $M_{\pi^0} = 135$~MeV) and the infinite volume limit ($M_\pi L
\rightarrow \infty$) using a combined fit in these three
variables~\cite{Bhattacharya:2016zcn,Bhattacharya:2015wna}.  Here, we
present results for four ensembles at lattice spacings $a\approx
0.127$ and $0.09$~fm with $M_\pi \approx 280$ and $170$~MeV. These
ensembles are labeled $a127m285$, $a094m280$, $a091m170$, and
$a091m170L$ and described in Table~\ref{tab:ens}.

In this work, we demonstrate that precise estimates for matrix
elements within nucleon states can be obtained by combining the
all-mode-averaging (AMA) error-reduction
technique~\cite{Bali:2009hu,Blum:2012uh} (Sec.~\ref{sec:AMA}) and the
coherent source sequential propagator
method~\cite{Bratt:2010jn,Yoon:2016dij}. A detailed analysis of 
excited-state contamination, comparing the variational method
and the 2-state fit to data at multiple source-sink separations 
$t_{\rm sep}$, was presented in~\cite{Yoon:2016dij} using the
$a094m280$ ensemble.\footnote{The label for this ensemble has been changed 
from the labeled $a081m315$ in~\protect\cite{Yoon:2016dij} to $a094m280$ 
because the estimate of the lattice spacing $a$ has been revised.}
In this work, we extend the 2-state fit results presented
there by doing the calculation at an additional value of the lattice
spacing ($a127m285$) and at a lighter pion mass on two ensembles with
different volumes ($a091m170$ and $a091m170L$).

The high statistics data allow us to perform a first analysis
including up to four states in fits to the two-point correlators and
three states in fits to the three-point functions. To obtain results
for the charges in the limit that the source-sink separation $\tsepi$,
we generate data at 4--5 values of $\tsep$ on each ensemble. Using
these data we perform a detailed comparison of results obtained using
2-state versus 3-state fits. Our final estimates of the charges are from
3-state fits.

The renormalization constants of the various quark bilinear operators
are calculated on three ensembles $a127m285$, $a094m280$, $a091m170$
in the RI-sMOM scheme and then converted to the $\overline{\text{MS}}$
scheme at $2$~GeV using 2-loop matching and 3-loop running.  Our final
estimates of the renormalized isovector charges of the nucleon in the
$\overline{\text{MS}}$ scheme at $2$~GeV are given in
Table~\ref{tab:FinalValues}.  Results for the connected part of the
flavor diagonal charges are given in Table~\ref{tab:FDCharges}.

Estimates of all three isovector charges, $g_A^{u-d}$, $g_S^{u-d}$ and
$g_T^{u-d}$ and of the flavor diagonal charges $g_{A,S,T}^{u,d}$ are
in very good agreement with similar high precision calculations done
using a 2+1+1-flavor clover-on-HISQ
formulation~\cite{Bhattacharya:2016zcn}.  Our estimates of $g_A$
obtained with heavy $u$ and $d$ quark masses corresponding to $M_\pi
\approx 280$ and $170$~MeV are within $5\%$ of the experimental
result, $1.276(3)$, from neutron beta
decay~\cite{Mendenhall:2012tz,Mund:2012fq}.

This paper is organized as follows. In Sec.~\ref{sec:Methodology}, we
describe the parameters of the gauge ensembles analyzed and the
various methods used to obtain high precision results. Two-state fits
to two- and three-point functions to extract the unrenormalized
charges are presented in Sec.~\ref{sec:ESC} along with a discussion of
our understanding of, and control over, excited-state
contamination. In Sec.~\ref{sec:3-state}, we extend the analysis to
include up to four states in fits to two-point functions and three
states in three-point correlation functions. The calculation of the
renormalization constants in the RI-sMOM scheme is discussed in
Sec.~\ref{sec:Zfac}. Our final renormalized estimates are given in
Sec.~\ref{sec:results} and compared to previous results obtained using
a 2+1+1-flavor clover-on-HISQ lattice formulation but with similar
statistics and lattice
parameters~\cite{Bhattacharya:2016zcn,Bhattacharya:2015wna} in
Sec.~\ref{sec:comparison}.  We end with conclusions in
Sec.~\ref{sec:conclusions}.

\section{Lattice Methodology}
\label{sec:Methodology}

A detailed description of the lattice methodology and our approach has
been presented in
Refs~\cite{Yoon:2016dij,Bhattacharya:2015wna,Bhattacharya:2016zcn}.
Here we reproduce the discussion necessary to establish the notation and
give details relevant to the analysis and the results.

The four ensembles of 2+1-flavor lattice QCD analyzed in this work
were generated by the JLab/W\&M collaboration~\cite{JLAB:2016} using
clover Wilson fermions and a tree-level tadpole-improved Symanzik
gauge action.  The update is carried out using the rational hybrid
Monte Carlo (RHMC) algorithm~\cite{Duane:1987de}.  One iteration of
stout smearing with the weight $\rho = 0.125$ for the staples is used
in the fermion action. A consequence of the stout smearing is that the
tadpole corrected tree-level clover coefficient $C_{\rm SW}$ used is
very close to the non-perturbative value determined, 
{\it a posteriori}, using the Schr\"odinger functional
method~\cite{JLAB:2016}.

The lattice parameters of the four ensembles are summarized in
Table~\ref{tab:ens}.  Estimates for the lattice spacing $a$ were
obtained using the Wilson-flow scale $w_0$ following the prescription
given in Ref.~\cite{Borsanyi:2012zs}.  We caution the reader that an
alternate estimate of $a$ for the ensemble we label $a127m280$ with
$a=0.127(2)$ and $M_\pi=285(6)$~MeV, has been quoted in
Ref.~\cite{Leskovec:2016lrm} to be $a=0.114(1)$ (and
$M_\pi=316(3)$~MeV since $aM_\pi=0.1834(5)$ is unchanged) using the
$\Upsilon(2s)-\Upsilon(1s)$ mass difference.  Thus, different
estimates of $a$ from this coarse ensemble may vary by $O(10\%)$
depending on the observable used to set them.\footnote{Good quantities
  to use to set the lattice scale $a$ are the ones that are least
  sensitive to the light quark masses and are easy to compute with
  high precision~\protect\cite{Sommer:2014mea}. Examples include the
  $\Upsilon(2s)-\Upsilon(1s)$ mass difference, the Wilson-flow scale
  $w_0$, and the length scales $r_0$ and $r_1$ extracted from the static quark
  potential. Differences in estimates of $a$ arise due to
  discretization errors that are taken care of by the final
  extrapolation of the results to the continuum limit.}  Similar but
smaller differences in $a$ obtained using different observables are
expected for the other three ensembles. Also, note that the ensemble
labeled $a094m280$ here was labeled $a081m315$ in
Ref.~\cite{Yoon:2016dij}. In this paper, we use estimates of $a$ and
$M_\pi$ primarily to label the ensembles and for comparing against
previous results with similar lattice parameters in
Sec.~\ref{sec:comparison}. For this reason, we postpone a more
detailed study of scale setting on these ensembles to future works.

The strange quark mass is first tuned in the 3-flavor theory by
requiring the quantity $(2M_{K^+}^2 - M_{\pi^0}^2)/M_{\Omega^-}^2$ to
equal its physical value 0.1678.  We choose this quantity since it is
independent of the light quark masses to lowest order in chiral
perturbation theory, i.e., the ratio depends only on the value of the
strange quark mass~\cite{Lin:2008pr} and can, therefore, be tuned in
the SU(3) symmetric limit.  The resulting value of $m_s$ is then kept
fixed as the light-quark masses are decreased in the (2+1)-flavor
theory to their physical values.  Further details involving the
generation of these gauge configurations will be presented in a
separate publication~\cite{JLAB:2016}.

The parameters used in the calculations of the two- and three-point
functions carried out on the four ensembles are given in
Table~\ref{tab:runs}. Analyzed configurations are separated by 6, 
alternating 4 and 6, 4, and 4 trajectories on the $a127m285$,
$a094m280$, $a091m170$ and $a091m170L$ ensembles, respectively. Note
that the $a094m280$ ensemble has been analyzed in 5 different ways
labeled as runs R1--R5 in~Table~\ref{tab:runs} to understand and
control excited-state contamination in nucleon matrix elements. As
discussed in Ref.~\cite{Yoon:2016dij}, and analyzed further here, the
five calculations give consistent results. Relevant details of the
lattice methods used and of the analyses carried out are summarized
next.

\begin{table*}
\begin{center}
\renewcommand{\arraystretch}{1.2} 
\begin{ruledtabular}
\begin{tabular}{l|cc|cccc|cc}
Ensemble ID    & $a$ (fm) & $M_\pi$ (MeV) &  $\beta$  & $C_{\rm SW}$  & $ am_{ud}$ & $am_s$    & $L^3\times T$    & $M_\pi L$   \\
\hline
a127m285       & 0.127(2) & 285(3)        &  6.1      & 1.24930971    & -0.2850    & -0.2450   & $32^3\times 96$  & 5.85        \\
\hline                                                                                               
a094m280       & 0.094(1) & 278(3)        &  6.3      & 1.20536588    & -0.2390    & -0.2050   & $32^3\times 64$  & 4.11        \\
\hline                                                                                              
a091m170       & 0.091(1) & 166(2)        &  6.3      & 1.20536588    & -0.2416    & -0.2050   & $48^3\times 96$  & 3.7        \\
\hline                                                                                               
a091m170L      & 0.091(1) & 172(6)        &  6.3      & 1.20536588    & -0.2416    & -0.2050   & $64^3\times 128$ & 5.08       \\
\end{tabular}
\end{ruledtabular}
\caption{Parameters of the 2+1 flavor lattices generated by the
  JLab/W\&M collaboration~\protect\cite{JLAB:2016} using clover-Wilson
  fermions and a tree-level tadpole-improved Symanzik gauge
  action. The lattice spacing $a$ is obtained using
  the Wilson-flow scale $w_0$ and is the dominant source of error
  in estimates of $M_\pi$.  The bare quark masses are defined as $am_i = (1/2\kappa_i - 4)$. 
  Note that the ensemble labeled $a094m280$ here was labeled $a081m315$ in 
  Ref.~\protect\cite{Yoon:2016dij}.}
\label{tab:ens}
\end{center}
\end{table*}

\begin{table*}
\centering
\begin{ruledtabular}
\begin{tabular}{l|c|c|c|c|c|c|c}
ID                      & Method   &  Analysis   &  Smearing Parameters                  & $t_{\rm sep}$       &  $N_{\rm conf}$ & $N_{\rm meas}^{\rm HP}$  & $N_{\rm meas}^{\rm LP}$ \\
\hline
C1: a127m285            & AMA      & 2-state     &  $\{5,60\}$                           & $8, 10,12,14$       & 1000  & 4000 &  128,000    \\
\hline                                                                                                                                  
C2: a094m280 (R1)       & AMA      & 2-state     &  $\{5,60\}$                           & $10,12, 14, 16, 18$ & 1005 & 3015  &  96,480    \\
C3: a094m280 (R2)       & LP       & VAR         &  $\{3,22\}$, $\{5,60\}$, $\{7,118\}$  & $12$                & 443  & 0     &  42,528    \\
C4: a094m280 (R3)       & AMA      & VAR         &  $\{5,46\}$, $\{7,91\}$, $\{9,150\}$  & $12$                & 443  & 1329  &  42,528    \\
C5: a094m280 (R4)       & AMA      & 2-state     &  $\{9,150\}$                          & $10,12, 14, 16, 18$ & 1005 & 3015  &  96,480    \\
C6: a094m280 (R5)       & AMA      & 2-state     &  $\{7,91\}$                           & $8, 10,12, 14, 16$  & 1005 & 3015  &  96,480    \\
\hline
C7: a091m170            & AMA      & 2-state     &  $\{7,91\}$                           & $8, 10,12, 14, 16$  & 629  & 2516  &  80,512    \\
\hline
C8: a091m170L           & AMA      & 2-state     &  $\{7,91\}$                           & $8, 10,12, 14, 16$  & 467  & 2335  &  74,720    \\
\end{tabular}
\end{ruledtabular}
\caption{Description of the ensembles and the lattice parameters used in the
  analyses. Results from the four
  runs, R1--R4, on the $a094m280$ ensemble were presented in Ref.~\protect\cite{Yoon:2016dij}. We 
  have extended the statistics in runs R1 and R4 and added R5 to further 
  understand the dependence of the estimates on the smearing size, the
  efficacy of the variational method and the 2-state fit to data at
  multiple source-sink separations $t_{\rm sep}$. The smearing
  parameters $\{\sigma, N_{\rm GS}\}$ are defined in the text. AMA
  indicates that the bias in the low-precision measurements (labeled
  LP) was corrected using high-precision measurements as described in
  Eq.~\protect\eqref{eq:2-3pt_AMA}.  VAR indicates that the full $3
  \times 3$ matrix of correlation functions with smearing sizes listed
  was calculated and a variational analysis performed to extract the
  ground state eigenvector as described in
  Sec.~\protect\ref{sec:var}. Analysis of data with multiple $t_{\rm sep}$ 
  to obtain the $\tsepi$ estimate is carried out using
  Eqs.~\protect\eqref{eq:2pt} and~\protect\eqref{eq:3pt}. }
  \label{tab:runs}
\end{table*}

\subsection{Correlation Functions}
\label{sec:CorrelationFunctions}

The interpolating operator $\chi$ used to create$/$annihilate the nucleon
state is taken to be
\begin{align}
 \chi(x) = \epsilon^{abc} \left[ {q_1^a}^T(x) C \gamma_5 
            \frac{(1 \pm \gamma_4)}{2} q_2^b(x) \right] q_1^c(x)
\label{eq:nucl_op}
\end{align}
with color indices denoted by $\{a, b, c\}$, charge conjugation matrix $C = \gamma_0 \gamma_2$, and
$q_1$ and $q_2$ the two different flavors of light quarks.
The non-relativistic projection $(1 \pm \gamma_4)/2$ is inserted to
improve the signal, with the plus and minus sign applied to the
forward and backward propagation in Euclidean time, respectively.

The two-point and three-point nucleon correlation functions at zero momentum 
are defined as 
\begin{align}
{\mathbf C}_{\alpha \beta}^{\text{2pt}}(t)
  &= \sum_{\mathbf{x}} 
   \langle 0 \vert \chi_\alpha(t, \mathbf{x}) \overline{\chi}_\beta(0, \mathbf{0}) 
   \vert 0 \rangle \,,  \nonumber \\
{\mathbf C}_{\Gamma; \alpha \beta}^{\text{3pt}}(t, \tau)
  &= \sum_{\mathbf{x}, \mathbf{x'}} 
  \langle 0 \vert \chi_\alpha(t, \mathbf{x}) \mathcal{O}_\Gamma(\tau, \mathbf{x'})
  \overline{\chi}_\beta(0, \mathbf{0}) 
   \vert 0 \rangle \,,
\label{eq:corr_funs}
\end{align}
where $\alpha$ and $\beta$ are the spinor indices. In writing
Eq.~\eqref{eq:corr_funs}, the source time slice has been translated to time
$t=0$; the sink time slice, at which a zero-momentum nucleon insertion
is made using the sequential source
method~\cite{Bratt:2010jn,Yoon:2016dij}, is at $t>0$ for forward
propagation; and $\tau$ is the time slice at which the bilinear
operator $\mathcal{O}_\Gamma^q(x) = \bar{q}(x) \Gamma q(x)$ is
inserted. The Dirac matrix $\Gamma$ is $1$, $\gamma_4$, $\gamma_i
\gamma_5$ and $\gamma_i \gamma_j$ for scalar (S), vector (V), axial
(A) and tensor (T) operators, respectively, with $\gamma_5 \equiv
\gamma_1 \gamma_2 \gamma_3 \gamma_4$.
In this work, subscripts $i$ and $j$ on gamma matrices run over $\{1,2,3\}$, 
with $i<j$. 

The nucleon charges $g_\Gamma^q$ are  defined as 
\begin{align}
 \langle N(p, s) \vert \mathcal{O}_\Gamma^q \vert N(p, s) \rangle
 = g_\Gamma^q \bar{u}_s(p) \Gamma u_s(p) \,,
\end{align}
where the normalization of the spinors in Euclidean space is 
\begin{equation}
\sum_{s} u_N(\vec p,s) \bar{u}_N(\vec{p},s) =
   \frac{E(\vec{p})\gamma_4-i\vec\gamma\cdot \vec{p} + M}{2 E(\vec{p})}.
\label{eq:spinor}
\end{equation}

To analyze the data, we construct the projected two- and three-point correlation functions
\begin{align}
C^{\text{2pt}}(t) & = {\langle \Tr [ \mathcal{P}_\text{2pt} {\mathbf C}^{\text{2pt}}(t) ] \rangle} 
 \label{eq:2pt_proj}
 \\
C_{\Gamma}^{\text{3pt}}(t, \tau)  & = \langle \Tr [ \mathcal{P}_{\rm 3pt} {\mathbf C}_{\Gamma}^{\text{3pt}}(t, \tau) ]\rangle \, .
 \label{eq:3pt_proj}
\end{align}
%
The projection operator $\mathcal{P}_\text{2pt} = (1+\gamma_4)/2$ is
used to project on to the positive parity contribution for the nucleon
propagating in the forward direction. For the connected three-point
contributions, $\mathcal{P}_{\rm 3pt} =
\mathcal{P}_\text{2pt}(1+i\gamma_5\gamma_3)$ is used.  Note that, at
zero-momentum, the $C_{\Gamma}^{\text{3pt}}(t, \tau)$ defined in
Eq.~\eqref{eq:3pt_proj} becomes zero unless $\Gamma = 1$, $\gamma_4$,
$\gamma_i \gamma_5$ and $\gamma_i \gamma_j$. 

The two- and three-point correlation functions defined in
Eq.~\eqref{eq:corr_funs} are constructed using quark propagators
obtained by inverting the clover Dirac matrix with gauge-invariant
Gaussian smeared sources. These smeared sources are generated by
applying $(1 + \sigma^2\nabla^2/(4N_{\rm GS}))^{N_{\rm GS}} $ to a
unit point source.  Here $\nabla^2$ is the three-dimensional Laplacian
operator and $N_{\rm GS}$ and $\sigma$ are smearing parameters that
are given in Table~\ref{tab:runs} for each calculation.  Throughout
this paper, the notation $S_i S_j$ will be used to denote a
calculation with source smearing $\sigma=i$ and sink smearing
$\sigma=j$.  Variations of the parameter $N_{\rm GS}$ over a large
range does not impact any of the results~\cite{Yoon:2016dij}, and it is
dropped from further discussions since our choice lies within this
range. Before constructing the smeared sources, the spatial gauge
links on the source time slice are smoothed by 20 hits of the stout
smearing procedure with weight $\rho = 0.08$. A more detailed
discussion of the efficacy of source smearing used in this study is
given in Ref.~\cite{Yoon:2016dij}.

%
\subsection{Behavior of the Correlation Functions}
\label{sec:2state}

Our goal is to extract the matrix elements of various bilinear
quark operators between ground state nucleons. The lattice operator
$\chi$, given in Eq.~\eqref{eq:nucl_op}, couples to the nucleon, all
its excitations and multiparticle states with the same quantum
numbers. The correlation functions, therefore, get contributions from
all these intermediate states. Using spectral decomposition, the
behavior of two- and three-point functions is given by the expansion:
\begin{align}
C^\text{2pt}
  &(t_f,t_i) = \nonumber \\
  &{|{\cal A}_0|}^2 e^{-aM_0 (t_f-t_i)} + {|{\cal A}_1|}^2 e^{-aM_1 (t_f-t_i)} + \nonumber \\ 
  &{|{\cal A}_2|}^2 e^{-aM_2 (t_f-t_i)} + {|{\cal A}_3|}^2 e^{-aM_3 (t_f-t_i)} + \ldots \,, 
\label{eq:2pt} \\
C^\text{3pt}_{\Gamma}&(t_f,\tau,t_i) = \nonumber\\
  & |{\cal A}_0|^2 \langle 0 | \mathcal{O}_\Gamma | 0 \rangle  e^{-aM_0 (t_f - t_i)} +{}\nonumber\\
  & |{\cal A}_1|^2 \langle 1 | \mathcal{O}_\Gamma | 1 \rangle  e^{-aM_1 (t_f - t_i)} +{}\nonumber\\
  & |{\cal A}_2|^2 \langle 2 | \mathcal{O}_\Gamma | 2 \rangle  e^{-aM_2 (t_f - t_i)} +{}\nonumber\\
  & {\cal A}_1{\cal A}_0^* \langle 1 | \mathcal{O}_\Gamma | 0 \rangle  e^{-aM_1 (t_f-\tau)} e^{-aM_0 (\tau-t_i)} +{}\nonumber\\
  & {\cal A}_0{\cal A}_1^* \langle 0 | \mathcal{O}_\Gamma | 1 \rangle  e^{-aM_0 (t_f-\tau)} e^{-aM_1 (\tau-t_i)} +{}\nonumber\\
  & {\cal A}_2{\cal A}_0^* \langle 2 | \mathcal{O}_\Gamma | 0 \rangle  e^{-aM_2 (t_f-\tau)} e^{-aM_0 (\tau-t_i)} +{}\nonumber\\
  & {\cal A}_0{\cal A}_2^* \langle 0 | \mathcal{O}_\Gamma | 2 \rangle  e^{-aM_0 (t_f-\tau)} e^{-aM_2 (\tau-t_i)} +{}\nonumber\\
  & {\cal A}_1{\cal A}_2^* \langle 1 | \mathcal{O}_\Gamma | 2 \rangle  e^{-aM_1 (t_f-\tau)} e^{-aM_2 (\tau-t_i)} +{}\nonumber\\
  & {\cal A}_2{\cal A}_1^* \langle 2 | \mathcal{O}_\Gamma | 1 \rangle  e^{-aM_2 (t_f-\tau)} e^{-aM_1 (\tau-t_i)} + \ldots \,,
\label{eq:3pt}
\end{align}
where we have shown all contributions from the ground state
$|0\rangle$ and the first three excited states $|1\rangle$,
$|2\rangle$ and $|3\rangle$ with masses $M_1$, $M_2$ and $M_3$ to the
two-point functions and from the first two excited states for the
three-point functions.  The analysis, using Eqs.~\eqref{eq:2pt}
and~\eqref{eq:3pt}, is called a ``2-state fit'' or ``3-state fit'' or
``4-state fit'' depending on the number of intermediate states
included.  The 2-state analysis (keeping one excited state) requires
extracting seven parameters ($M_0$, $M_1$, ${\cal A}_0$, ${\cal A}_1$,
$\langle 0 | \mathcal{O}_\Gamma | 0 \rangle$, $ \langle 1 |
\mathcal{O}_\Gamma | 0 \rangle$ and $ \langle 1 | \mathcal{O}_\Gamma |
1 \rangle$) from fits to the two- and three-point functions. The
3-state analysis introduces five additional parameters: $M_2$, ${\cal
  A}_2$, $\langle 0 | \mathcal{O}_\Gamma | 2 \rangle$, $\langle 1 |
\mathcal{O}_\Gamma | 2 \rangle$ and $\langle 2 | \mathcal{O}_\Gamma |
2 \rangle$.  These simultaneous fits to data at multiple values of $\tsep$
provide estimates of the charges in the limit $\tsepi$. Throughout
this paper, values of $t,\ \tau$ and $\tsep$ are in lattice units
unless explicitly stated. 

Nine of the twelve parameters in the 3-state analysis---the three
masses $M_0$, $M_1$ and $M_2$ and the six matrix elements $\langle i |
\mathcal{O}_\Gamma | j \rangle$---are independent of the details of
the interpolating operator. Our goal is to obtain their values by
removing the discretization errors and the higher excited-state
contaminations.  The amplitudes ${\cal A}_i$ depend on the choice of
the interpolating nucleon operator and/or the smearing parameters used
to generate the smeared sources.  It is evident from
Eqs.~\eqref{eq:2pt} and~\eqref{eq:3pt} that the ratio of the
amplitudes, ${\cal A}_i/{\cal A}_0$, is the quantity to minimize in
order to reduce excited-state contamination as it determines the
relative size of the overlap of the nucleon operator with the excited
states. A detailed analysis of how it can be reduced by tuning the
smearing size $\sigma$ and a comparison of the efficacy with a
variational analysis (run R2 and R3), described in Sec.~\ref{sec:var},
was presented in Ref.~\cite{Yoon:2016dij} using the $a094m280$
ensemble. We present an update on the comparison using renormalized
charges obtained from fits with the full covariance matrix in
Sec.~\ref{sec:comparison}.

We extract the charges $g_S$ and $g_V$ ($g_A$ and $g_T$) from the real
(imaginary) part of the three-point function with operator insertion
at zero momentum.  In the 2-state fits discussed in
Sec.~\ref{sec:2-state}, we first estimate the four parameters, $M_0$,
$M_1$, ${\cal A}_0$ and ${\cal A}_1$ from the two-point function
data. The results for all four ensembles and for three selected fit
ranges investigated are collected in Table~\ref{tab:2ptfits}. These
are then used as inputs in the extraction of matrix elements from fits
to the three-point data.  For the insertion of each operator
$\mathcal{O}_\Gamma = \mathcal{O}_{A,S,T,V}$, extraction of the three
matrix elements $\langle 0 | \mathcal{O}_\Gamma | 0 \rangle$, $
\langle 1 | \mathcal{O}_\Gamma | 0 \rangle$ and $ \langle 1 |
\mathcal{O}_\Gamma | 1 \rangle$ is done by making one overall fit to
the data versus the operator insertion time $\tau$ and the various
source-sink separations $t_{\rm sep}$ using Eq.~\eqref{eq:3pt}.  In
these fits, we neglect the data on $\tskip$ time slices on either end
adjacent to the source and the sink for each $t_{\rm sep} $ to reduce
the contributions of the neglected higher excited states.  Fits to
both the two- and three-point data are done within the same single
elimination jackknife process to estimate the errors.  The same
procedure is then followed in the 3-state analysis described in
Sec.~\ref{sec:3-state}.

In this study, we demonstrate that stable estimates for the masses,
mass-gaps and the charges $g_{\Gamma} \equiv \langle 0 |
\mathcal{O}_\Gamma | 0 \rangle$ can be obtained with $O(10^5)$
measurements. The errors in the other matrix elements are large,
nevertheless certain qualitative features can be established.

\subsection{The variational Method}
\label{sec:var}

One can also reduce excited-state contamination by constructing the two-
and three-point correlation functions incorporating a variational analysis
(see~\cite{Dragos:2015ocy,Dragos:2016rtx} and references therein for
previous use of the variational method for calculating nucleon matrix
elements).  To implement this method on the $a094m280$ ensemble, we
constructed correlation functions using quark propagators with three
different smearing sizes $\sigma_i$ that are summarized under runs R2
and R3 in Table~\ref{tab:runs} but with a single $t_{\rm sep} = 12
\approx 1$~fm.  The two-point correlation function for the nucleon at
any given time separation $t$ is then a $3 \times 3$ matrix
$G_{ij}(t)$ made up of correlation functions with source smearing
$S_i$ and sink smearing $ S_j$.  The best overlap with the ground
state is given by the eigenvector corresponding to the largest
eigenvalue $\lambda_0$ obtained from the generalized eigenvalue
relation~\cite{Fox:1981xz}:
\begin{eqnarray}
G(t+\Delta t) u_i = \lambda_i G(t) u_i \,, 
\label{eq:EV}
\end{eqnarray}
where $u_i$ are the eigenvectors with eigenvalues $\lambda_i$. 
The matrix $G$ should be symmetric up to statistical fluctuations, so
we symmetrize it by averaging the off-diagonal matrix
elements. 
Our final analysis for the calculation of the ground state
eigenvector $u_0$ was done with $t=6$ and $\Delta t=3$ as discussed
in~\cite{Yoon:2016dij}.

Similarly, in our variational analysis, the three-point function data
$C_\Gamma^{\rm 3pt} (\tau,t_{\rm sep})$, from which various charges
are extracted, are $3\times 3$ matrices $G_\Gamma^{\rm 3pt}
(\tau,t_{\rm sep})$. The ground state estimate is obtained by
projecting these matrices using the ground state vector $u_0$
estimated from the two-point variational analysis, i.e., $u_0^{T}
G^{\rm 3pt}(\tau,t_{\rm sep}) u_0$. This projected correlation
function is expected to have smaller excited state contamination
compared to the correlation function with single smearings.  Since the
variational correlation function has been calculated at a single
$t_{\rm sep}=12$, we analyze it using only 2-state fits. Note that the
contribution of the matrix element $ \langle 1 | \mathcal{O}_\Gamma |
1 \rangle$ cannot be isolated from $ \langle 0 | \mathcal{O}_\Gamma |
0 \rangle$ from fits to data with a single value of
$\tsep$. Consequently, our variational estimates of the charges,
collected in Table~\ref{tab:gren-1}, include the contamination from
the $ \langle 1 | \mathcal{O}_\Gamma |1 \rangle$ transition unlike
results from multistate fits to data at a number of values of $\tsep$.

\subsection{The AMA Method for High Statistics}
\label{sec:AMA}

The high statistics calculation on the four ensembles was carried out
using the all-mode-averaging (AMA)
technique~\cite{Bali:2009hu,Blum:2012uh} and the coherent sequential
source method~\cite{Bratt:2010jn,Yoon:2016dij}.  To implement these
methods, we choose at random four time slices separated by $T/4=24$ on
each configuration of the $a127m285$ and $a091m170$ ensembles and on five
time slices separated by twenty-five time slices on the $a091m170L$
ensemble.  On the $a094m280$ lattices, we choose three time slices
separated by $ 21$ time slices on each configuration and staggered
these by $9$ time slices between successive configurations to reduce 
correlations.

On each of these time slices, we choose $N_{\rm LP}=32$ randomly
selected source locations from which low-precision (LP) evaluation of
the quark propagator is carried out. The resulting LP estimates for two-
and three-point functions from these sources are, {\it a priori}, biased
due to the low-precision inversion of the Dirac matrix. To remove this bias,
we selected an additional source point on each of the time slices from
which a high-precision (HP) and LP measurement of the correlation
functions was carried out. The total number of measurements made on
each ensemble are given in Table~\ref{tab:runs}.

On each configuration, the bias corrected two- and three-point function data 
are constructed first using the HP and the LP correlators as
\begin{align}
 C^\text{AMA}& 
 = \frac{1}{N_\text{LP}} \sum_{i=1}^{N_\text{LP}} 
    C_\text{LP}(\mathbf{x}_i^\text{LP}) \nonumber \\
  +& \frac{1}{N_\text{HP}} \sum_{i=1}^{N_\text{HP}} \left[
    C_\text{HP}(\mathbf{x}_i^\text{HP})
    - C_\text{LP}(\mathbf{x}_i^\text{HP})
    \right] \,,
  \label{eq:2-3pt_AMA}
\end{align}
where $C_\text{LP}$ and $C_\text{HP}$ are the two- and three-point
correlation functions calculated in LP and HP,
respectively. Correlators from the two kinds of source positions
$\mathbf{x}_i^\text{LP}$ and $\mathbf{x}_i^\text{HP}$, are assumed to
be translated to a common point when defining
Eq.~\eqref{eq:2-3pt_AMA}.  The bias in the LP estimate (first term) is
corrected by the second term provided the LP approximation is
covariant under lattice translations, which is true for the two- and
three-point functions. The contribution to the overall error by the second
term is small provided the HP and LP calculations from the same source
point are correlated. To estimate errors, the measurements on each
configuration are first averaged and then the single elimination
Jackknife procedure is carried out over these configuration averages.

We used the adaptive multigrid algorithm for inverting the Dirac
matrix~\cite{Babich:2010qb} and set the low-accuracy stopping
criterion $r_{\rm LP} \equiv |{\rm residue}|_{\rm LP}/|{\rm source}| =
10^{-3}$ and the HP criterion to the analogous $r_{\rm HP} = 10^{-10}$.  We have
compared the AMA and LP estimates for both the two- and three-point
correlation functions themselves and for the fit parameters $M_i$,
${\cal A}_i$, and the matrix elements $\langle i | \mathcal{O}_\Gamma
| j \rangle$. In all cases, we find the difference between the AMA and
LP estimates is a tiny fraction (few percent) of the error in either
measurement~\cite{Yoon:2016dij}.  In short, based on all the
calculations we have carried out, possible bias in the LP measurements
with $r_{\rm LP} = 10^{-3}$ as the stopping criteria in the adaptive
multigrid inverter is much smaller than the statistical errors
estimated from $O(10^5)$ measurements.

%
%

\subsection{Statistics}
\label{sec:stats}

The total number of LP and HP measurements and the values of
source-sink separations $t_{\rm sep}$ analyzed are given in
Table~\ref{tab:runs}.  Our statistical tests show that correlations
between measurements are reduced by choosing the source points
randomly within and between configurations~\cite{Yoon:2016dij}. Also,
using the coherent source method for constructing the sequential
propagators from the sink points to reduce computational cost does not
significantly increase the errors~\cite{Bratt:2010jn,Yoon:2016dij}.

On all the ensembles, we first estimate the masses $M_i$ and 
the amplitudes ${\cal A}_i$ using the 2-, 3- or 4-state fits to the
two-point function data and then use these as inputs in the extraction
of matrix elements from fits to the three-point data.  Both of these fits,
to two- and three-point data, are done within the same Jackknife process to
take into account correlations in the estimation of errors.  We
performed both correlated and uncorrelated fits to the nucleon two- and
three-point function data. In all cases, correlated and uncorrelated fits
gave overlapping estimates.  For the final results, we use fits
minimizing the correlated $\chi^2/{\rm d.o.f.}$. 

We find that the central values from the 3-state fits are consistent
with those from the 2-state fits, and the error estimates are
comparable.  Our final quoted estimates are from 4-state fits to the
two-point data and a 3-state fit to the three-point data with the matrix
element $\langle 2 | \mathcal{O}_S | 2 \rangle $ set to zero as
discussed in Sec.~\ref{sec:3-state}.  Our overall conclusion is that
to obtain the isovector charges $g_A$ and $g_T$ with 1\% uncertainty
(or 2\% uncertainty at the physical pion mass and after extrapolation
to the continuum limit) will require $O(10^6)$ measurements on each
ensemble. Approximately five times larger statistics are needed to
extract $g_S$ with the same precision.


\begin{figure*}[tb]
  \subfigure{
     \includegraphics[width=0.48\linewidth]{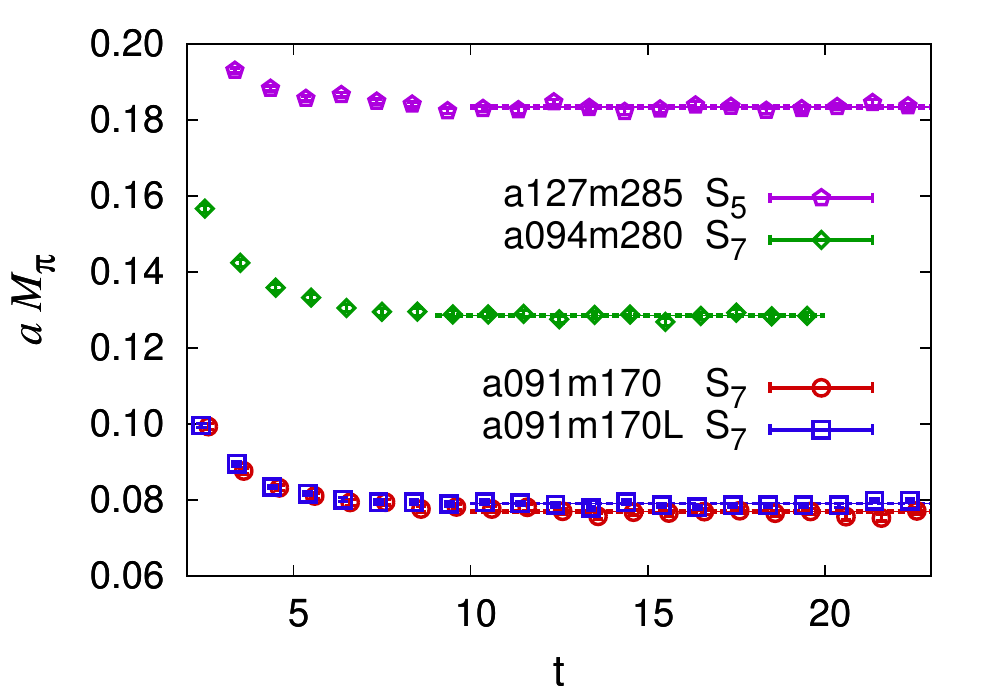}
     \includegraphics[width=0.48\linewidth]{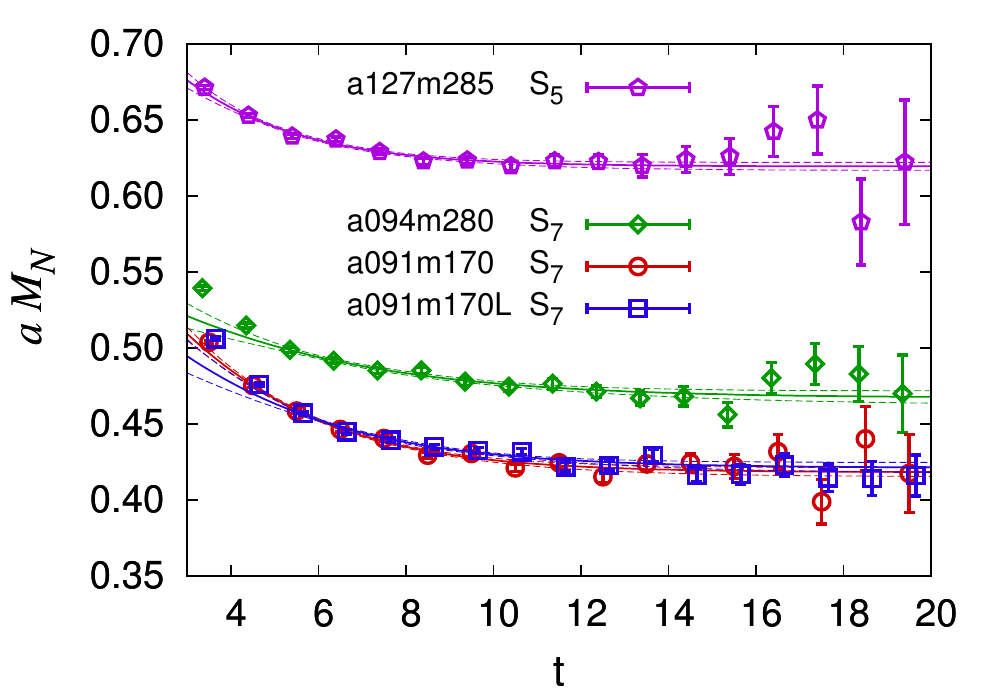}
  }
 \caption{Data for the the effective mass of the pion (left) and the
   nucleon (nucleon) obtained using 2-state fits to the zero-momentum two-point correlation
   functions. The ensemble ID and smearing size are specified in the
   labels.  The 2-state fits to the nucleon are made with our ``best''
   value of $\tmin$ described in the text.
}
\label{fig:2pt_CandD}
\end{figure*}

\begin{figure*}[tb]
  \subfigure{
     \includegraphics[width=0.48\linewidth]{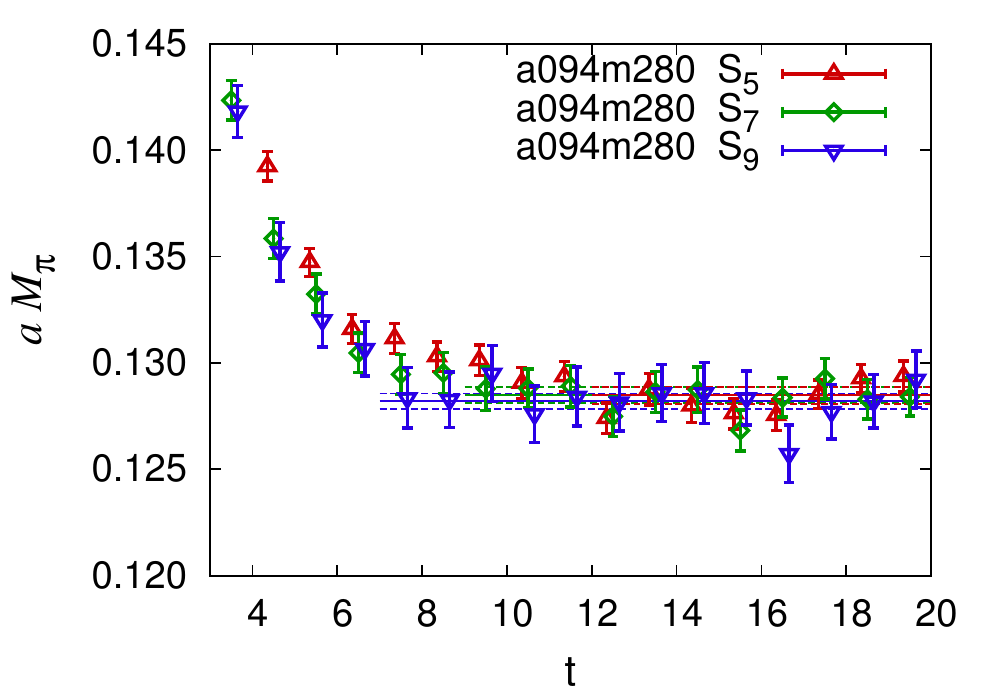}
     \includegraphics[width=0.48\linewidth]{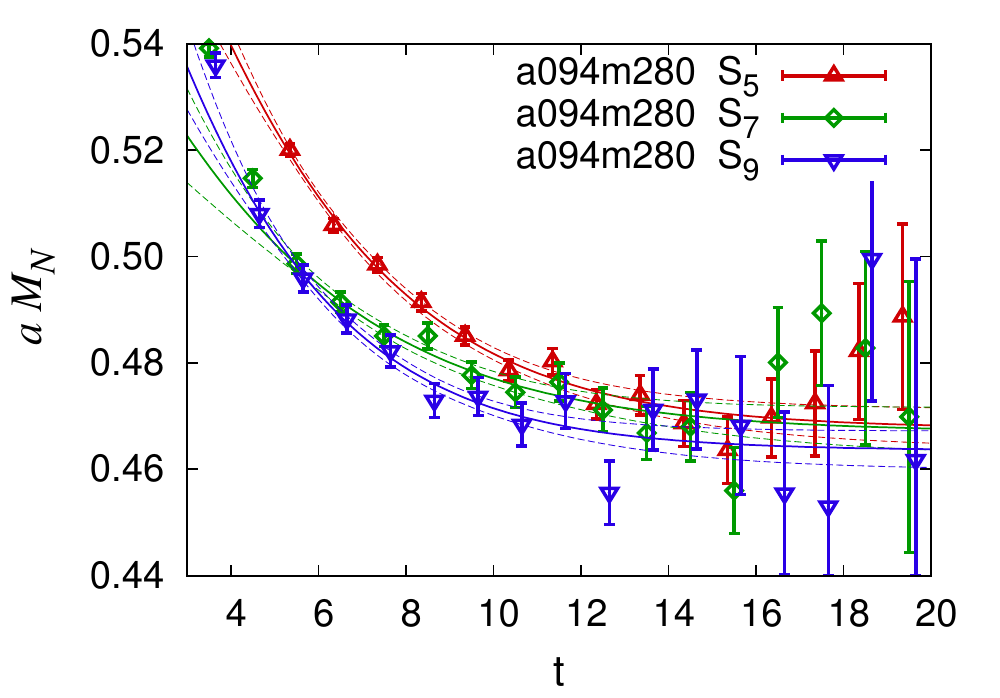}
  }
 \caption{Comparison of the effective mass of the pion (left) and the
   nucleon (right) obtained using 2-state fits to the zero-momentum
   two-point correlation functions on the $a094m280$ ensemble for three
   different smearings, $S_5 S_5$, $S_7 S_7$ and, $S_9 S_9$. The
   2-state fits to the nucleon are made with our ``best'' value of
   $\tmin$ described in the text.  }
\label{fig:2pt_D5S579}
\end{figure*}

\begin{figure*}[tb]
\centering
  \subfigure{
    \includegraphics[width=0.9\linewidth,trim={0      0.01cm 0 0},clip]{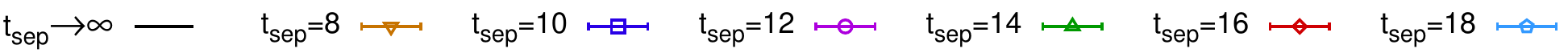}  
  }
\\
\vspace{-0.5cm}
  \subfigure{
    \includegraphics[width=0.371\linewidth,trim={0      0.2cm 0 0},clip]{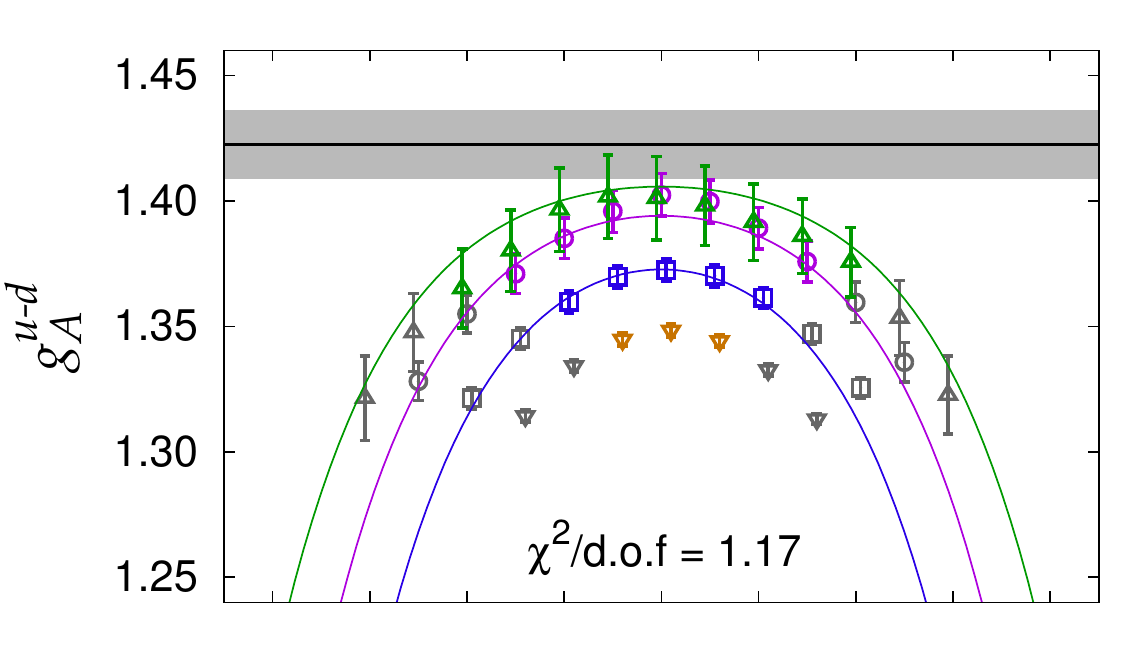}  \hspace{0.15in}
    \includegraphics[width=0.342\linewidth,trim={0.9cm  0.2cm 0 0},clip]{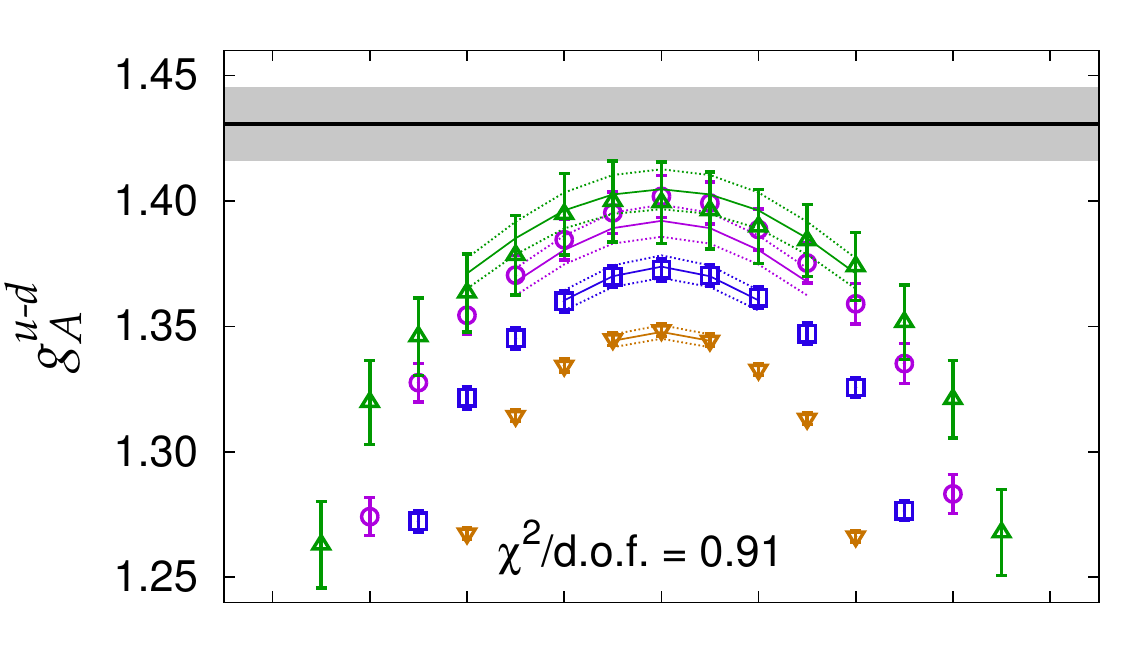}
    \includegraphics[width=0.08\linewidth,trim={0.0cm  0.0cm 0 0},clip]{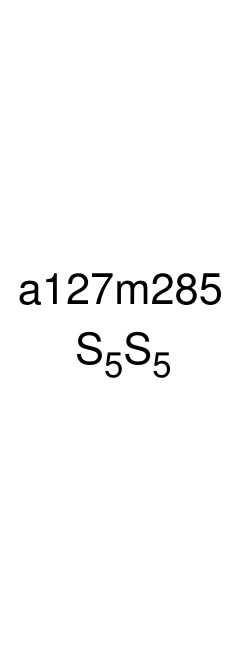}
  } 
\\
\vspace{-0.5cm}
  \subfigure{
    \includegraphics[width=0.371\linewidth,trim={0      0.2cm 0 0},clip]{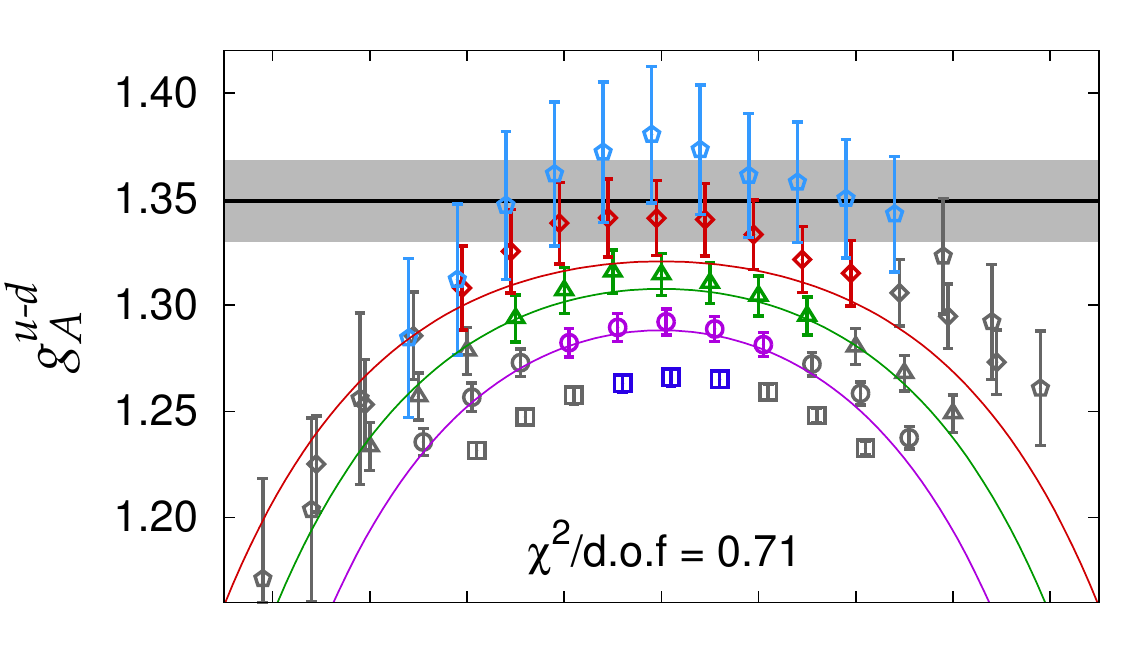}  \hspace{0.15in}
    \includegraphics[width=0.342\linewidth,trim={0.9cm  0.2cm 0 0},clip]{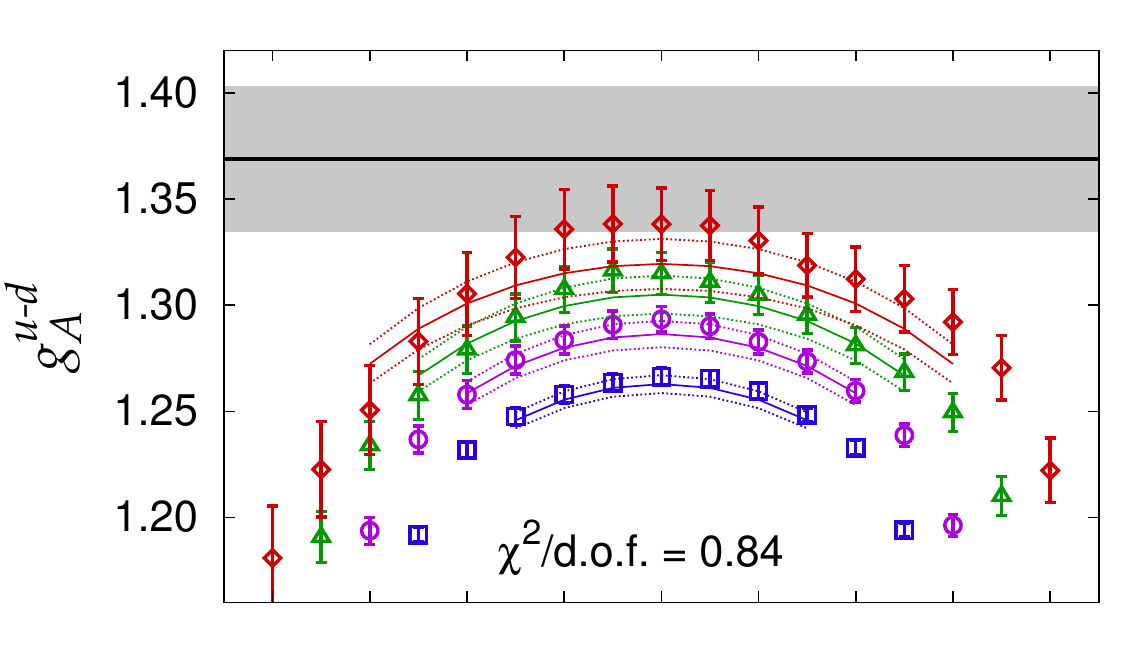}
    \includegraphics[width=0.08\linewidth,trim={0.0cm  0.0cm 0 0},clip]{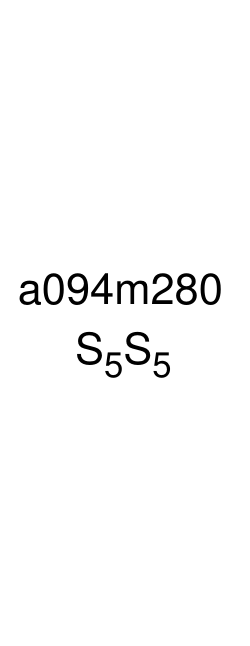}
  }
\\
\vspace{-0.5cm}
  \subfigure{
    \includegraphics[width=0.371\linewidth,trim={0      0.2cm 0 0},clip]{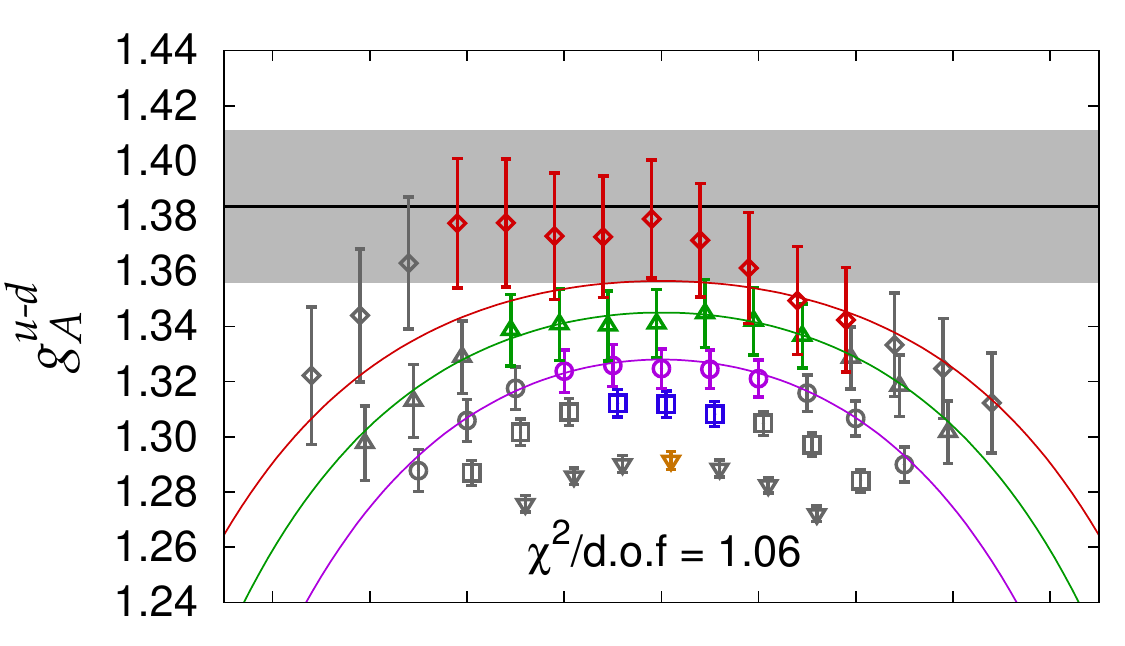}  \hspace{0.15in}
    \includegraphics[width=0.342\linewidth,trim={0.9cm  0.2cm 0 0},clip]{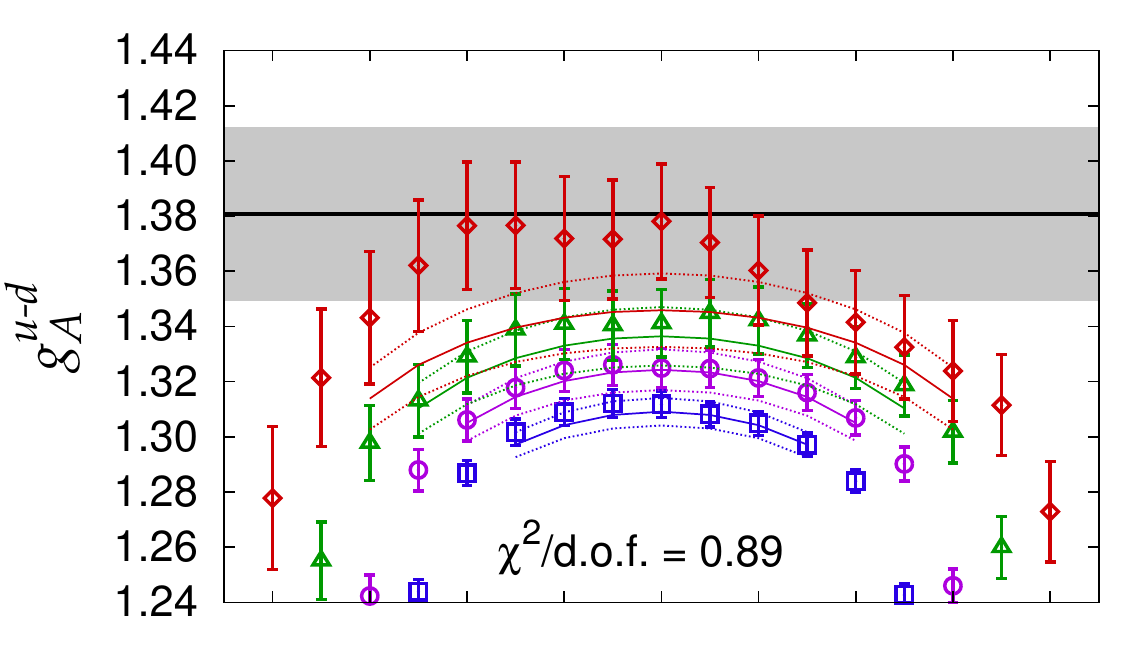}
    \includegraphics[width=0.08\linewidth,trim={0.0cm  0.0cm 0 0},clip]{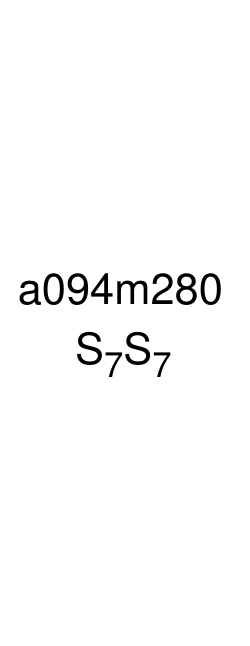}
  }
\\
\vspace{-0.5cm}
  \subfigure{
    \includegraphics[width=0.371\linewidth,trim={0      0.2cm 0 0},clip]{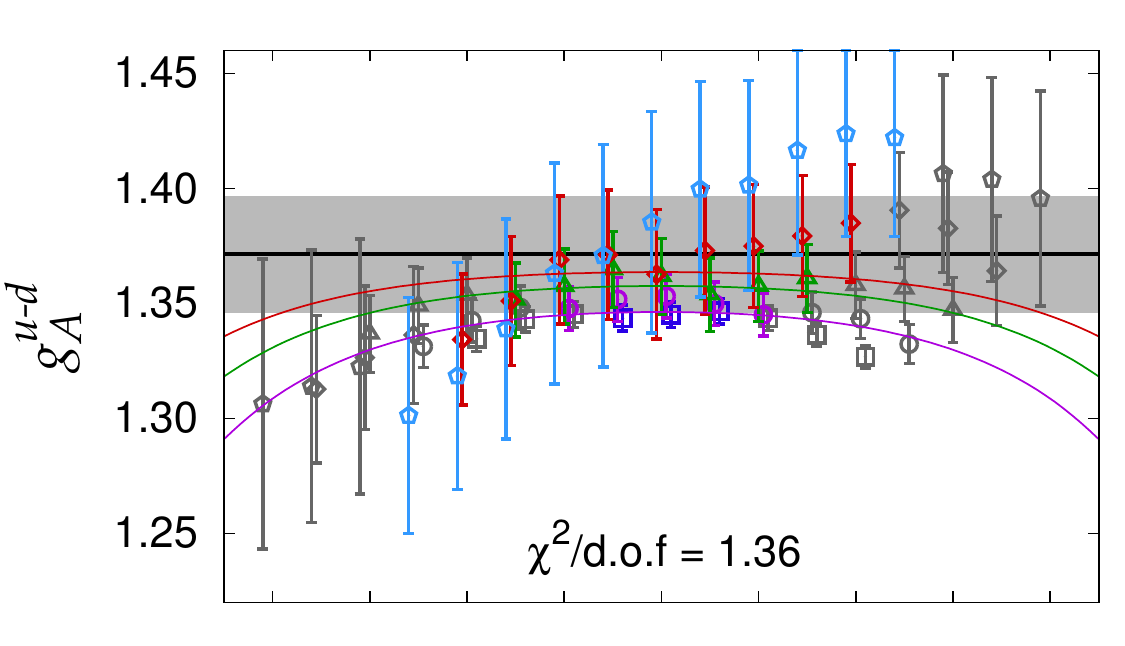}  \hspace{0.15in}
    \includegraphics[width=0.342\linewidth,trim={0.9cm  0.2cm 0 0},clip]{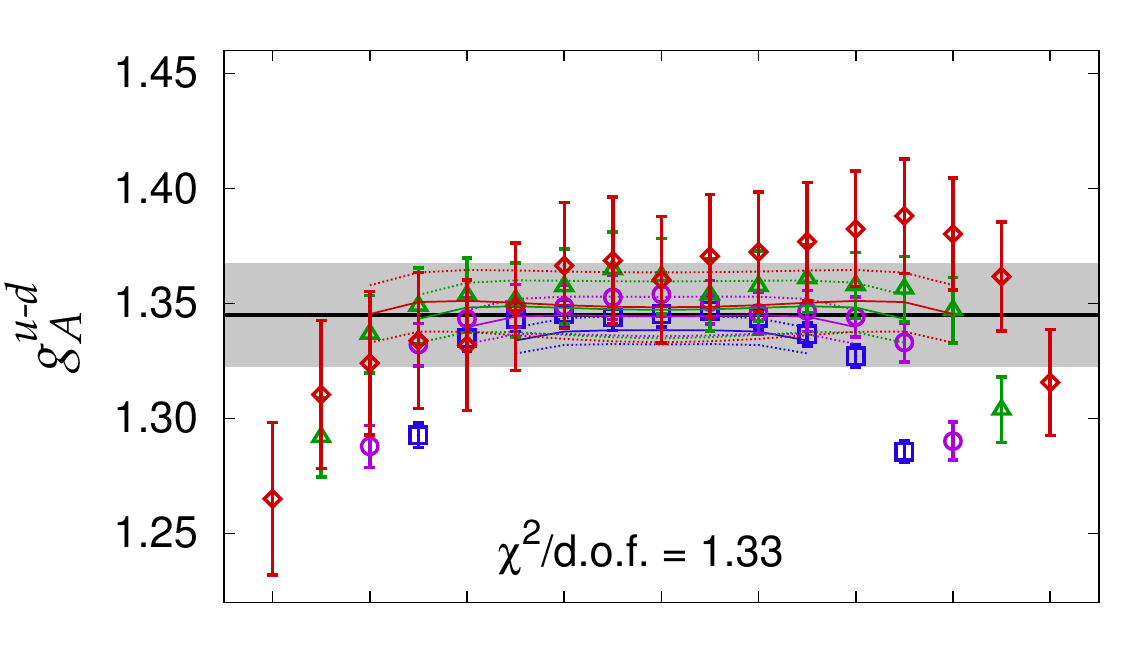}
    \includegraphics[width=0.08\linewidth,trim={0.0cm  0.0cm 0 0},clip]{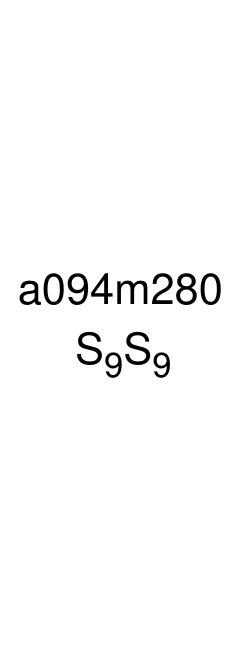}
  }
\\
\vspace{-0.5cm}
  \subfigure{
    \includegraphics[width=0.371\linewidth,trim={0      0.2cm 0 0},clip]{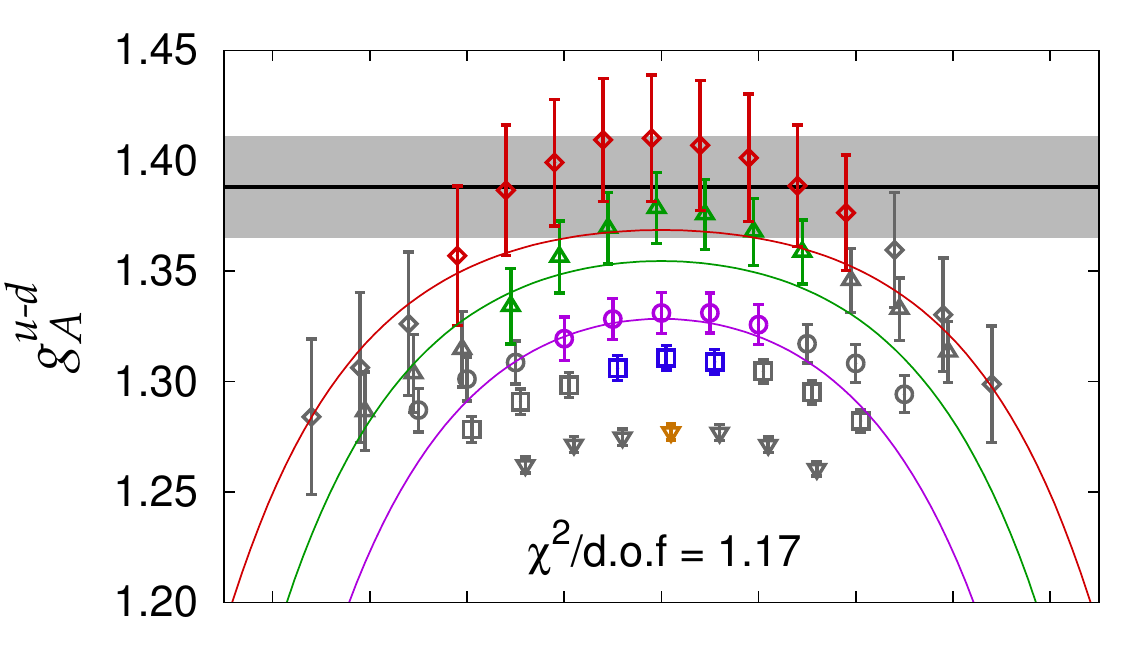}  \hspace{0.15in}
    \includegraphics[width=0.342\linewidth,trim={0.9cm  0.2cm 0 0},clip]{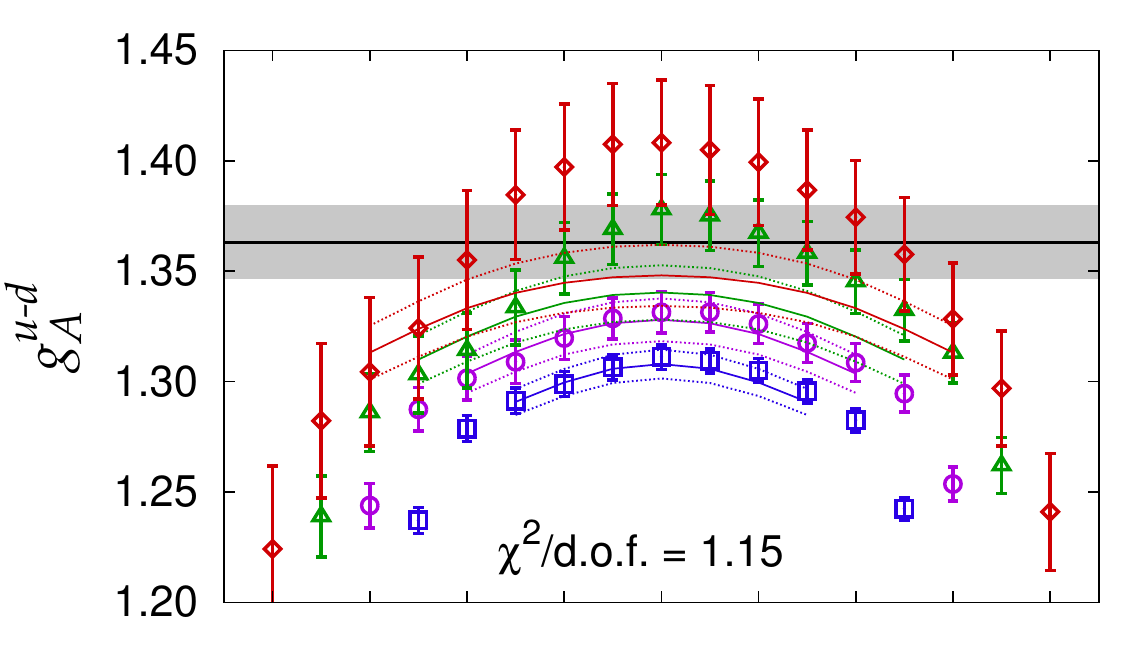}
    \includegraphics[width=0.08\linewidth,trim={0.0cm  0.0cm 0 0},clip]{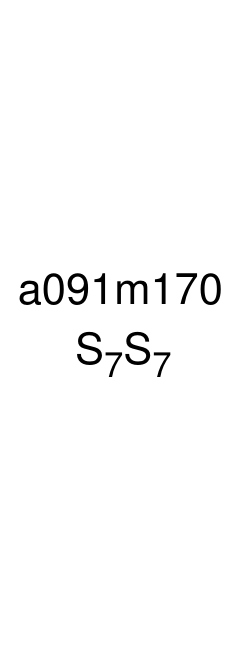}
  }
\\
\vspace{-0.5cm}
  \subfigure{
    \includegraphics[width=0.371\linewidth,trim={0      0.2cm 0 0},clip]{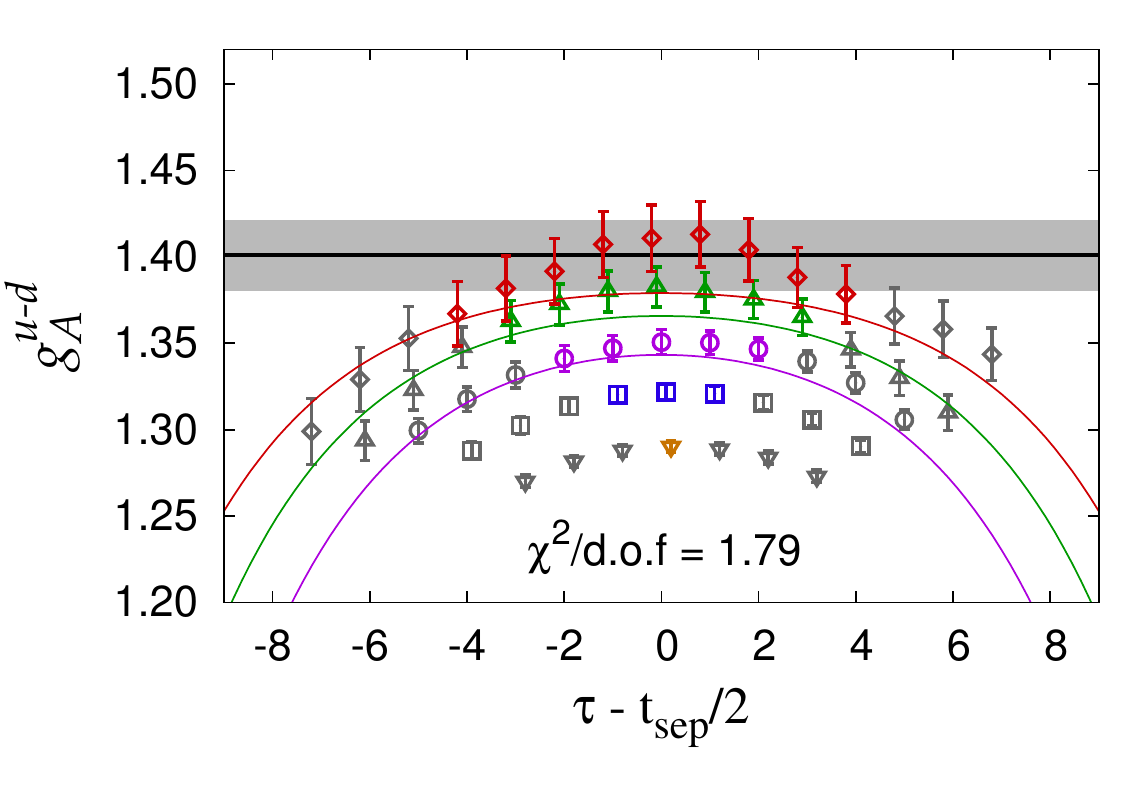}  \hspace{0.15in}
    \includegraphics[width=0.342\linewidth,trim={0.9cm  0.2cm 0 0},clip]{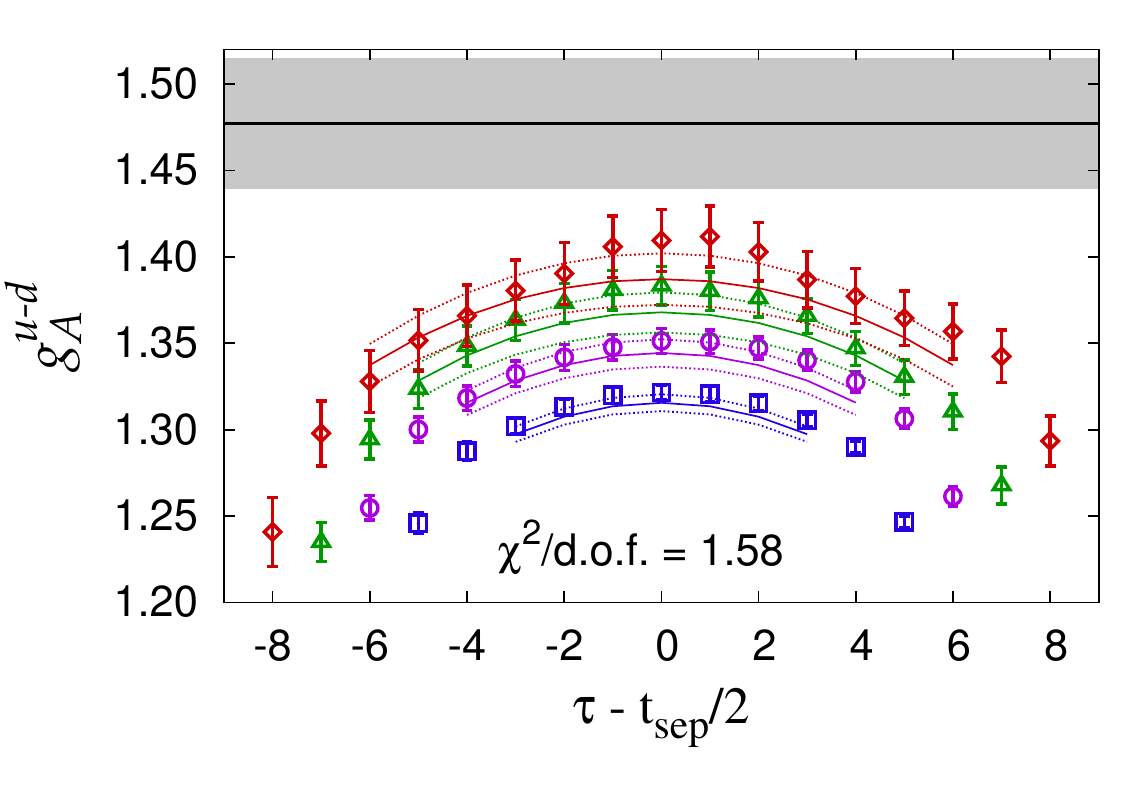}
    \includegraphics[width=0.08\linewidth,trim={0.0cm  0.0cm 0 0},clip]{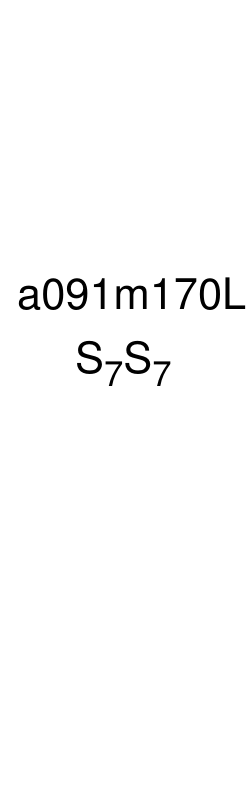}
  }
\vspace{-0.4cm}
\caption{Two- (left) and three-state (right) fits to $g_A^{u-d}$ from
  the 6 simulations on the 4 ensembles as described in the text.
  (Left) Data not included in the fits based on $\tau_{\rm skip}^{\rm best}$ 
  are shown in grey. (Right) Lines showing the fits are limited to points fit.  }
\label{fig:gA6}
\end{figure*}

\begin{figure*}[tb]
\centering
  \subfigure{
    \includegraphics[width=0.9\linewidth,trim={0      0.01cm 0 0},clip]{crop-legend_h}  
  }
\\
\vspace{-0.5cm}
  \subfigure{
    \includegraphics[width=0.371\linewidth,trim={0      0.2cm 0 0},clip]{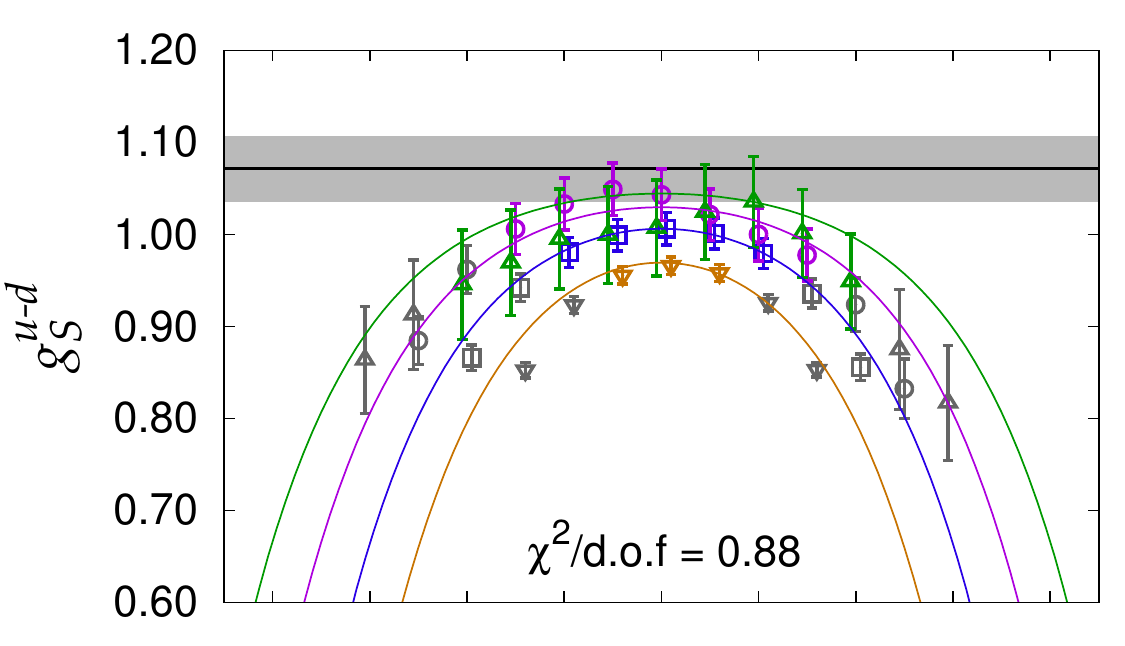}  \hspace{0.1in}
    \includegraphics[width=0.342\linewidth,trim={0.9cm  0.2cm 0 0},clip]{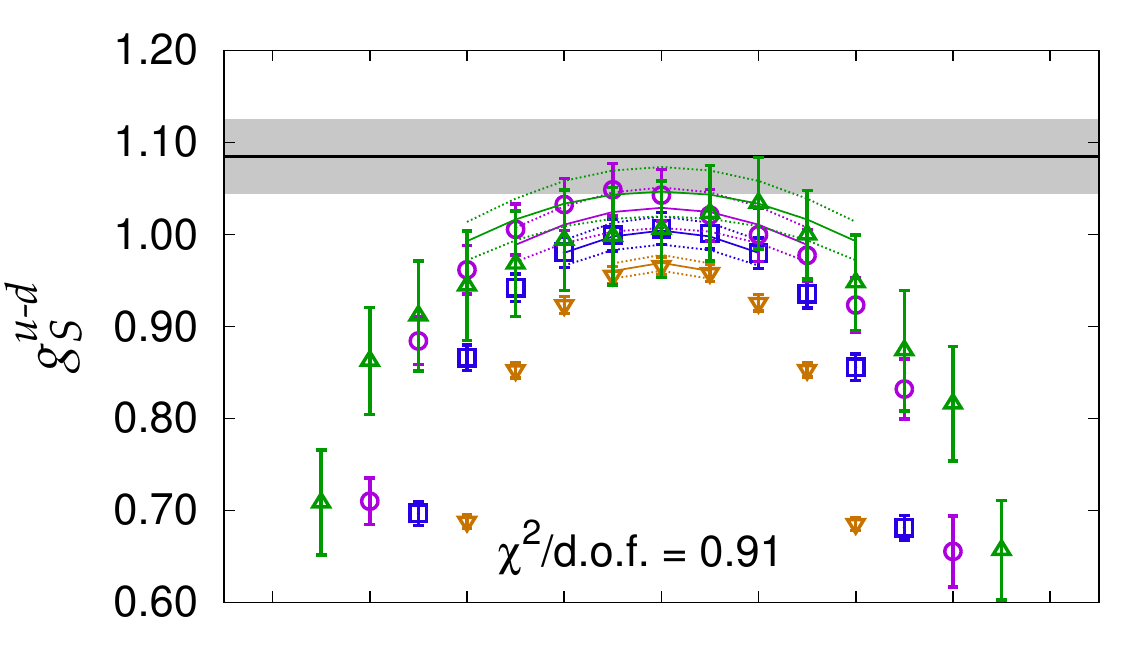}
    \includegraphics[width=0.08\linewidth,trim={0.0cm  0.0cm 0 0},clip]{labels/lab_C13}
  }
\\
\vspace{-0.4cm}
  \subfigure{
    \includegraphics[width=0.371\linewidth,trim={0      0.2cm 0 0},clip]{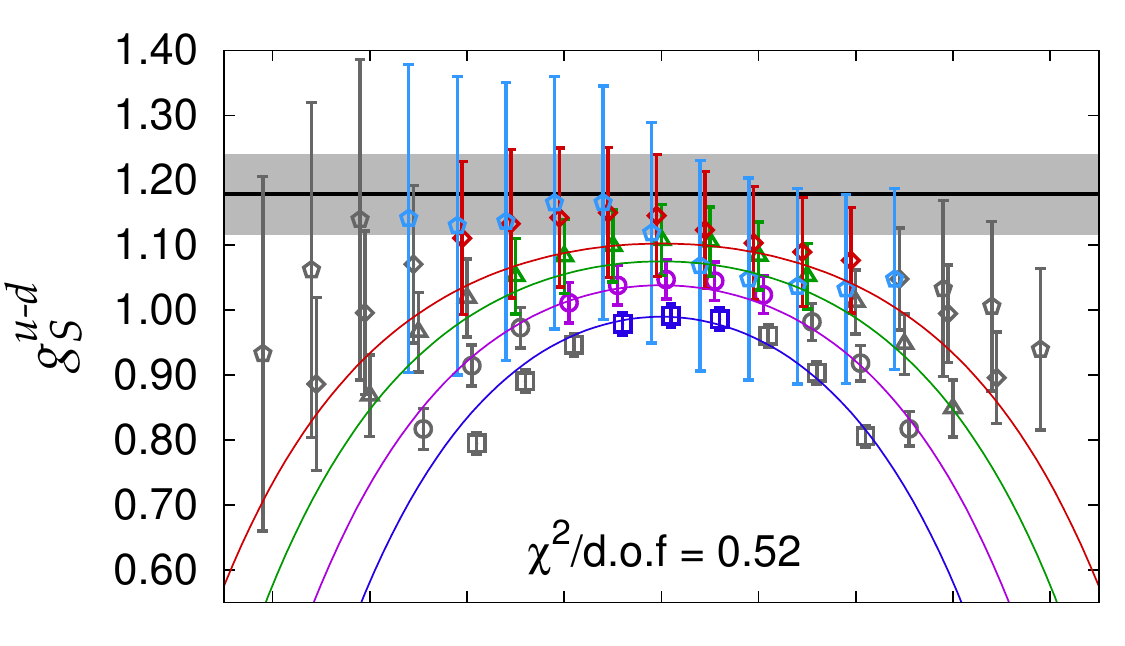}  \hspace{0.1in}
    \includegraphics[width=0.342\linewidth,trim={0.9cm  0.2cm 0 0},clip]{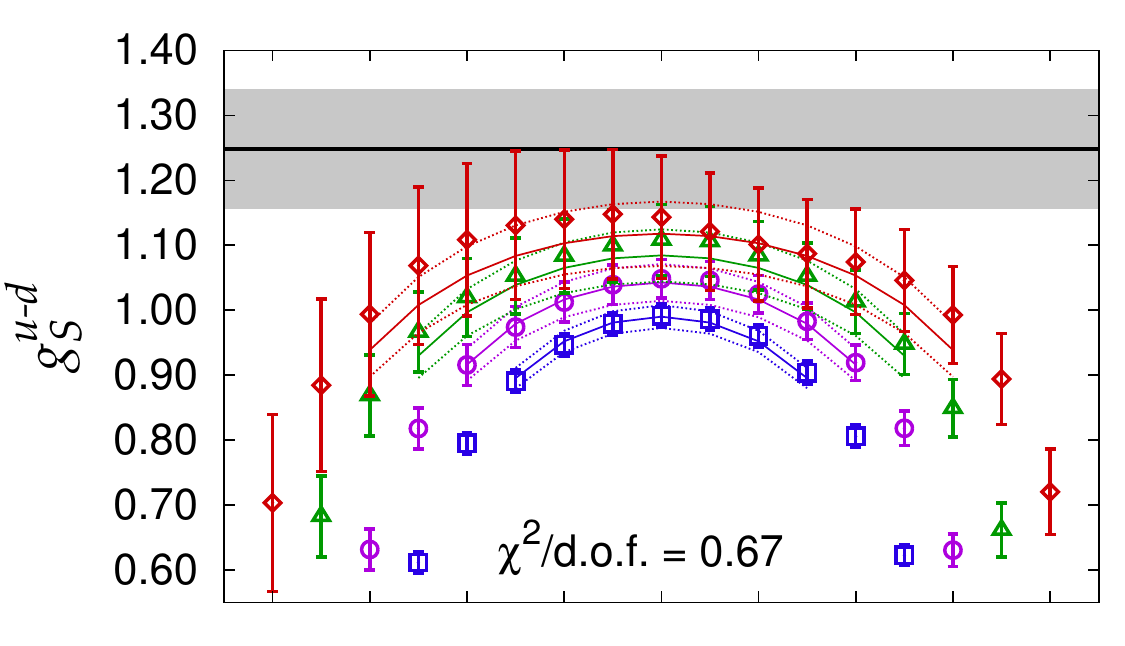}
    \includegraphics[width=0.08\linewidth,trim={0.0cm  0.0cm 0 0},clip]{labels/lab_D5_5}
  }
\\
\vspace{-0.4cm}
  \subfigure{
    \includegraphics[width=0.371\linewidth,trim={0      0.2cm 0 0},clip]{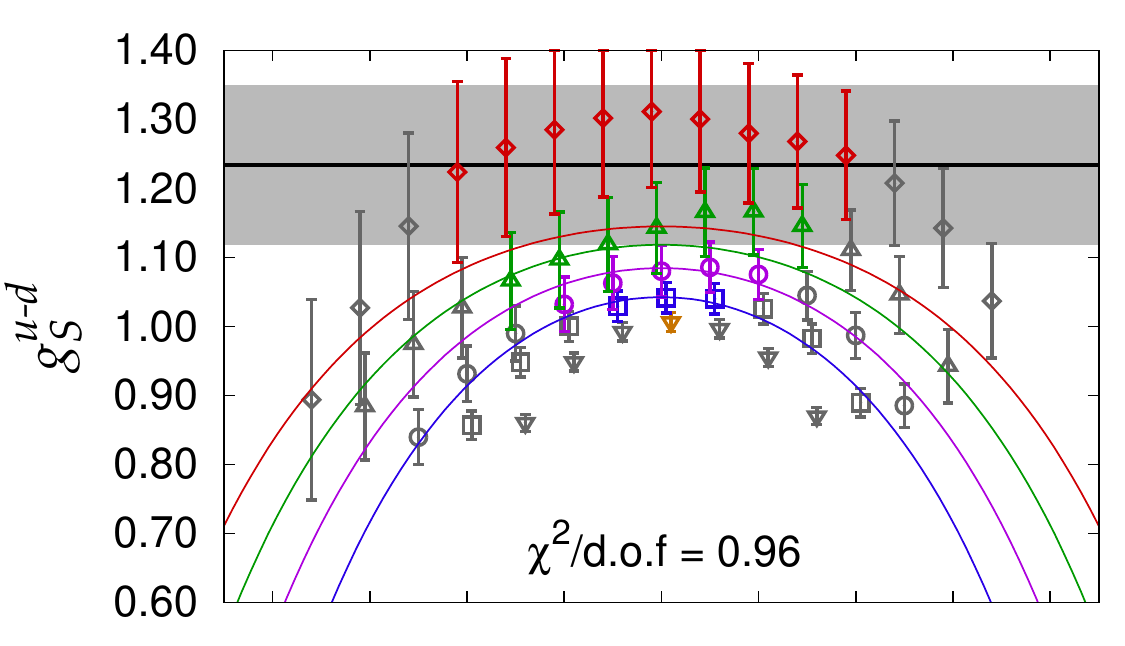}  \hspace{0.1in}
    \includegraphics[width=0.342\linewidth,trim={0.9cm  0.2cm 0 0},clip]{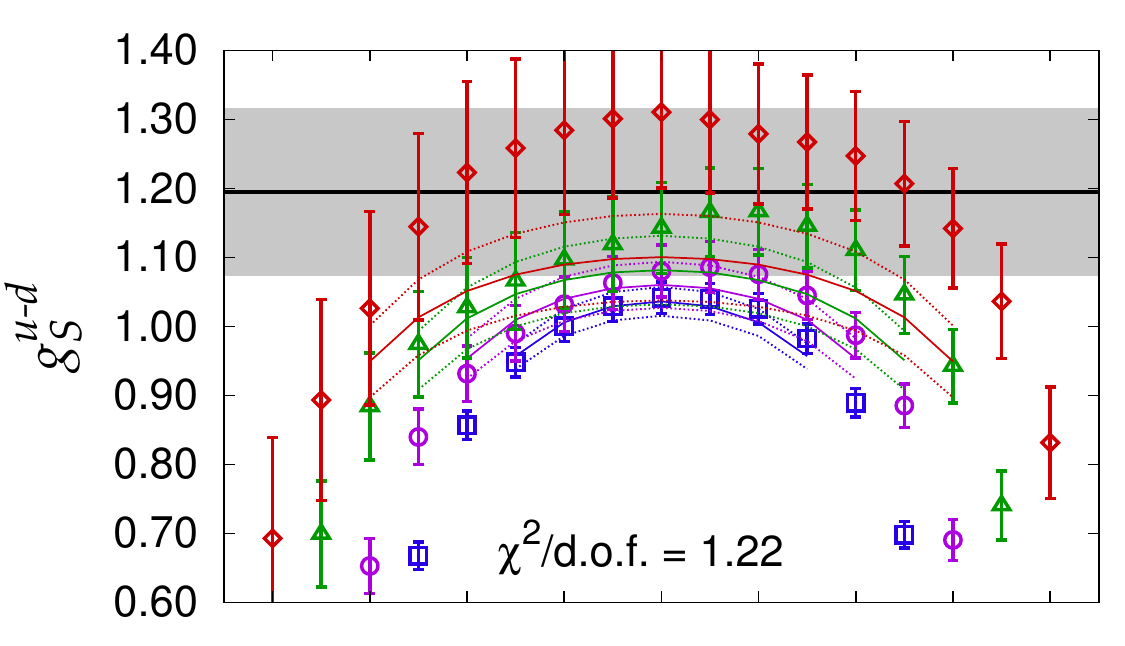}
    \includegraphics[width=0.08\linewidth,trim={0.0cm  0.0cm 0 0},clip]{labels/lab_D5_7}
  }
\\
\vspace{-0.4cm}
  \subfigure{
    \includegraphics[width=0.371\linewidth,trim={0      0.2cm 0 0},clip]{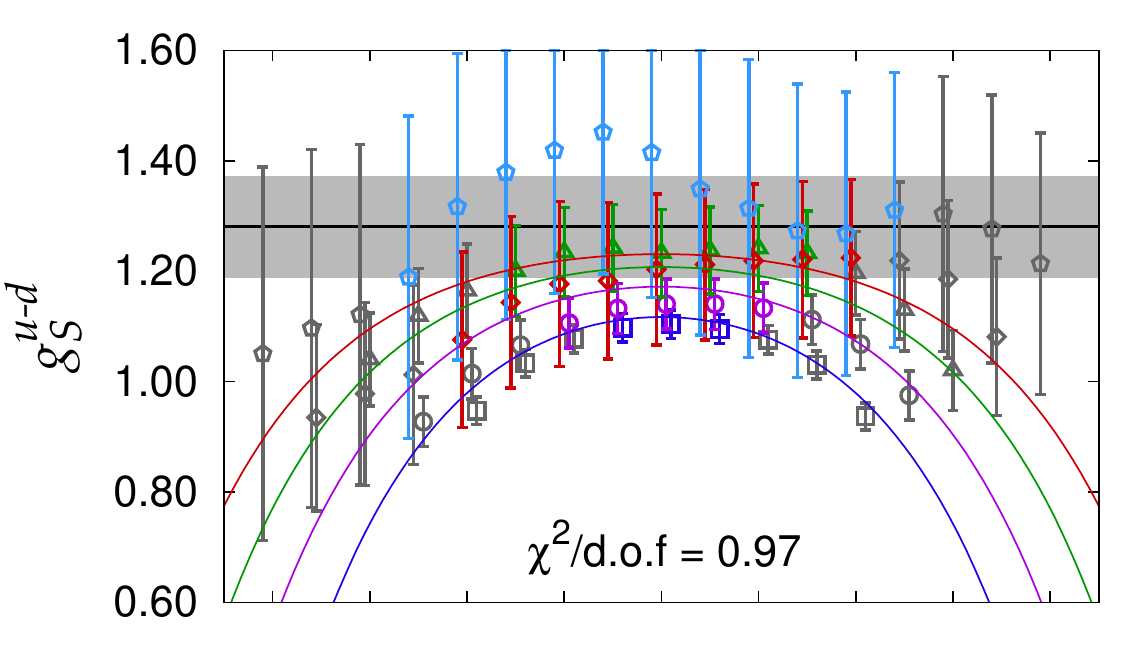}  \hspace{0.1in}
    \includegraphics[width=0.342\linewidth,trim={0.9cm  0.2cm 0 0},clip]{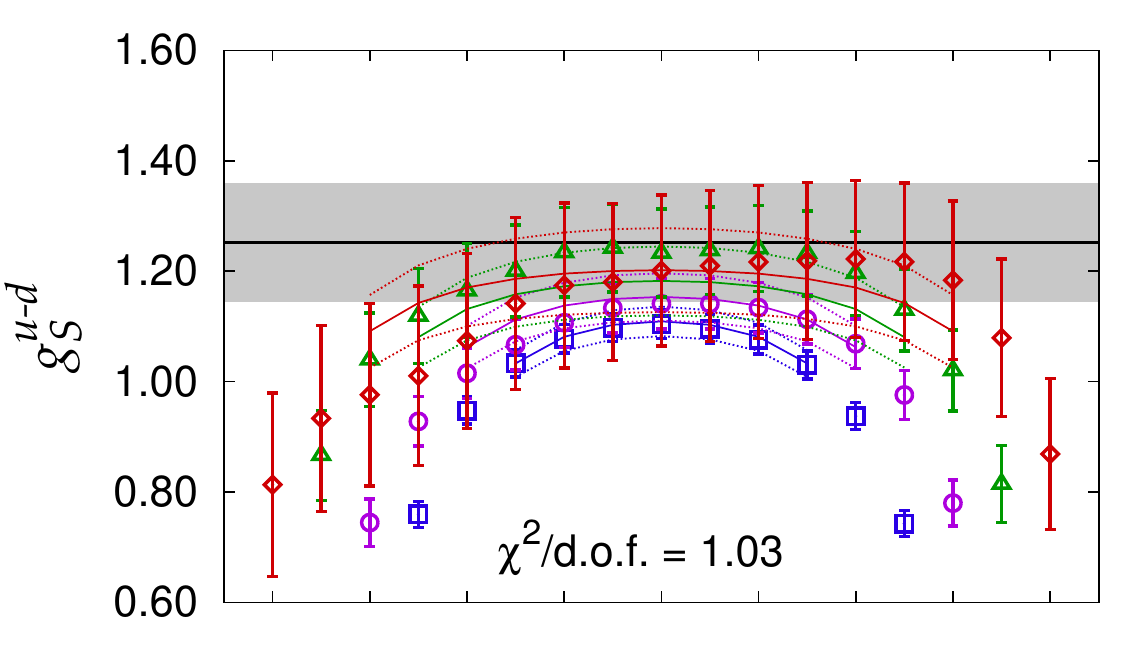}
    \includegraphics[width=0.08\linewidth,trim={0.0cm  0.0cm 0 0},clip]{labels/lab_D5_9}
  }
\\
\vspace{-0.4cm}
  \subfigure{
    \includegraphics[width=0.371\linewidth,trim={0      0.2cm 0 0},clip]{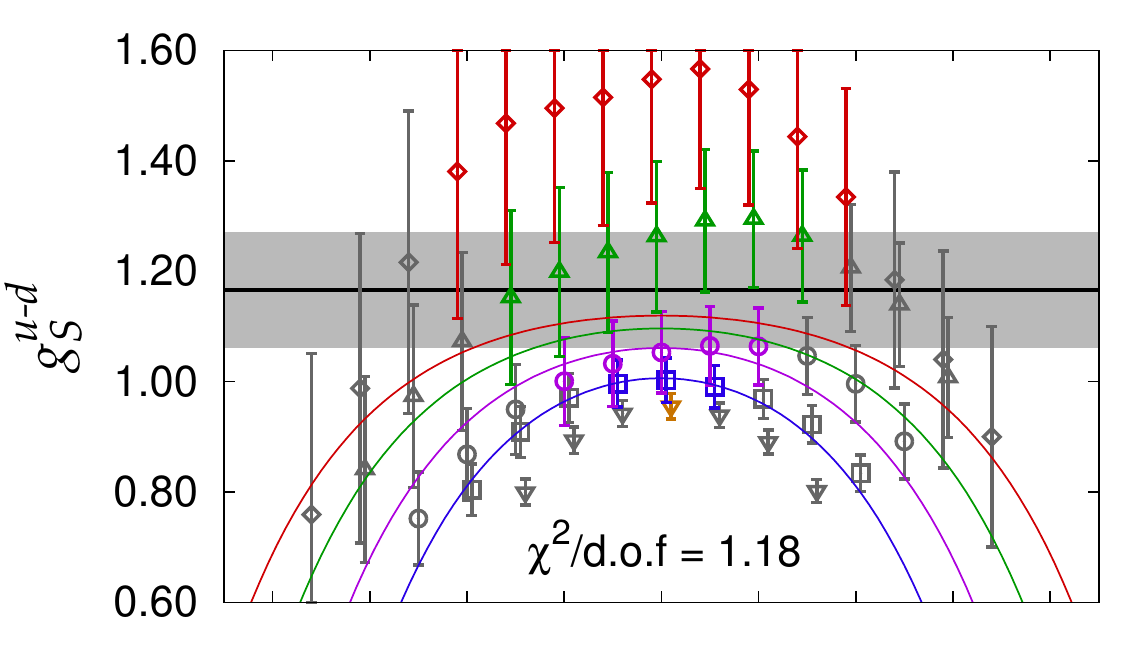}  \hspace{0.1in}
    \includegraphics[width=0.342\linewidth,trim={0.9cm  0.2cm 0 0},clip]{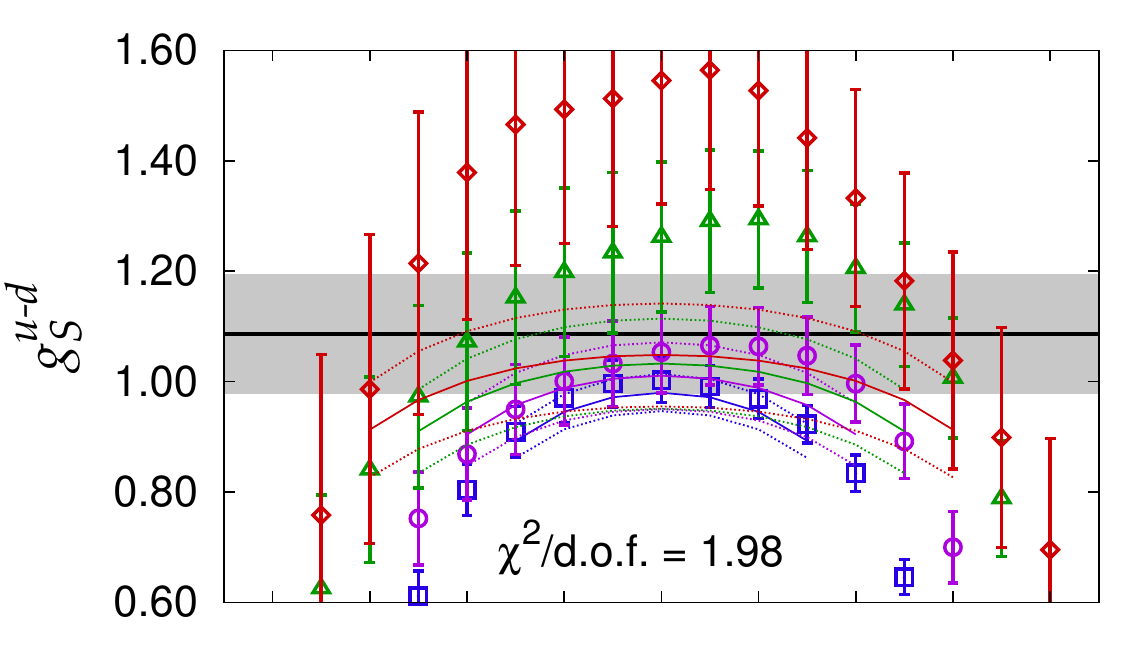}
    \includegraphics[width=0.08\linewidth,trim={0.0cm  0.0cm 0 0},clip]{labels/lab_D6}
  }
\\
\vspace{-0.4cm}
  \subfigure{
    \includegraphics[width=0.371\linewidth,trim={0      0.2cm 0 0},clip]{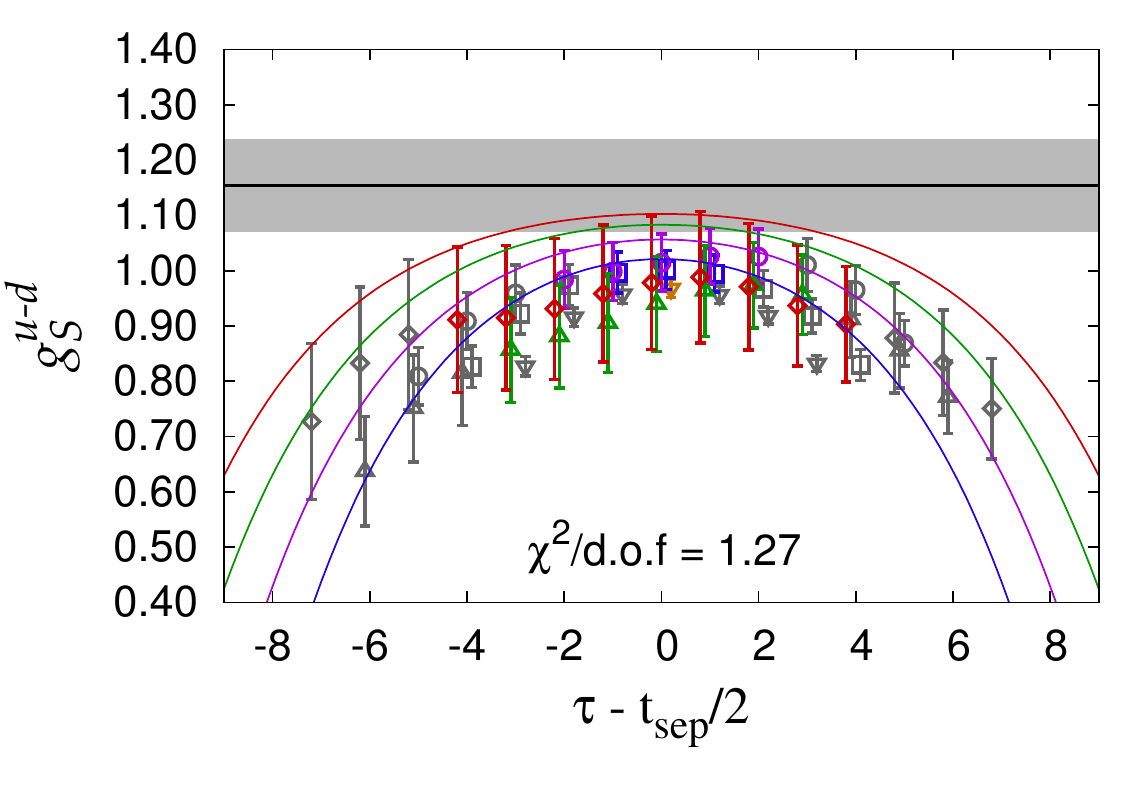}  \hspace{0.1in}
    \includegraphics[width=0.342\linewidth,trim={0.9cm  0.2cm 0 0},clip]{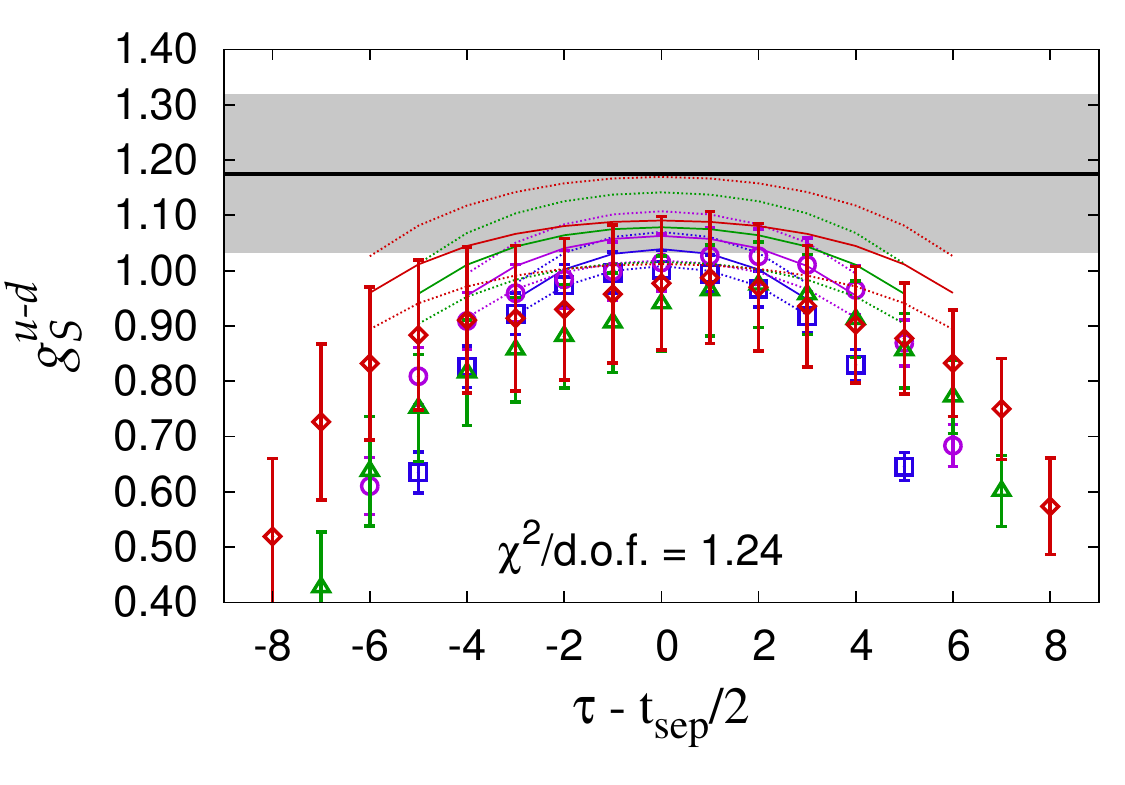}
    \includegraphics[width=0.08\linewidth,trim={0.0cm  0.0cm 0 0},clip]{labels/lab_D7}
  }
\vspace{-0.4cm}
\caption{Two- (left) and three-state (right) fits to $g_S^{u-d}$
   from the 6 simulations on the 4 ensembles as described in the text. 
  (Left) Data not included in the fits based on $\tau_{\rm skip}^{\rm best}$ 
  are shown in grey. (Right) Lines showing the fits are limited to points fit.  }
\label{fig:gS6}
\end{figure*}

\begin{figure*}[tb]
\centering
  \subfigure{
    \includegraphics[width=0.9\linewidth,trim={0      0.01cm 0 0},clip]{crop-legend_h}  
  }
\\
\vspace{-0.5cm}
  \subfigure{
    \includegraphics[width=0.371\linewidth,trim={0      0.2cm 0 0},clip]{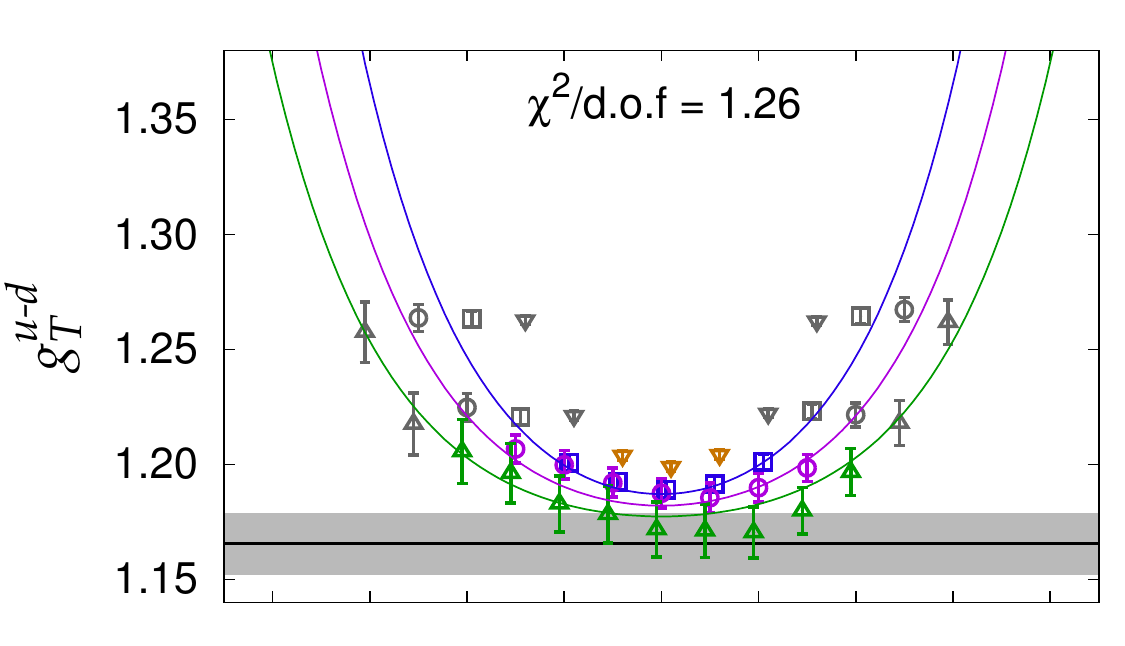}  \hspace{0.1in}
    \includegraphics[width=0.342\linewidth,trim={0.9cm  0.2cm 0 0},clip]{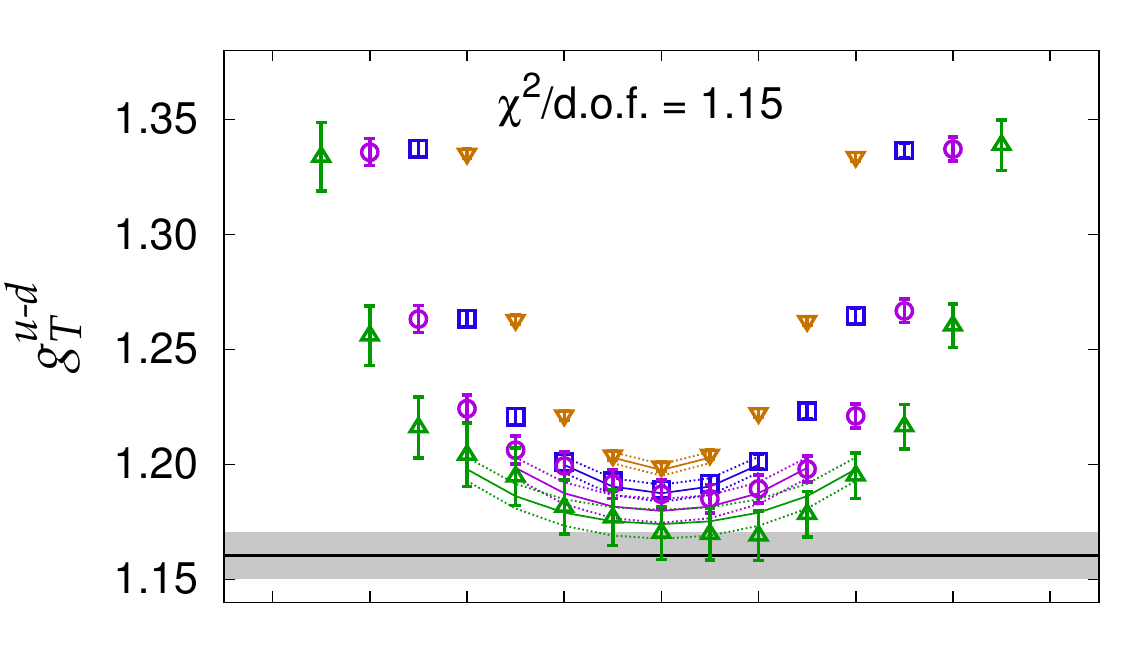}
    \includegraphics[width=0.08\linewidth,trim={0.0cm  0.0cm 0 0},clip]{labels/lab_C13}
  }
\\
\vspace{-0.4cm}
  \subfigure{
    \includegraphics[width=0.371\linewidth,trim={0      0.2cm 0 0},clip]{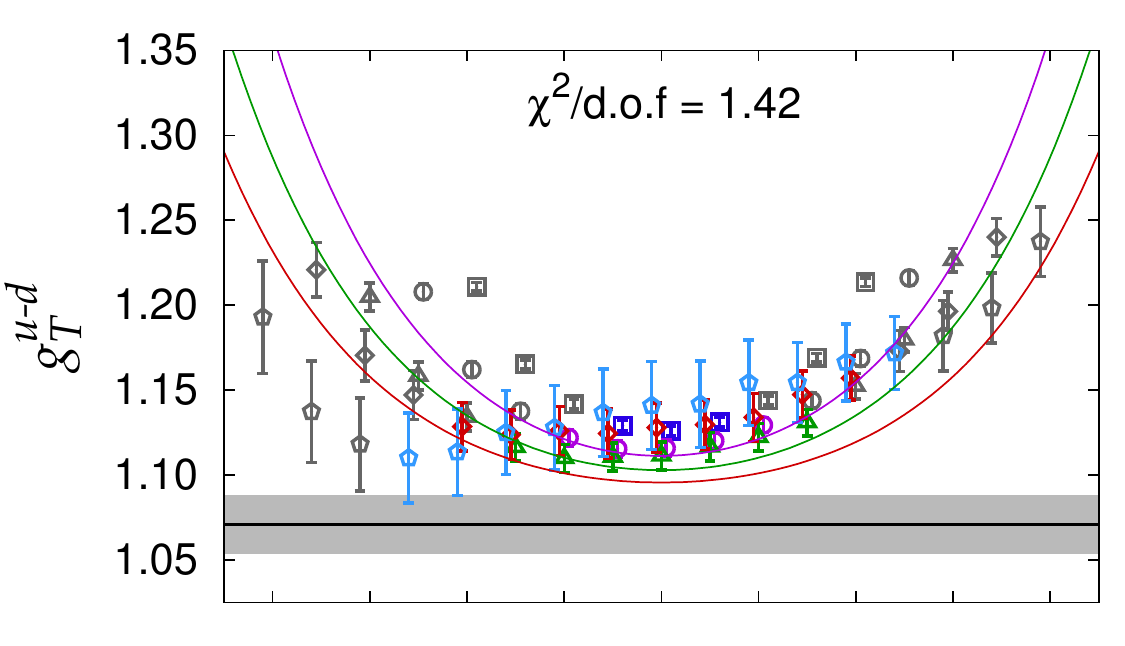}  \hspace{0.1in}
    \includegraphics[width=0.342\linewidth,trim={0.9cm  0.2cm 0 0},clip]{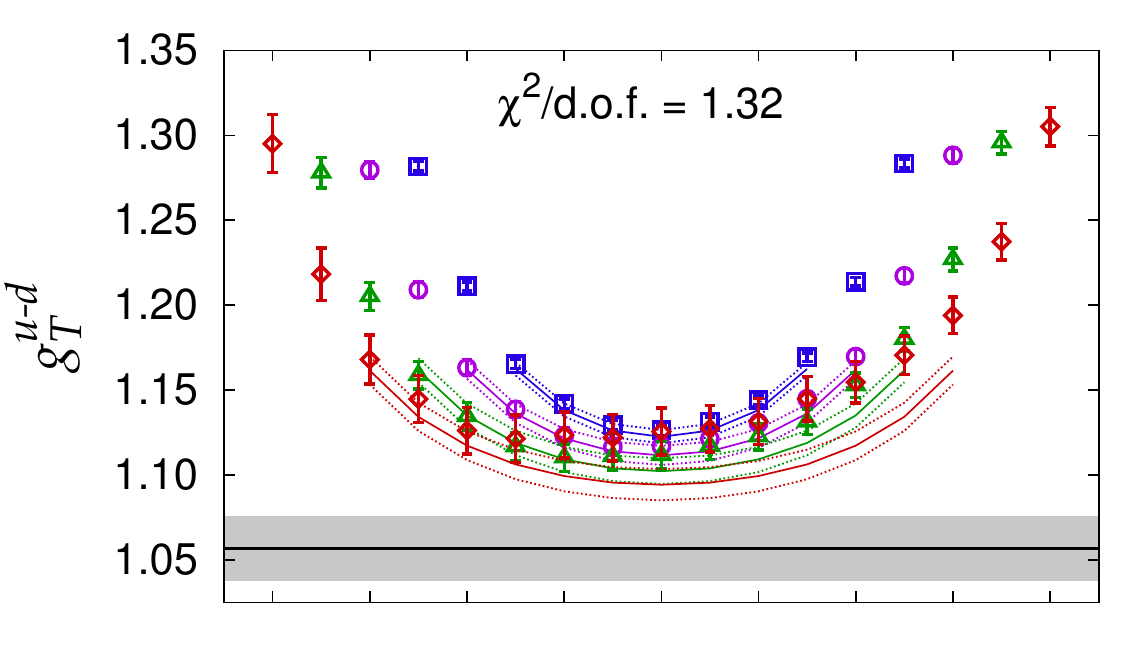}
    \includegraphics[width=0.08\linewidth,trim={0.0cm  0.0cm 0 0},clip]{labels/lab_D5_5}
  }
\\
\vspace{-0.4cm}
  \subfigure{
    \includegraphics[width=0.371\linewidth,trim={0      0.2cm 0 0},clip]{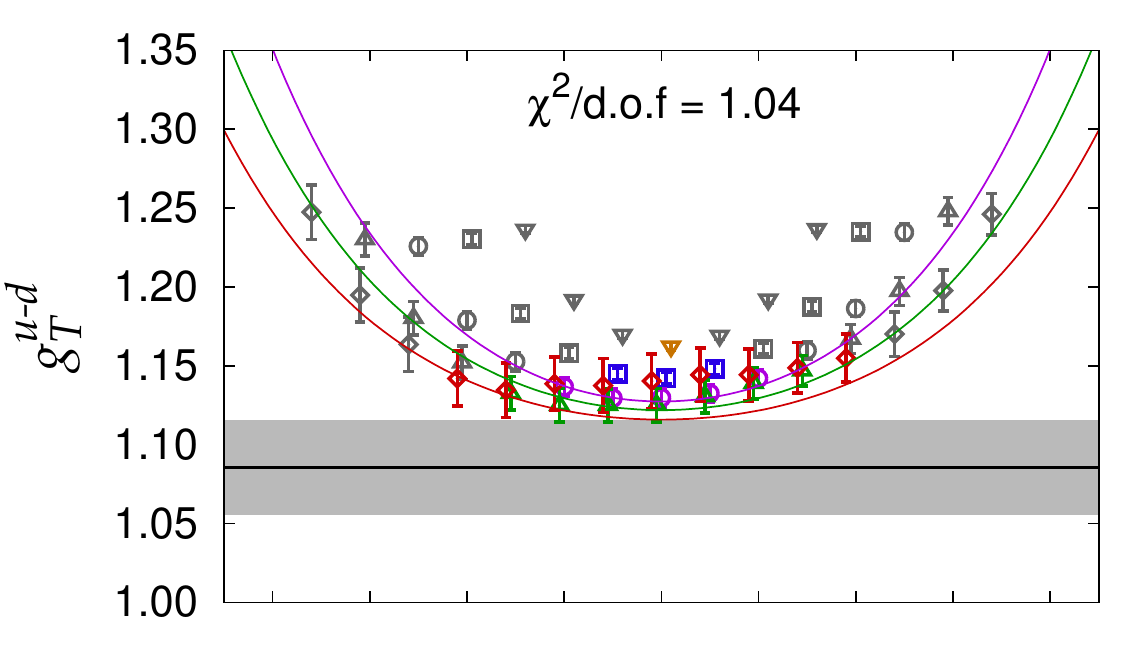}  \hspace{0.1in}
    \includegraphics[width=0.342\linewidth,trim={0.9cm  0.2cm 0 0},clip]{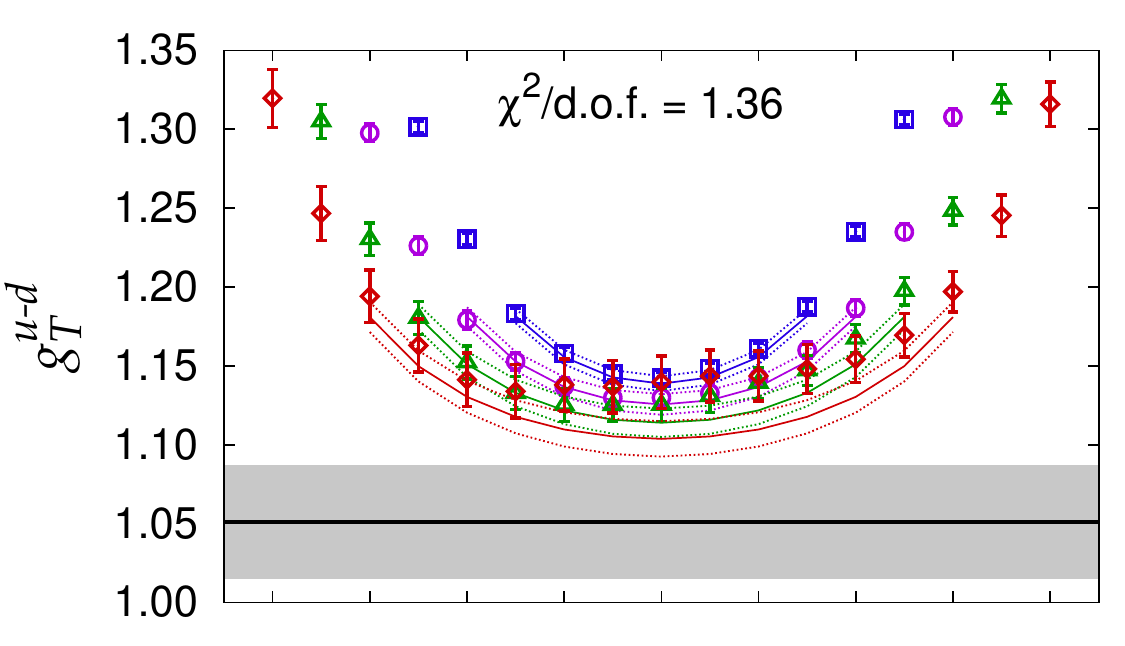}
    \includegraphics[width=0.08\linewidth,trim={0.0cm  0.0cm 0 0},clip]{labels/lab_D5_7}
  }
\\
\vspace{-0.4cm}
  \subfigure{
    \includegraphics[width=0.371\linewidth,trim={0      0.2cm 0 0},clip]{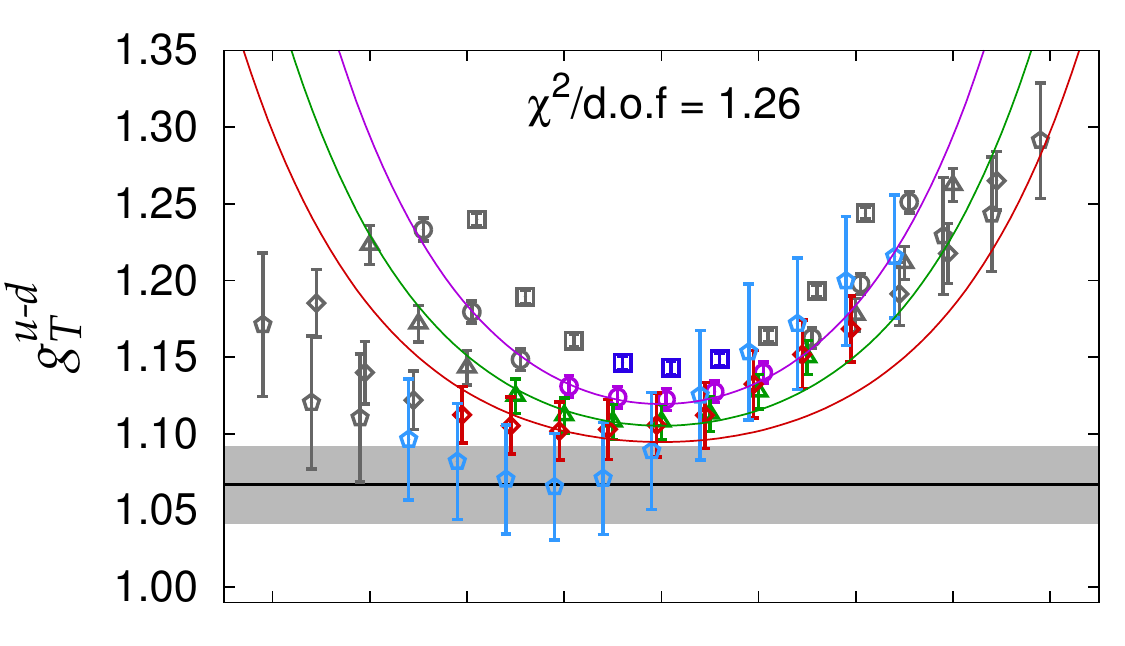}  \hspace{0.1in}
    \includegraphics[width=0.342\linewidth,trim={0.9cm  0.2cm 0 0},clip]{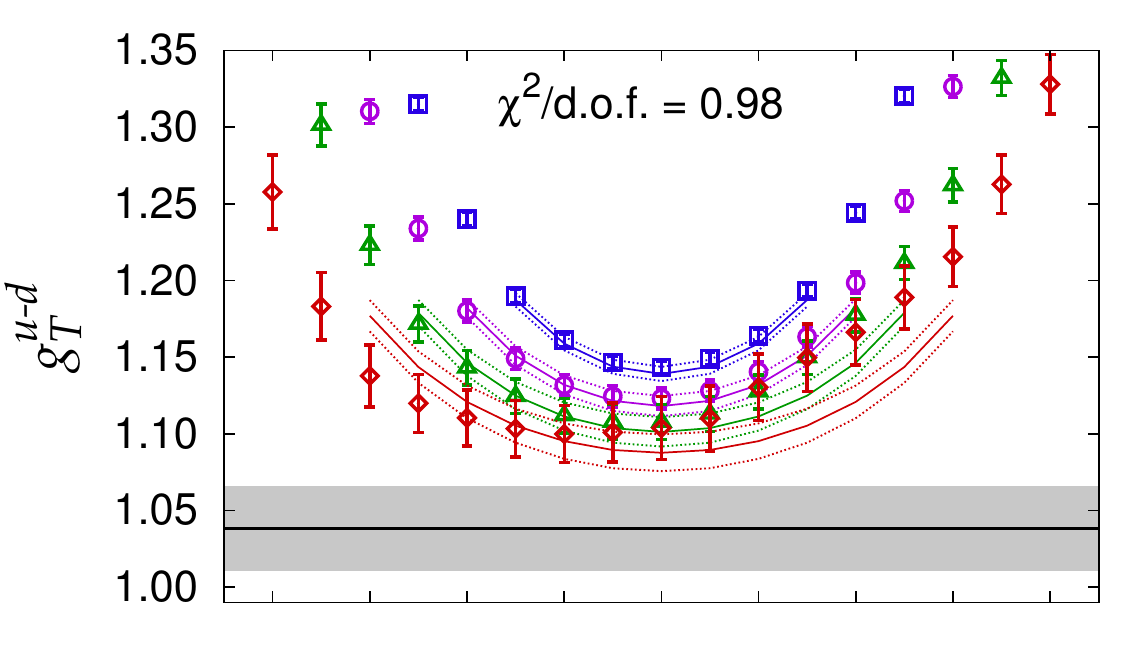}
    \includegraphics[width=0.08\linewidth,trim={0.0cm  0.0cm 0 0},clip]{labels/lab_D5_9}
  }
\\
\vspace{-0.4cm}
  \subfigure{
    \includegraphics[width=0.371\linewidth,trim={0      0.2cm 0 0},clip]{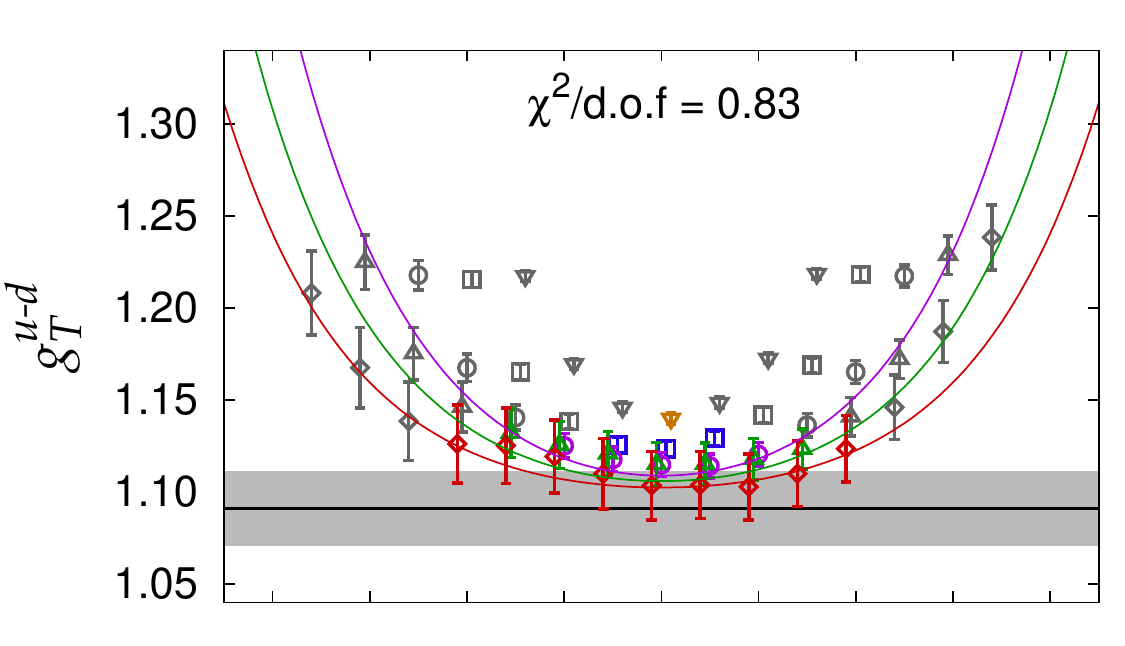}  \hspace{0.1in}
    \includegraphics[width=0.342\linewidth,trim={0.9cm  0.2cm 0 0},clip]{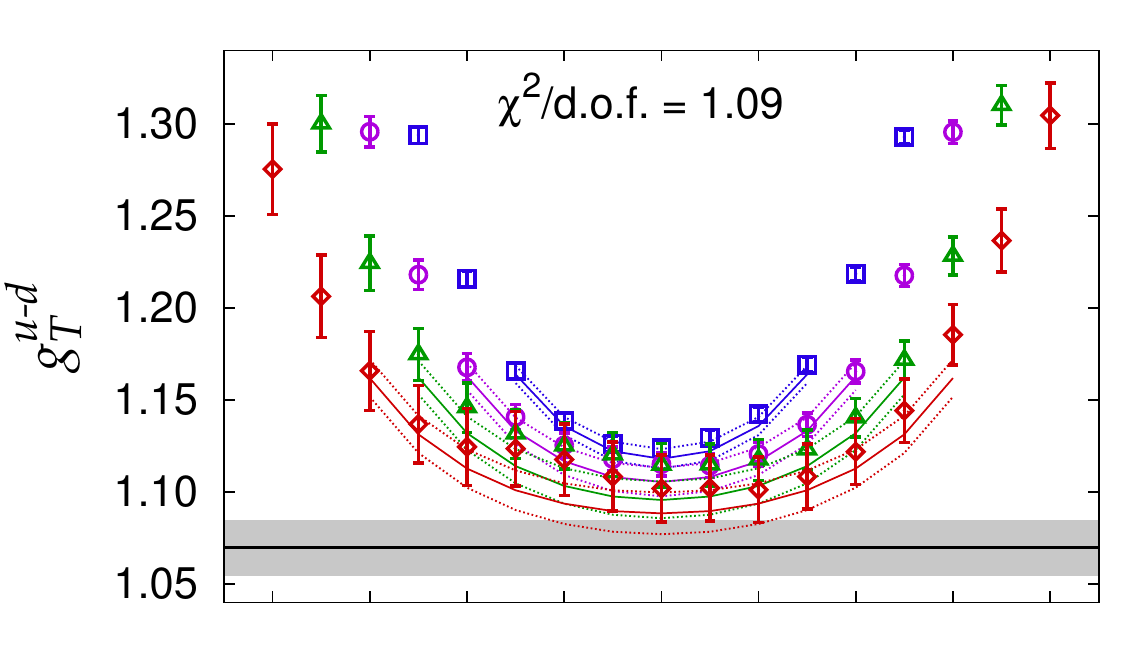}
    \includegraphics[width=0.08\linewidth,trim={0.0cm  0.0cm 0 0},clip]{labels/lab_D6}
  }
\\
\vspace{-0.4cm}
  \subfigure{
    \includegraphics[width=0.371\linewidth,trim={0      0.2cm 0 0},clip]{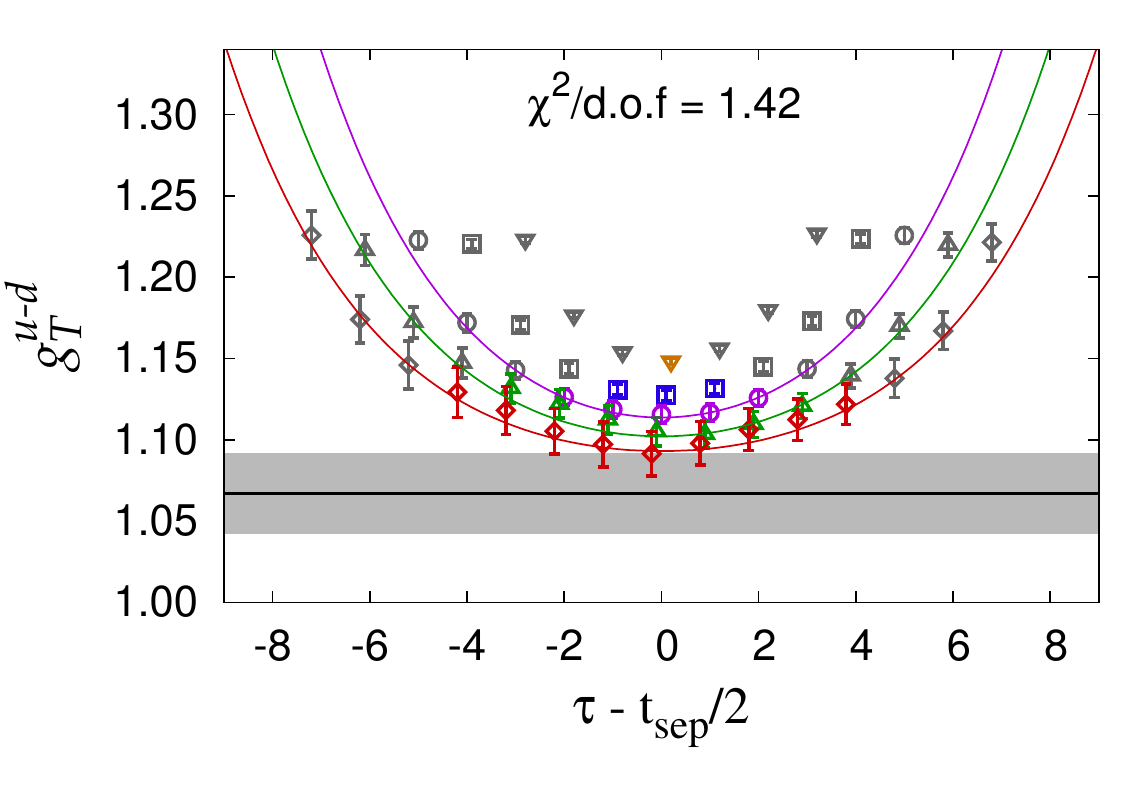}  \hspace{0.1in}
    \includegraphics[width=0.342\linewidth,trim={0.9cm  0.2cm 0 0},clip]{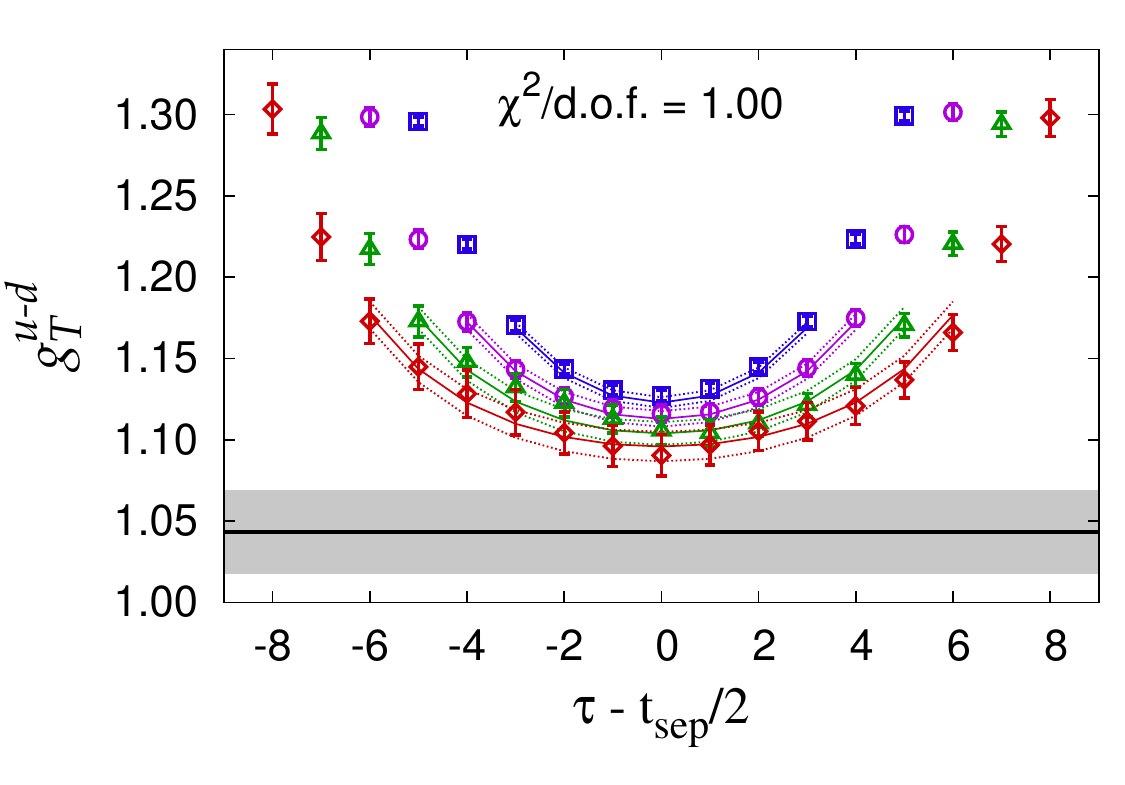}
    \includegraphics[width=0.08\linewidth,trim={0.0cm  0.0cm 0 0},clip]{labels/lab_D7}
  }
\vspace{-0.4cm}
\caption{Two- (left) and three-state (right) fits to $g_T^{u-d}$ 
   from the 6 simulations on the 4 ensembles as described in the text. 
  (Left) Data not included in the fits based on $\tau_{\rm skip}^{\rm best}$ 
  are shown in grey. (Right) Lines showing the fits are limited to points fit.  }
\label{fig:gT6}
\end{figure*}

\begin{figure*}[tb]
\centering
  \subfigure{
    \includegraphics[width=0.9\linewidth,trim={0      0.01cm 0 0},clip]{crop-legend_h}  
  }
\\
\vspace{-0.5cm}
  \subfigure{
    \includegraphics[width=0.371\linewidth,trim={0      0.2cm 0 0},clip]{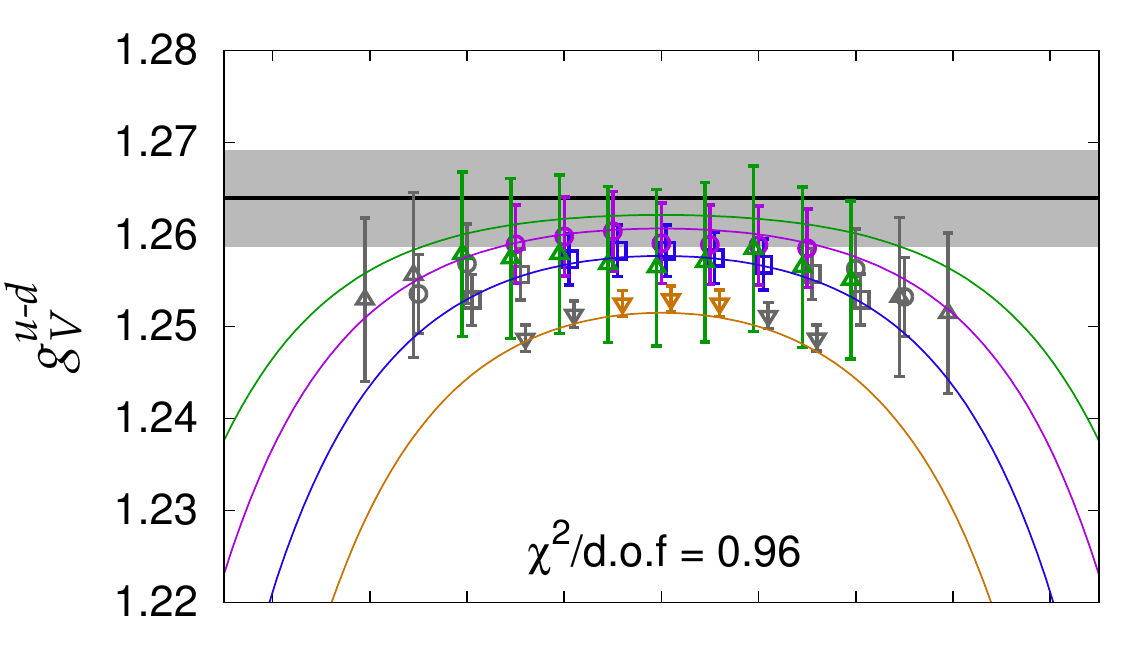}  \hspace{0.1in}
    \includegraphics[width=0.342\linewidth,trim={0.9cm  0.2cm 0 0},clip]{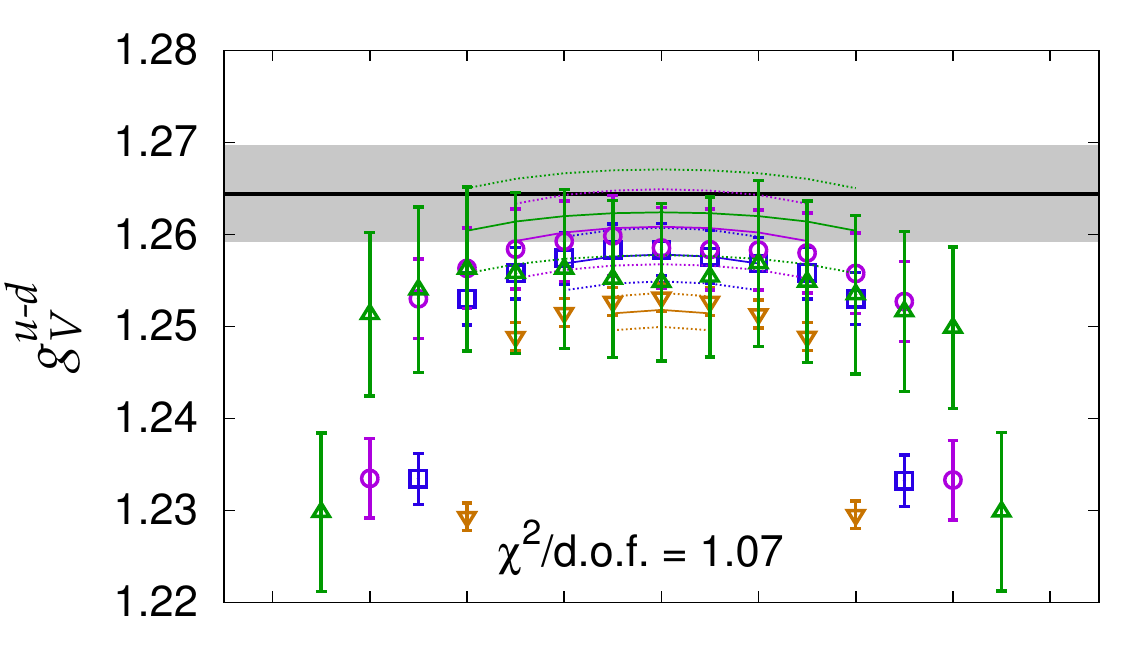}
    \includegraphics[width=0.08\linewidth,trim={0.0cm  0.0cm 0 0},clip]{labels/lab_C13}
  }
\\
\vspace{-0.4cm}
  \subfigure{
    \includegraphics[width=0.371\linewidth,trim={0      0.2cm 0 0},clip]{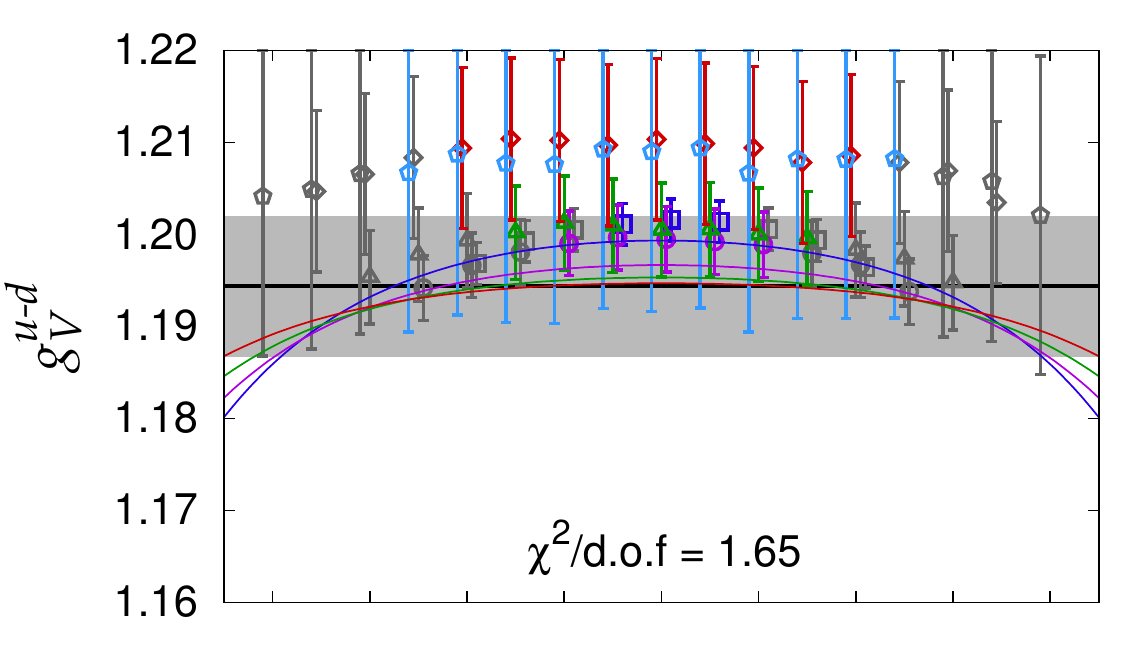}  \hspace{0.1in}
    \includegraphics[width=0.342\linewidth,trim={0.9cm  0.2cm 0 0},clip]{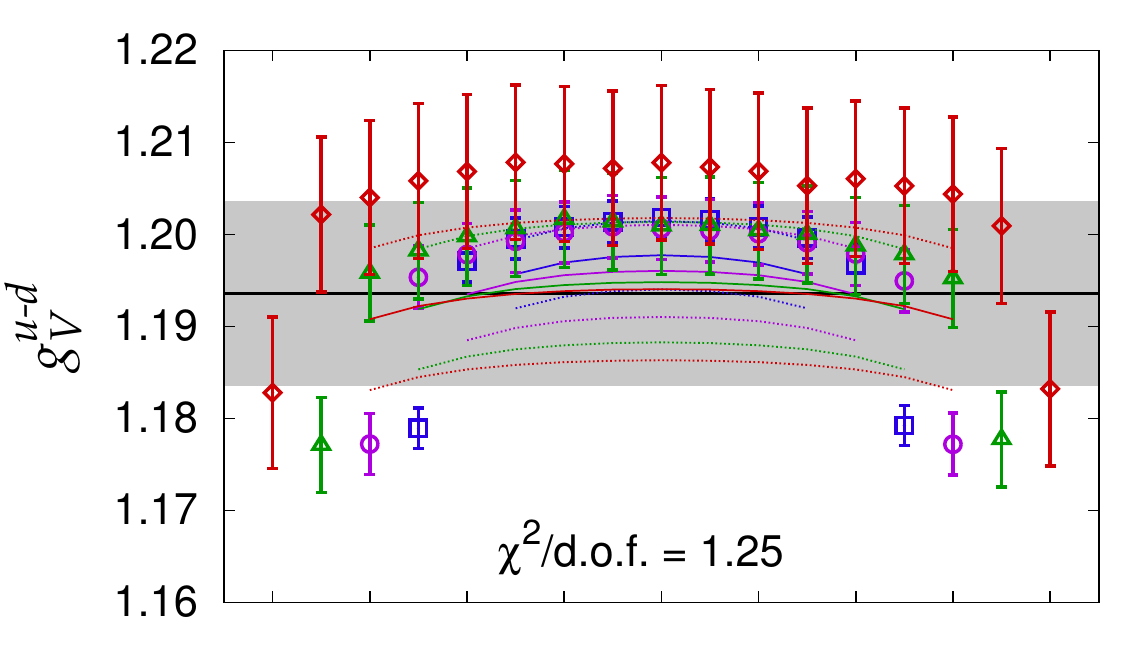}
    \includegraphics[width=0.08\linewidth,trim={0.0cm  0.0cm 0 0},clip]{labels/lab_D5_5}
  }
\\
\vspace{-0.4cm}
  \subfigure{
    \includegraphics[width=0.371\linewidth,trim={0      0.2cm 0 0},clip]{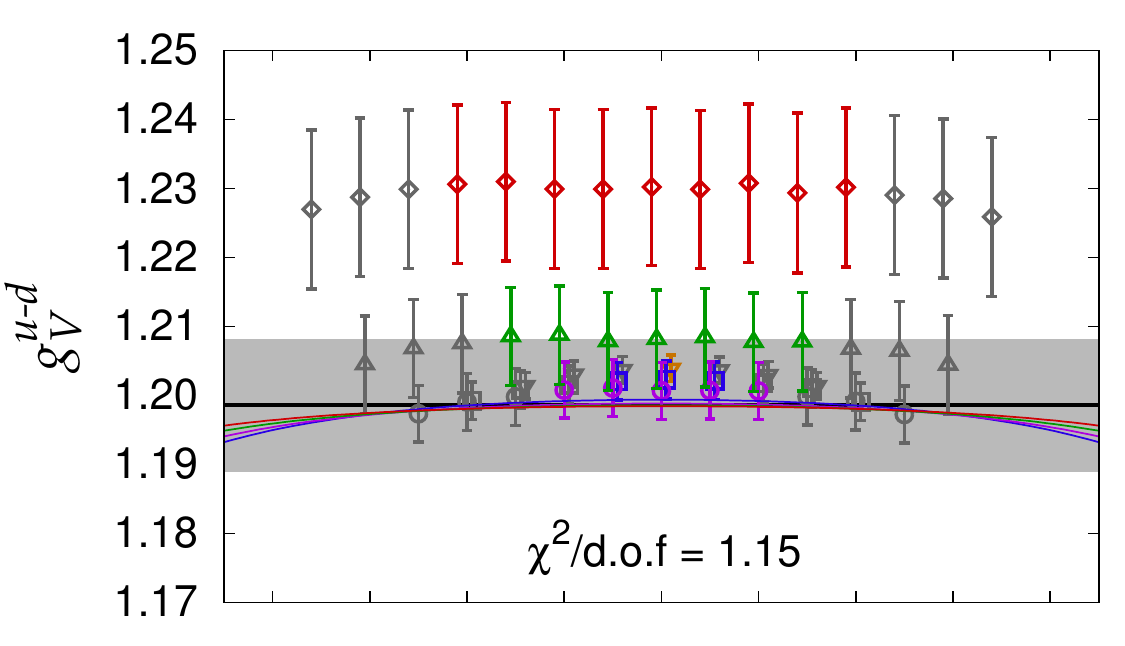}  \hspace{0.1in}
    \includegraphics[width=0.342\linewidth,trim={0.9cm  0.2cm 0 0},clip]{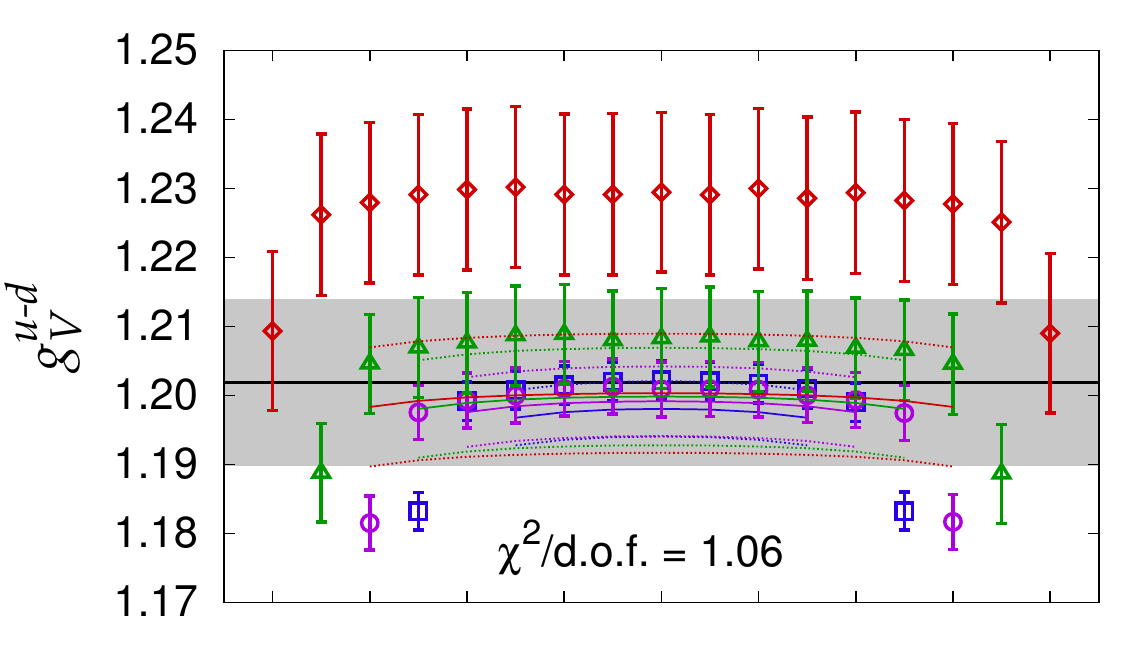}
    \includegraphics[width=0.08\linewidth,trim={0.0cm  0.0cm 0 0},clip]{labels/lab_D5_7}
  }
\\
\vspace{-0.4cm}
  \subfigure{
    \includegraphics[width=0.371\linewidth,trim={0      0.2cm 0 0},clip]{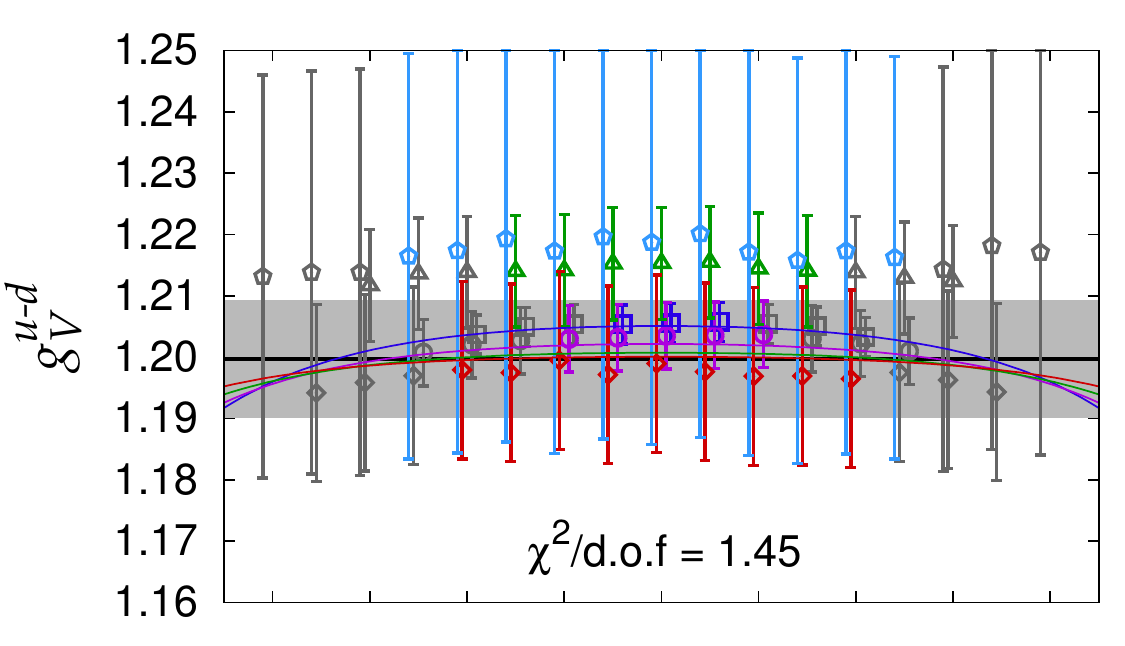}  \hspace{0.1in}
    \includegraphics[width=0.342\linewidth,trim={0.9cm  0.2cm 0 0},clip]{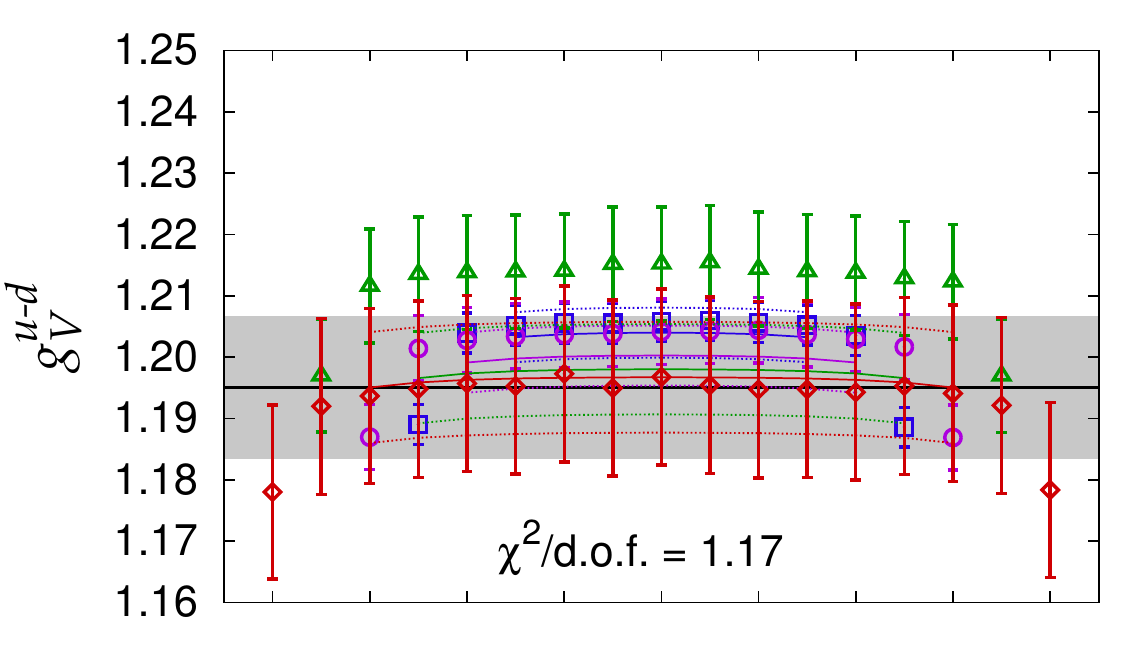}
    \includegraphics[width=0.08\linewidth,trim={0.0cm  0.0cm 0 0},clip]{labels/lab_D5_9}
  }
\\
\vspace{-0.4cm}
  \subfigure{
    \includegraphics[width=0.371\linewidth,trim={0      0.2cm 0 0},clip]{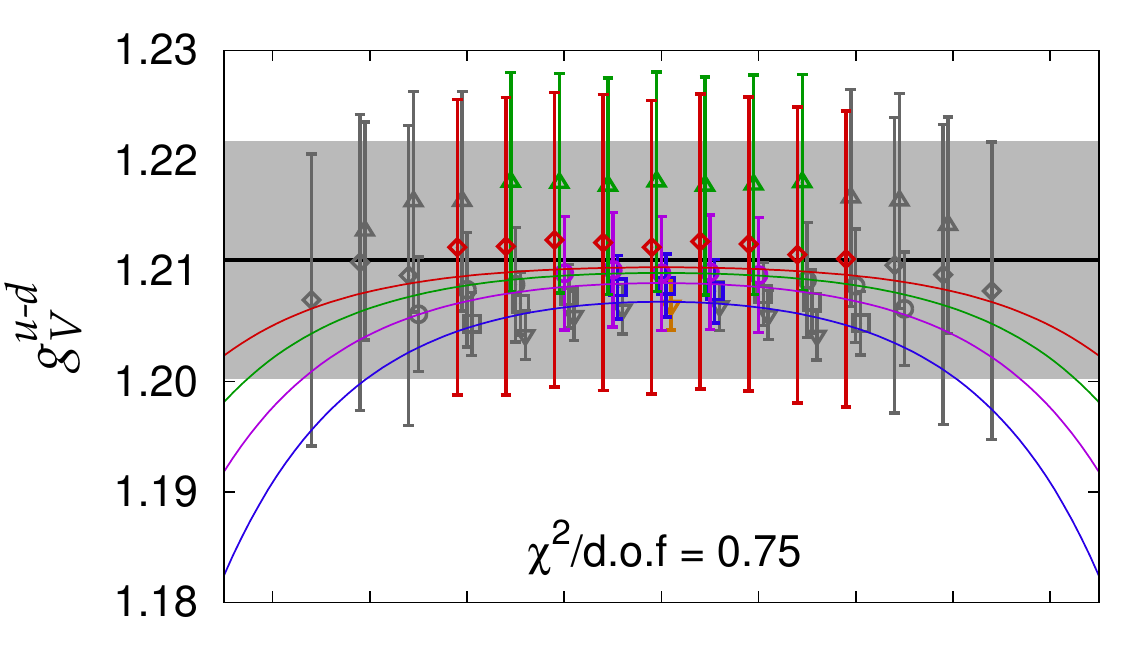}  \hspace{0.1in}
    \includegraphics[width=0.342\linewidth,trim={0.9cm  0.2cm 0 0},clip]{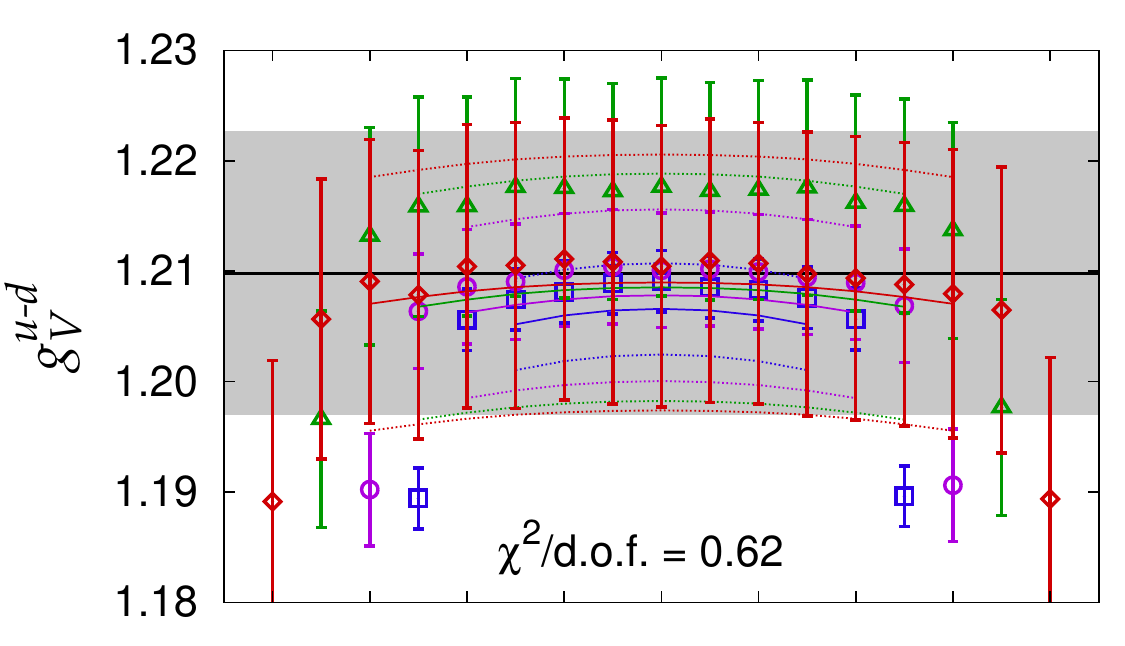}
    \includegraphics[width=0.08\linewidth,trim={0.0cm  0.0cm 0 0},clip]{labels/lab_D6}
  }
\\
\vspace{-0.4cm}
  \subfigure{
    \includegraphics[width=0.371\linewidth,trim={0      0.2cm 0 0},clip]{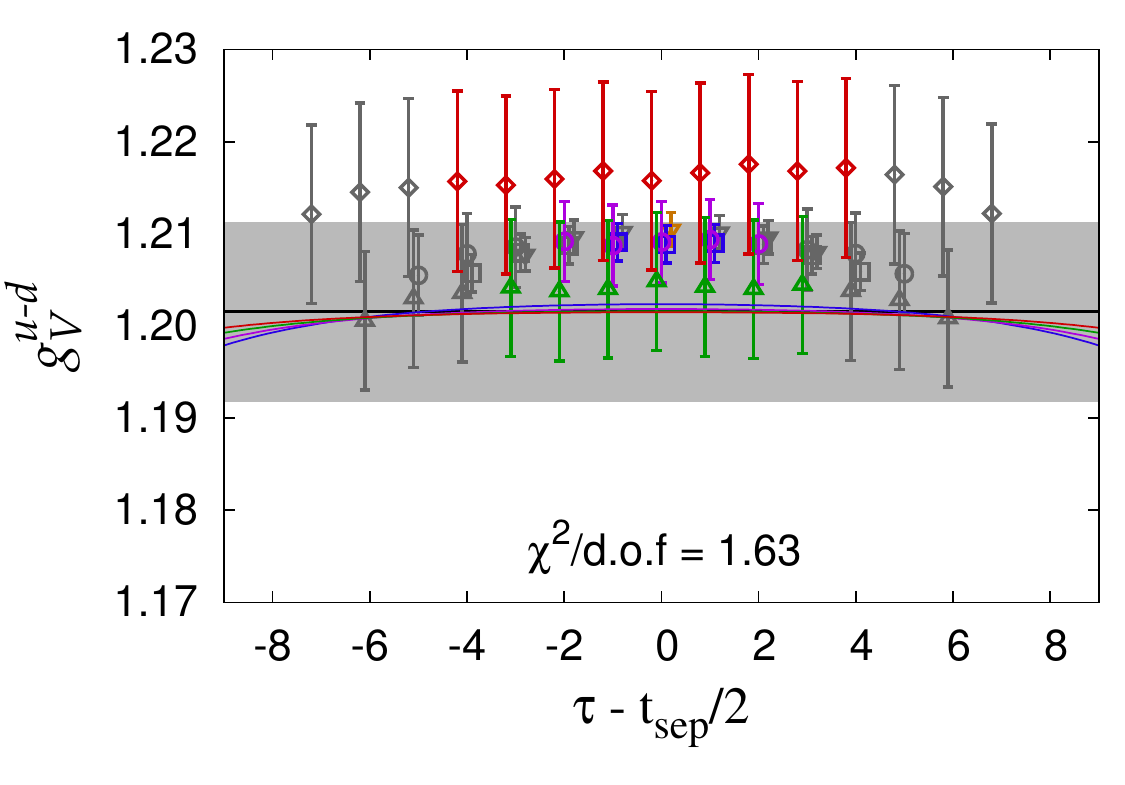}  \hspace{0.1in}
    \includegraphics[width=0.342\linewidth,trim={0.9cm  0.2cm 0 0},clip]{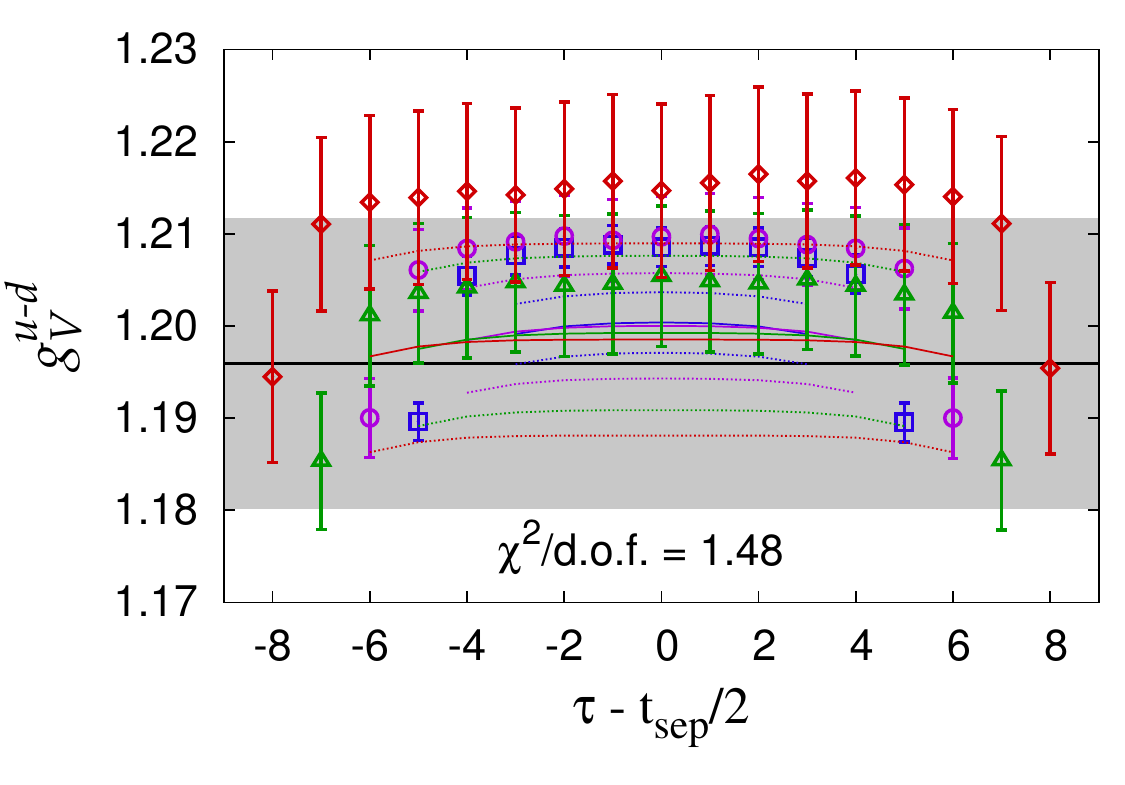}
    \includegraphics[width=0.08\linewidth,trim={0.0cm  0.0cm 0 0},clip]{labels/lab_D7}
  }
\vspace{-0.4cm}
\caption{Two- (left) and three-state (right) fits to $g_V^{u-d}$ 
   from the 6 simulations on the 4 ensembles as described in the text. 
  (Left) Data not included in the fits based on $\tau_{\rm skip}^{\rm best}$ 
  are shown in grey. (Right) Lines showing the fits are limited to points fit.  }
\label{fig:gV6}
\end{figure*}

\begin{figure*}[tb]
\centering
  \subfigure{
    \includegraphics[width=0.85\linewidth,trim={0      0.01cm 0 0},clip]{crop-legend_h}  
  }
\\
\vspace{-0.5cm}
  \subfigure{
    \includegraphics[width=0.33\linewidth]{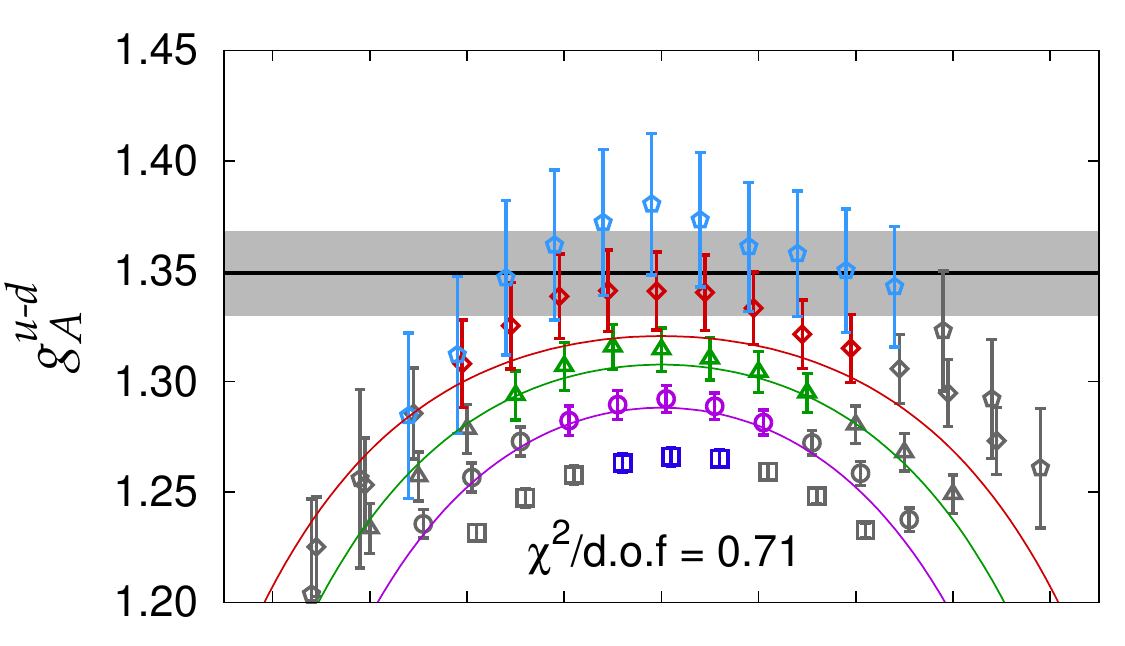}
    \includegraphics[width=0.33\linewidth]{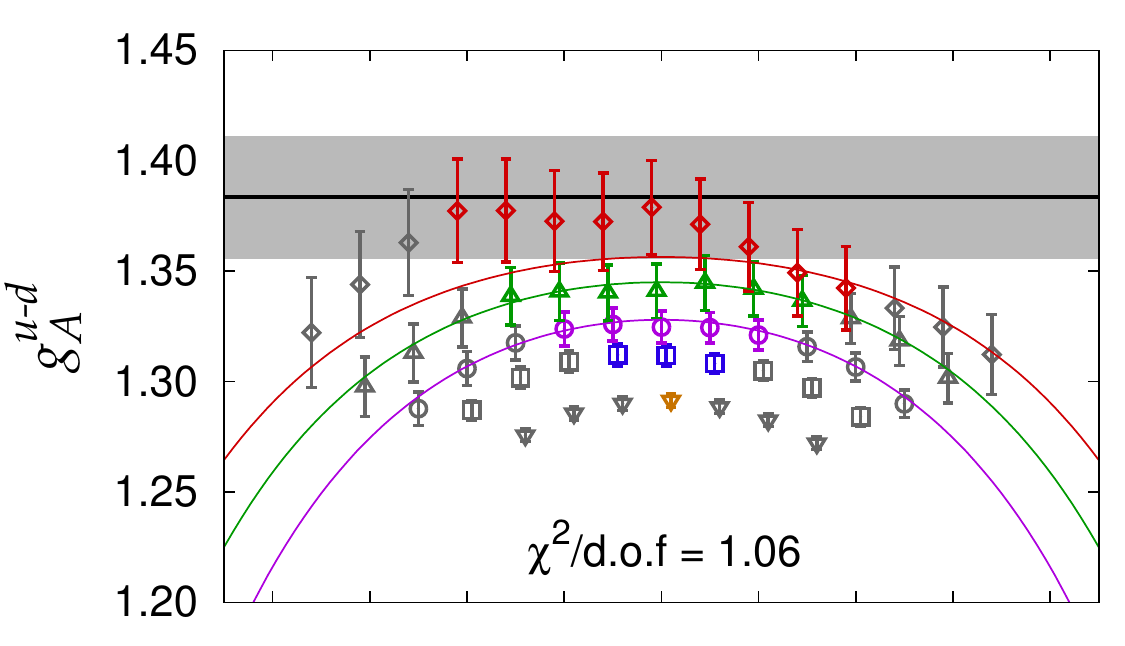}
    \includegraphics[width=0.33\linewidth]{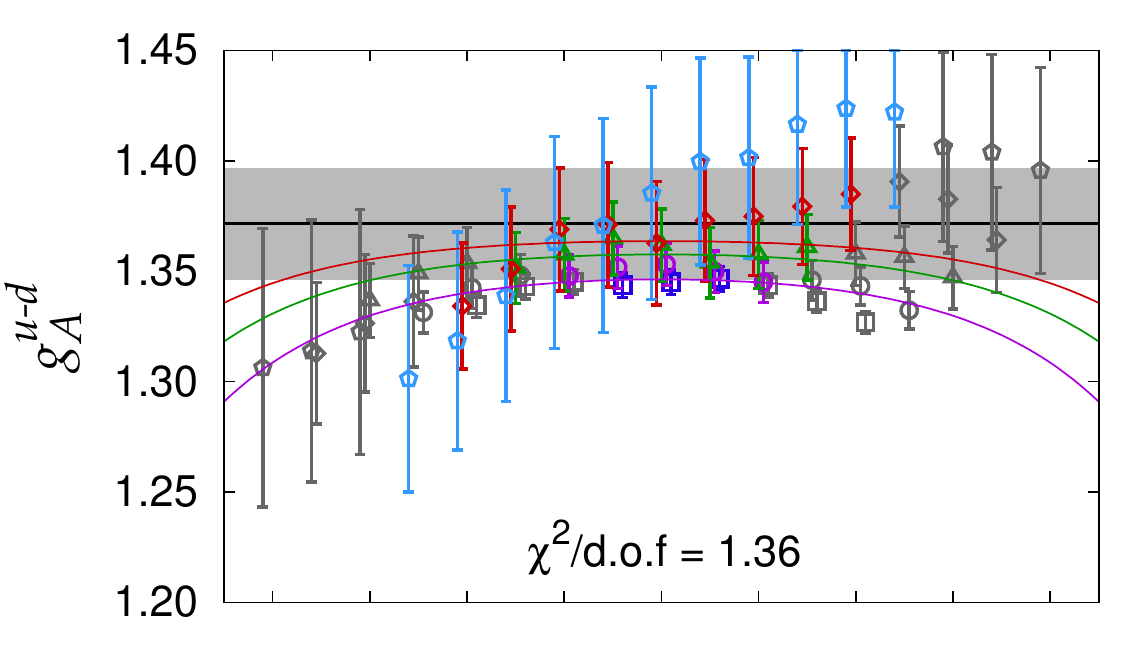}
  }
  \subfigure{
    \includegraphics[width=0.33\linewidth]{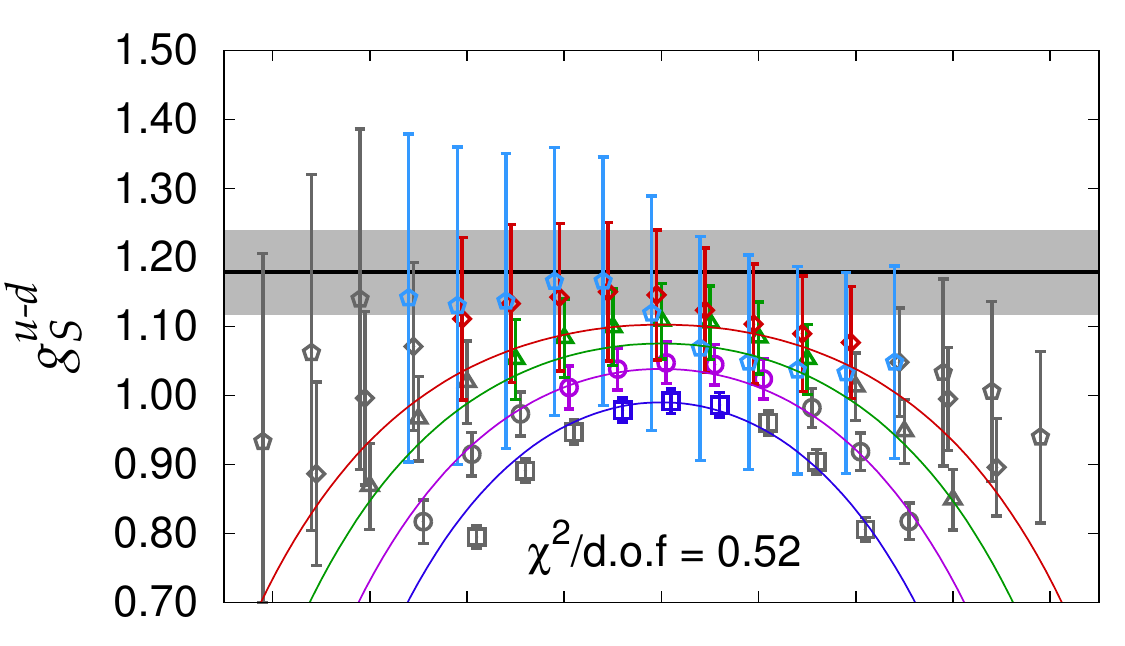}
    \includegraphics[width=0.33\linewidth]{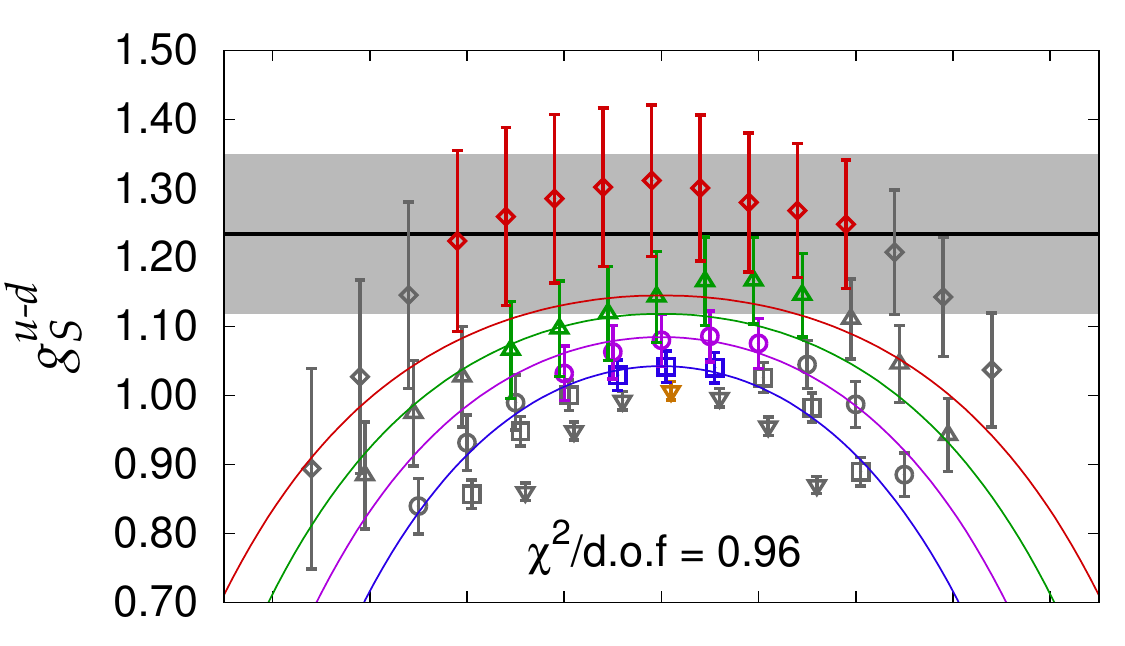}
    \includegraphics[width=0.33\linewidth]{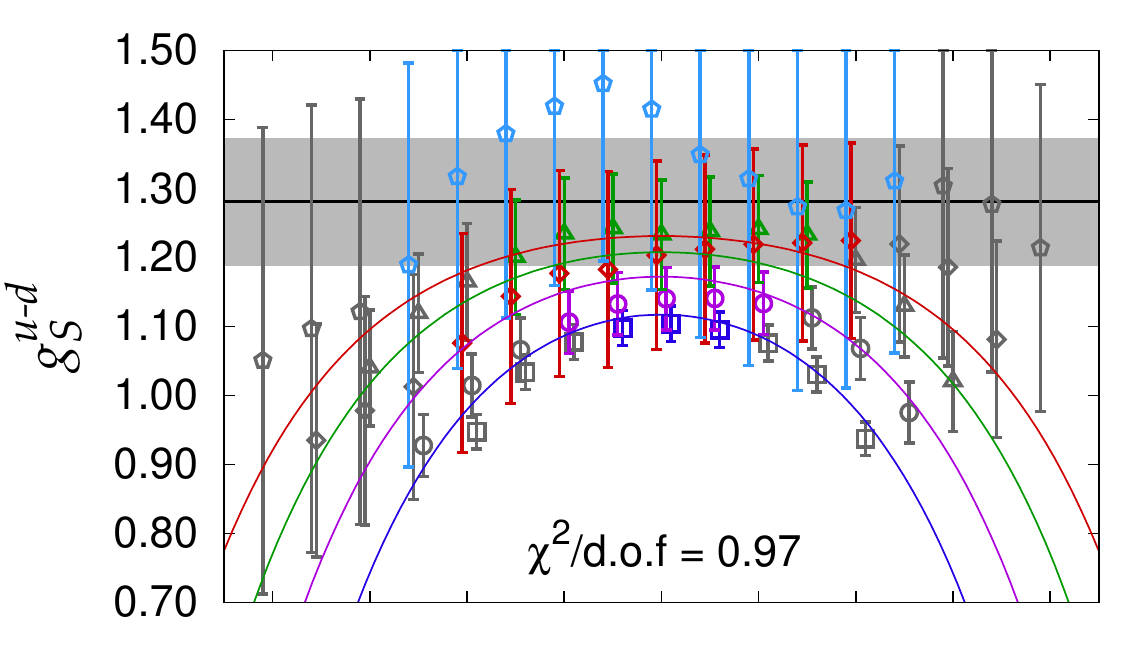}
  }
  \subfigure{
    \includegraphics[width=0.33\linewidth]{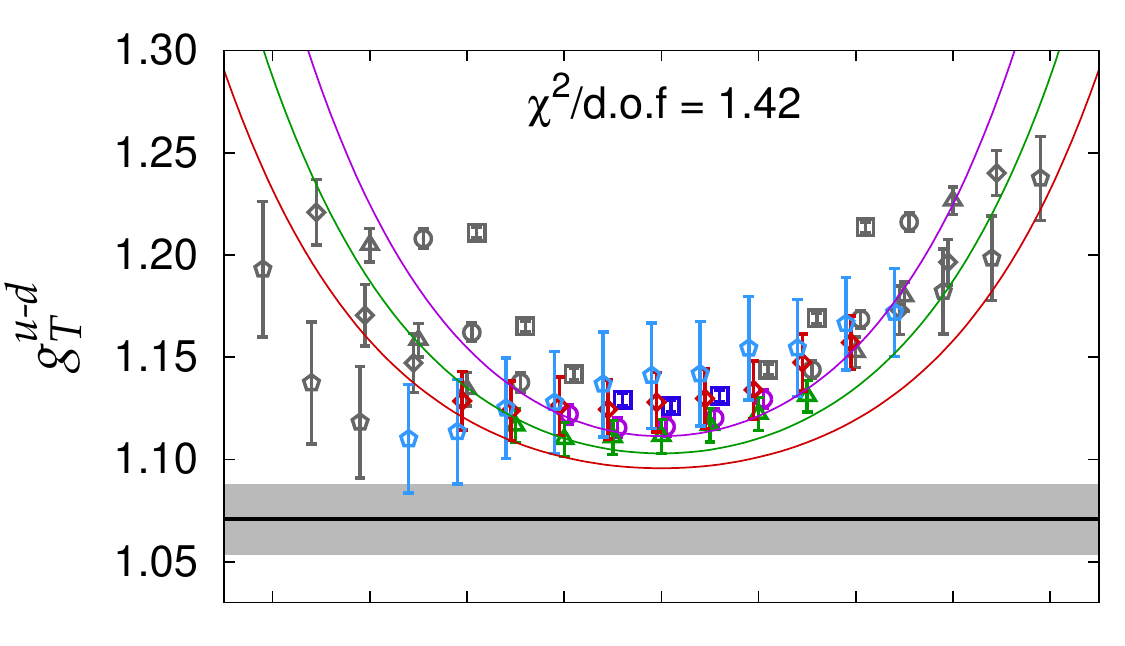}
    \includegraphics[width=0.33\linewidth]{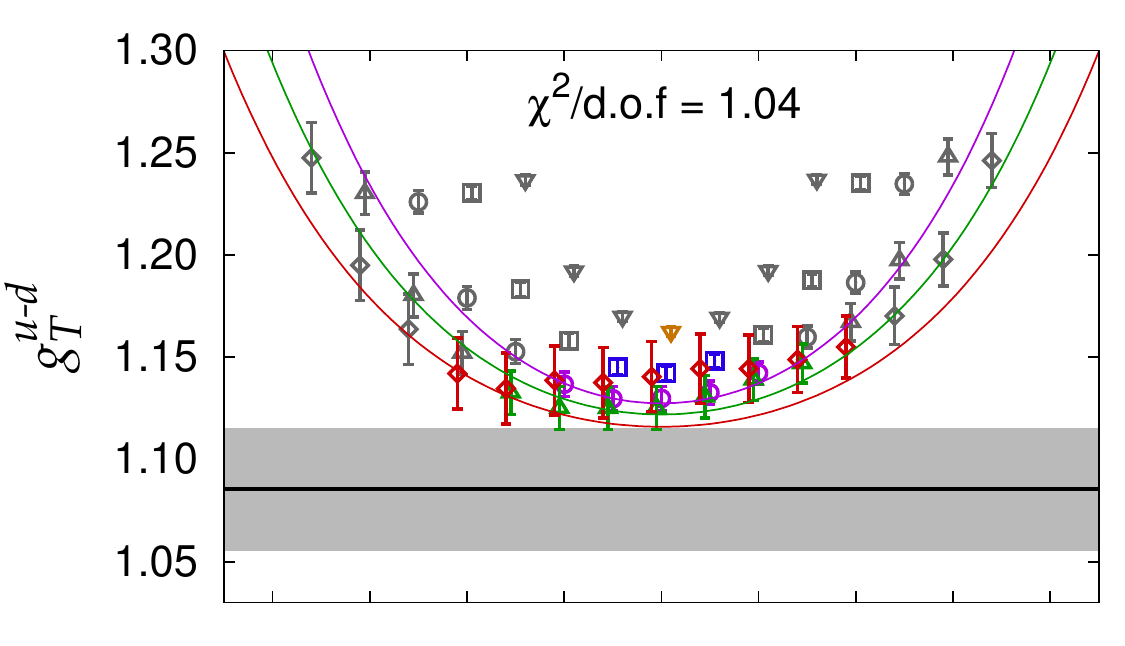}
    \includegraphics[width=0.33\linewidth]{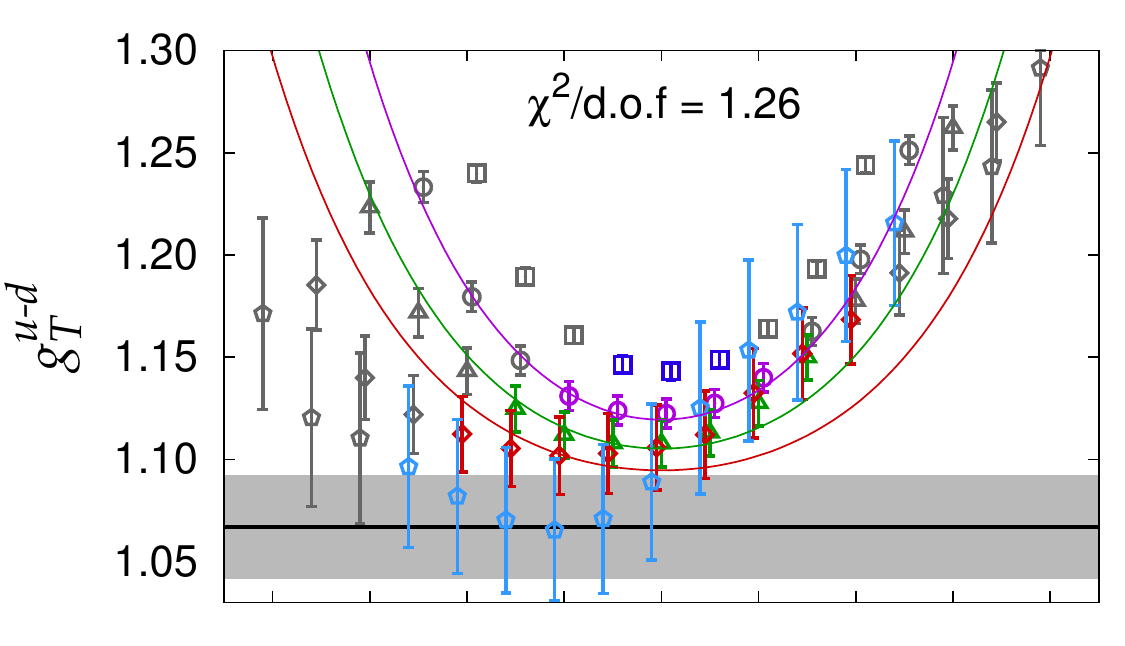}
  }
  \subfigure{
    \includegraphics[width=0.33\linewidth]{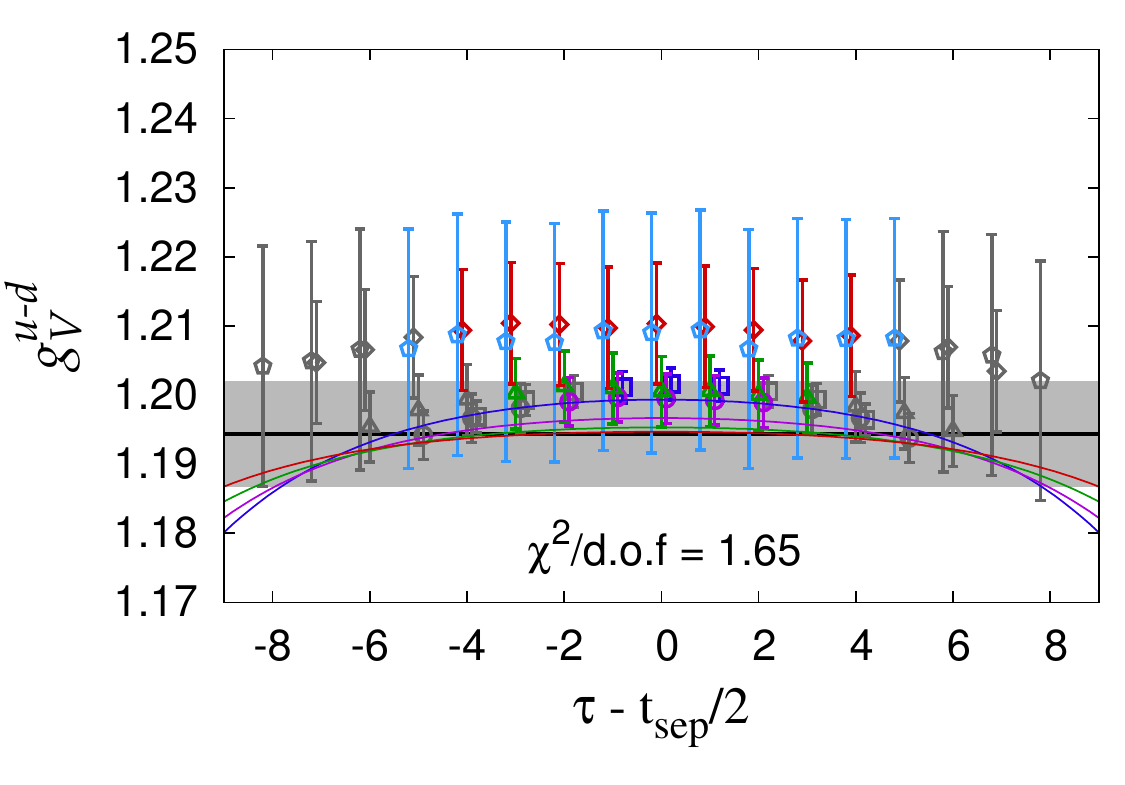}
    \includegraphics[width=0.33\linewidth]{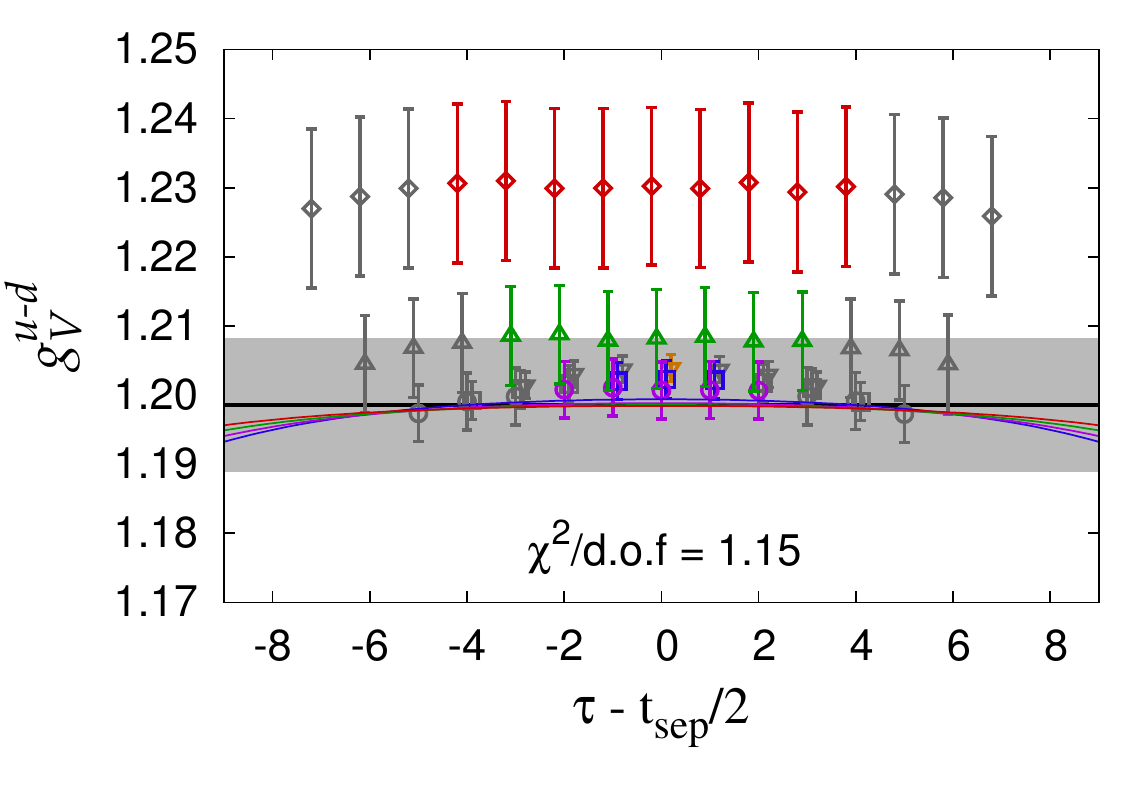}
    \includegraphics[width=0.33\linewidth]{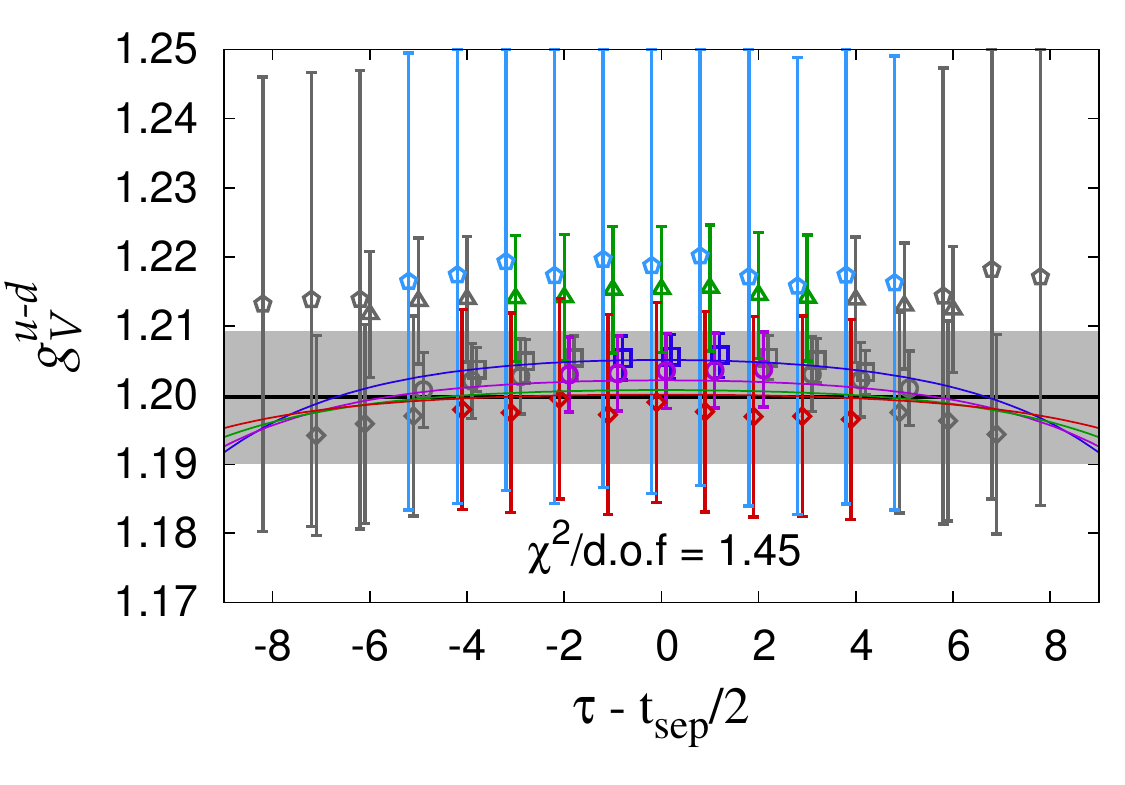}
  }
 \caption{Comparison of the 2-state fits to the data for
   $g_{A,S,T,V}^{u-d}$ for the three calculations on the $a094m280$
   ensemble with different smearing parameter $\sigma$.  Each row
   shows results for the $S_5 S_5$ (left) $S_7 S_7$ (middle) and $S_9
   S_9$ (right) calculations.  }
\label{fig:3smearingCOMP}
\end{figure*}


\section{Fits and Excited-state contamination}
\label{sec:ESC}

To understand and control excited-state contamination we present
analyses using 2-, 3- and 4-state fits to the two-point functions and 2-
and 3-state fits to the three-point functions. We find that to get
reliable estimates of the masses $M_i$ and amplitudes ${\cal A}_i$ for
the first $N$ states we need to include $N+1$ states in the fit to the
two-point function. For this reason, we analyze the correlation
functions with the following combinations $\{\Ntwo,\Nthree\} =
\{2,2\}$, $\{3,2\}$, $\{4,2\}$, $\{3,3\}$ and $\{4,3\}$ where the
first (second) value is the number of states included in fits to the
two-point (three-point) functions. In each case, the methodology employed in
the analysis is the same except that when using three and higher state
fits to two-point functions we introduce non-uniform priors.

In each fit, to understand and quantify the excited-state
contamination, there are three parameters that we optimize: (i) the
starting time slice $\tmin$ used in fits to the two-point data; (ii)
the number of time slices $\tskip$, adjacent to the source and sink,
skipped in fits to the three-point functions; and (iii) the values of
$\tsep$ at which data are collected used in the fits. The final values
of these parameters, chosen on the basis of the $\chi^2$ and the
stability of the fits, represent a compromise between statistical
precision and reducing excited-state contamination. In general, we
reduce the value of $\tmin$ and $\tskip$ and enlarge the number of
$\tsep$ values included when increasing the number of states in the
fit ansatz.  For example, we set $\tmin=2$ in 4-state fits to the
two-point functions.  Even then, in the case of 4-state fits, only
about eight points contribute to determining the six excited-state
parameters since the plateau in the effective-mass plot starts at $t
\approx 10$ as shown in Figs.~\ref{fig:2pt_CandD} and~\ref{fig:2pt_D5S579}.

Our focus is on obtaining estimates for the charges in the $\tsep \to
\infty$ limit for the six calculations labeled as $\{C1, C2, C5, C6,
C7, C8\}$ in Table~\ref{tab:runs}. Two overall caveats that will be
made explicit at appropriate places are: the statistics in the case of
the $a091m170L$ ensemble ($C8$) are insufficient as the
auto-correlations between configurations are large. Similarly, the
errors in the data for $g_S$ are much larger than for $g_{A,T}$,
consequently the fits used to extract $g_S$ are much less stable.  In
all cases, the fits and the error analysis presented here are based on
using the full covariance matrix.

\subsection{Analysis using 2-state fits}
\label{sec:2-state}

The selection of the best combination of $\tmin$, $\tskip$ and $\tsep$ 
for the quoted results using 2-state fits was carried out as follows:
\begin{itemize}
\item
Step 1: Fits to the two-point correlators using the full covariance
matrix for different values of $\tmin$ were made. The best value 
minimizing the correlated $\chi^2/{\rm d.o.f.}$ was
determined to be $\tminb=\{4,5,5,4,4,5\}$ for the six calculations.
\item
Step 2: Using these values of $\tminb$, we determined the four
parameters $aM_0$, $aM_1$, ${\cal A}_0^2$ and ${\cal A}_1^2$. Then,
2-state fits to the three-point data were performed for the three sets of
$\tsep$, labeled A, B and C, in Table~\ref{tab:tsep}.  Since the
pattern of excited-state contamination is different in the various
charges, the best set was determined separately for each charge. Our
final results are based on Fit B for the scalar and vector charges,
and Fit C for the axial and tensor charges as discussed below.
\item
Step 3: Repeat Step 2 for the three cases $\tskip=\{1,2\},\ \{2,3\}$
and $\{3,4\}$ for the $a \approx \{0.127,0.09\}$ ensembles, respectively. All
three cases gave overlapping results. For our quoted 2-state results,
we use $\{3,4\}$ as it removes the most points adjacent to the source
and sink that have the largest excited-state contamination.
\item
Step 4: To quantify the dependence on $\tmin$, we repeat Steps 1--3 for $\tmin = \tminb \pm 1$. 
\end{itemize}
Results for $aM_0$, $aM_1$, ${\cal A}_0^2$ and ${\cal A}_1^2$ for
$\tminb$ and $\tminb \pm 1$ are given in Table~\ref{tab:2ptfits}. 
Results for the three matrix elements for our best choice of 
$\tmin$, $\tskip$ and $\tsep$ 
are given in Table~\ref{tab:BareME}. Estimates of the ratios of
the unrenormalized charges, $g_{A,S,T}/g_V$, are given in
Table~\ref{tab:BareRatios}.

\begin{table}
\centering
\begin{ruledtabular}
\begin{tabular}{c|ccc}
             &  Fit A           &  Fit B               & Fit C             \\
\hline                                                                         
$a127m285$   &  $\{8,10,12\}$   &  $\{8,10,12,14\}$    & $\{10,12,14\}$     \\
\hline                                                                           
$a094m280$,  &                  &                      &                    \\
$a091m170$,  &  $\{10,12,14\}$  &  $\{10,12,14,16\}$   & $\{12,14,16\}$     \\
$a091m170L$  &                  &                      &                    \\
\end{tabular}
\end{ruledtabular}
\caption{The three fits, defined by the values of $\tsep$ used in the
  2-state fit, investigated to quantify the stability of the $\tsep
  \to \infty$ estimate.  }
\label{tab:tsep}
\end{table}


%
\begin{table*}
\centering
\begin{ruledtabular}
\begin{tabular}{c|c|c|cccc|c|c}
ID          & Type    & Fit Range & $aM_0$       & $aM_1$      & ${\cal A}_0^2$  & ${\cal A}_1^2$ & ${\cal A}_1^2/{\cal A}_0^2$  & $\chi^2/$d.o.f.   \\
\hline
            & $S_5 S_5$ & 3--20 &  0.6206(20) & 1.099(45) &  3.51(7)e-08  &  2.19(11)e-08 &  0.624(28) &  1.26  \\
$a127m285$  & $S_5 S_5$ & 4--20 &  0.6193(27) & 1.048(71) &  3.46(10)e-08 &  1.98(23)e-08 &  0.572(55) &  1.31  \\
            & $S_5 S_5$ & 5--20 &  0.6181(31) & 0.980(85) &  3.40(13)e-08 &  1.66(29)e-08 &  0.487(77) &  1.36  \\
\hline
            & $S_5 S_5$ & 4--20 &  0.4721(25) & 0.851(28) &  2.86(9)e-08  &  3.41(13)e-08 &  1.195(41) &  1.33  \\
$a094m280$  & $S_5 S_5$ & 5--20 &  0.4676(37) & 0.776(39) &  2.67(15)e-08 &  2.92(16)e-08 &  1.096(57) &  0.96  \\
            & $S_5 S_5$ & 6--20 &  0.4643(51) & 0.724(50) &  2.51(22)e-08 &  2.62(19)e-08 &  1.042(90) &  0.91  \\
\hline
            & $S_7 S_7$ & 4--20 &  0.4711(30) & 0.864(65) &  5.55(21)e-10 &  3.91(40)e-10 &  0.705(59) &  1.08  \\
$a094m280$  & $S_7 S_7$ & 5--20 &  0.4670(48) & 0.739(77) &  5.19(40)e-10 &  3.02(29)e-10 &  0.583(66) &  0.9  \\
            & $S_7 S_7$ & 6--20 &  0.4654(65) & 0.70(10)  &  5.04(58)e-10 &  2.86(37)e-10 &  0.568(92) &  0.97  \\
\hline
            & $S_9 S_9$ & 3--20 &  0.4672(25) & 0.925(47) &  4.52(13)e-12 &  3.83(24)e-12 &  0.847(45) &  0.94  \\
$a094m280$  & $S_9 S_9$ & 4--20 &  0.4635(37) & 0.809(65) &  4.29(21)e-12 &  3.02(28)e-12 &  0.705(58) &  0.73  \\
            & $S_9 S_9$ & 5--20 &  0.4642(40) &  0.83(10) &  4.33(24)e-12 &  3.21(75)e-12 &   0.74(15) &  0.78  \\
\hline
            & $S_7 S_7$ & 3--22 &  0.4209(24) & 0.859(29) &  4.67(12)e-10 &  4.90(14)e-10 &  1.050(27) &  1.42  \\
$a091m170$  & $S_7 S_7$ & 4--22 &  0.4180(30) & 0.802(37) &  4.50(16)e-10 &  4.41(22)e-10 &  0.981(43) &  1.25  \\
            & $S_7 S_7$ & 5--22 &  0.4183(32) & 0.808(55) &  4.51(18)e-10 &  4.49(55)e-10 &  1.00(10)  &  1.34  \\
\hline
            & $S_7 S_7$ & 4--22 &  0.4252(19) & 0.895(44) &  4.87(10)e-10 &  4.80(44)e-10 &  0.985(76) &  1.75  \\
$a091m170L$ & $S_7 S_7$ & 5--22 &  0.4210(36) & 0.755(72) &  4.59(24)e-10 &  3.35(40)e-10 &  0.730(62) &  1.46  \\
            & $S_7 S_7$ & 6--22 &   0.410(11) & 0.584(79) &  3.73(85)e-10 &  2.82(50)e-10 &  0.76(30)  &  1.1  \\
\end{tabular}
\end{ruledtabular}
\caption{Estimates of the nucleon masses $M_0$ and $M_1$ and the amplitudes
  ${\cal A}_0$ and ${\cal A}_1$ extracted from fits to the two-point
  correlation functions data using the 2-state ansatz given
  in~\eqref{eq:2pt}. For the $a094m280$ ensemble, we give the
  estimates from the three runs with different smearing parameters described in
  Table~\protect\ref{tab:runs}. The notation $S_5 S_5$ labels a nucleon
  correlation function with source and sink constructed using smearing
  parameter $\sigma=5$.  We give estimates for three different fit ranges 
  $\tminb$ and $\tminb \pm 1$, expressed in lattice units, as described in the text. 
  }
  \label{tab:2ptfits}
\end{table*}

\begin{table*}
\centering
\begin{ruledtabular}
\begin{tabular}{c|c|ccc|ccc|ccc}
         &      &  \multicolumn{3}{c|} {Axial}  &   \multicolumn{3}{c|} {Scalar}  &   \multicolumn{3}{c} {Tensor}     \\
ID       & Type & $\langle 0 | \mathcal{O}_A | 0 \rangle$ & $\langle 0 | \mathcal{O}_A | 1 \rangle$  & $\langle 1 | \mathcal{O}_A | 1 \rangle$
                & $\langle 0 | \mathcal{O}_S | 0 \rangle$ & $\langle 0 | \mathcal{O}_S | 1 \rangle$  & $\langle 1 | \mathcal{O}_S | 1 \rangle$
                & $\langle 0 | \mathcal{O}_T | 0 \rangle$ & $\langle 0 | \mathcal{O}_T | 1 \rangle$  & $\langle 1 | \mathcal{O}_T | 1 \rangle$ \\
\hline
a127m285  & $S_5 S_5$ &  1.423(14)  &  -0.179(21)  &   -0.9(2.4) &   1.07(4)  &  -0.35(4)  &   0.6(1.1)  &  1.166(13)  &  0.182(16)  &  -0.2(1.2)   \\
\hline     
a094m280  & $S_5 S_5$ &  1.349(19)  &  -0.130(20)  &    0.6(0.7) &   1.18(6)  &  -0.42(5)  &   1.0(0.8)  &  1.071(17)  &  0.157(15)  &  0.7(4)   \\
a094m280  & $S_7 S_7$ &  1.384(28)  &  -0.111(36)  &    0.3(1.3) &   1.23(12) &  -0.52(12) &   1.4(1.7)  &  1.085(30)  &  0.221(36)  &  0.0(0.8)   \\
a094m280  & $S_9 S_9$ &  1.372(25)  &  -0.026(39)  &   -0.4(2.5) &   1.28(9)  &  -0.42(9)  &  -0.6(3.3)  &  1.067(25)  &  0.276(28)  &  0.6(1.6)   \\
\hline   
a091m170  & $S_7 S_7$ &  1.388(23)  &  -0.133(33)  &   -2.1(2.6) &   1.17(11) &  -0.48(7)  &   0.1(3.9)  &  1.091(20)  &  0.154(22)  &  -0.2(1.7)   \\
\hline   
a091m170L & $S_7 S_7$ &  1.401(20)  &  -0.118(26)  &   -1.0(2.4) &   1.15(8)  &  -0.44(8)  &   1.4(2.2)  &  1.067(25)  &  0.235(23)  &  0.5(8)   \\
\end{tabular}
\end{ruledtabular}
\caption{Estimates of the three matrix elements $\langle 0 |
  \mathcal{O}_\Gamma | 0 \rangle$, $\langle 0 | \mathcal{O}_\Gamma | 1
  \rangle$, $\langle 1 | \mathcal{O}_\Gamma | 1 \rangle$ for the three
  isovector operators obtained using the 2-state fit to the
  three-point correlators with our ``best'' choices of $\tmin$,
  $\tsep$ and $\tskip$.  }
\label{tab:BareME}
\end{table*}

\begin{table}
\centering
\begin{ruledtabular}
\begin{tabular}{c|c|ccc}
ID        & Type      &  $g_A/g_V$   &    $g_S/g_V$  &    $g_T/g_V$ \\
\hline
a127m285  & $S_5 S_5$ &  1.125(11)   &  0.848(27)    &  0.922(12)  \\
\hline                   
a094m280  & $S_5 S_5$ &  1.130(17)   &  0.987(50)    &  0.897(15)  \\
a094m280  & $S_7 S_7$ &  1.154(24)   &  1.030(95)    &  0.906(27)  \\
a094m280  & $S_9 S_9$ &  1.143(22)   &  1.068(76)    &  0.889(22)  \\
\hline                   
a091m170  & $S_7 S_7$ &  1.146(21)   &  0.963(84)    &  0.901(16)  \\
\hline                   
a091m170L & $S_7 S_7$ &  1.166(18)   &  0.960(69)    &  0.888(21)  \\
\end{tabular}
\end{ruledtabular}
\caption{Estimates of the ratios of the unrenormalized isovector charges $g_{A,S,T}/g_V$ with our ``best''
  choices of $\tmin$, $\tsep$ and $\tskip$ in the 2-state fits. }
\label{tab:BareRatios}
\end{table}

The quality of the data for the two-point functions 
on the four ensembles is illustrated by plotting the effective mass, 
\begin{equation}
aM_{N,eff}(t+0.5)   = \ln \frac{C^{\rm 2pt}(t)}{C^{\rm 2pt}(t+1)} \,,
\label{eq:effmass}
\end{equation}
for the pion and the nucleon in Fig.~\ref{fig:2pt_CandD}.  As
expected, the signal in the pion does not degrade with $t$, whereas
that for the proton becomes noisy by $t=20$, with 1--2~$\sigma$
fluctuations in $M_{N,eff}$ apparent already at $t \approx 16$. The
onset of a plateau indicates that the ground-state pion mass can be
extracted using 1-state fits to data at $t > 10$.  In practice, the
ground state mass is largely determined from the region $ 10 \lesssim t
\lesssim 16$, while the excited state masses and amplitudes are
determined from the region $t \lesssim 10$. The value of $\tmin$ is,
therefore, adjusted depending on the number of states included in the
fit.

To assess the statistical quality of the data, the auto-correlation
function was calculated using two quantities that have reasonable
estimates on each configuration: (i) the pion two-point correlator at
$t=14$ and (ii) the three-point correlation function at the midpoint in
$\tau$ for $\tsep=12$.  Autocorrelations increase as $M_\pi$ or 
$a$ is decreased or $L$ is increased. In particular, the data from the $a091m170L$ ensemble showed
significant auto-correlations. In this case, the 467 configurations
consist of four streams with roughly 170, 100, 100 and 100
configurations. These are too few to even determine the
auto-correlation time reliably. For the other ensembles, the
auto-correlation function falls to $<1/e$ by 1--2 configurations, and
binning the data by a factor of two did not change the Jackknife error
estimates. Our overall conclusion is that much larger statistics are
needed to get reliable error estimates on the $a091m170L$ ensemble and
it is very likely that the quoted $1\sigma$ errors for this ensemble,
evaluated without taking into account auto-correlations, are
underestimates.

To exhibit the dependence of the two-point correlation functions on the
smearing size given by $\sigma$, we show a comparison of the effective mass for the
pion and the nucleon for the $S_5 S_5$, $S_7 S_7$ and $S_9 S_9$
correlators on the $a094m280$ ensemble in
Fig.~\ref{fig:2pt_D5S579}. We find that the errors in the
effective-mass data (and in the raw two-point functions) increase with
the smearing size $\sigma$ for both the pion and the proton. The onset
of the plateau in both states, however, occurs at earlier times with
larger $\sigma$. Thus, the relative reduction in the excited-state
contamination in the correlation functions with larger smearing
$\sigma$ has to be balanced against the increase in statistical
noise. Based on these trends on the $a094m280$ ensemble, our
compromise choice for the three ensembles at $a \approx 0.09$~fm is
$\sigma=7$ and for the $a127m285$ ensemble it is $\sigma=5$. In
physical units, this choice corresponds to setting the size of the
smearing parameter $\sigma \approx 0.65$~fm.

Our final estimates for the four parameters $aM_0$, $aM_1$, ${\cal
  A}_0^2$, ${\cal A}_1^2$ and the ratio ${\cal A}_1^2 / {\cal A}_0^2$
for three values of $\tmin$ are given in Table~\ref{tab:2ptfits}. In
addition to minimizing $\chi^2/{\rm d.o.f.}$, we required $\lesssim 1 \sigma$
stability in the value of $M_0$ under the variation $\tminb \pm 1$ as
criteria for choosing our best $\tmin$. With the selected
$\tminb=\{4,5,5,4,4,5\}$, we find that $aM_1$ is also consistent
within $2\sigma$ except on the $a091m170L$ ensemble which, as
stated above, requires much higher statistics.

To illustrate the three-point function data and the size of
excited-state contamination, we plot an ``effective'' charge,
\begin{equation}
g_\Gamma^\text{eff}(t_\text{sep},\tau) =
\frac{C^\text{3pt}_\Gamma(t_\text{sep},\tau) }{
C^\text{2pt}_\text{fit}(t_\text{sep})} \, , 
\end{equation}
i.e., the ratio of the three-point function to the n-state fit that
describes the two-point function. This ratio converges to $g_\Gamma$
as the time separations $\tau$ and $t_\text{sep}-\tau$ become large
provided the fit to the two-point function,
$C^\text{2pt}_\text{fit}(t_\text{sep})$, gives the ground state.  Our
methodology for taking into account excited-state contamination and
obtaining estimates of the charges from data with $\tsep$ in the
limited range 1--1.5~fm is described next.

The data and the 2-state fits to the ratio of the three- to two-point
functions using our best choice of $\tmin$, $\tsep$ and $\tskip$ are
shown in Figs.~\ref{fig:gA6}, ~\ref{fig:gS6}, ~\ref{fig:gT6}
and~\ref{fig:gV6} for the four isovector charges. In the right panels
of these figures, we show the 3-state fits, discussed in
Sec.~\ref{sec:3-state}, to facilitate comparison.

The magnitude of the excited-state contamination as a function of
$\tsep$ and the smearing parameter $\sigma$ is different for the four
charges.  The dependence on $\sigma$ is exhibited in
Fig.~\ref{fig:3smearingCOMP} for the $S_5 S_5$, $S_7 S_7$ and $S_9
S_9$ calculations on the $a094m280$ ensemble. In $g_A^{u-d}$, the
magnitude of the excited-state contamination, measured as the
difference between the data at the central values of $\tau$ for
$\tsep=12$ (about 1~fm) and the $\tsep \to \infty$ estimate, is about
10\%, 5\% and 3\% for the $S_5 S_5$, $S_7 S_7$ and $S_9 S_9$
calculations, respectively. The pattern in $g_S^{u-d}$ is similar,
however, the reduction in the contamination with $\sigma$ is
smaller. For $g_T^{u-d}$, the overall variation with $\tsep$ and
between the three estimates is $\le 5\%$. The vector charge
$g_V^{u-d}$ shows insignificant excited-state contamination and no
detectable dependence on $\sigma$.  On the other hand, the errors in
individual data points increase with the smearing $\sigma$ for all
four charges.  

We use the data with the three values of the smearing $\sigma$ on the
$a094m280$ ensemble to test whether the 2-state fit gives equally
reliable $\tsep \to \infty$ estimates in spite of differences in the
excited-state contamination.  We find that the three estimates are
consistent within $1\sigma$ for all four charges as shown in
Fig.~\ref{fig:3smearingCOMP}. However, because the magnitude of the
excited-state effect is different in the four charges $g_{A,S,T,V}$,
we do not uniformly use the same set of values of $\tsep$ in our final
2-state fit, but tune them for each case.

Based on the data shown in these figures and on the results of the
fits with our best choices of the input fit parameters given in
Tables~\ref{tab:BareME} and~\ref{tab:BareRatios}, we evaluate below the
excited-state effect in each of the charges and the efficacy of the
2-state fit in providing $\tsep \to \infty$ estimates.
\begin{itemize}
\item
The data for the axial charge $g_A^{u-d}$ shown in Fig.~\ref{fig:gA6}
converges to the $\tsep \to \infty$ value from below and at the
central values of $\tau$ show up to 10\% variation with $\tsep$ due to
excited-state contamination.  We, therefore, use Fit C based on data
with the larger values of $\tsep$. In all but the $a127m285$ case, the
data at $\tsep \gtrsim 16$ lies above the result of the fit. 
The errors in these data, however, are large.\footnote{We do not include the
  data at $\tsep=18$ in the fits as these have been obtained only for the 
  $S_5 S_5$ and $S_9 S_9$ calculations on the $a094m280$ ensemble.} Thus,
to confirm the $\tsepi$ estimate, requires additional high precision
data for $\tsep \ge 16$.

The data with $S_9 S_9$ correlators on the $a094m280$ ensemble show
the least excited-state effect: the estimates at the central values of
$\tau$ show only a tiny increase with $\tsep$ and their error bands overlap. 
As a result, the matrix elements $\langle 0 | \mathcal{O}_T |
1 \rangle$ and $\langle 1 | \mathcal{O}_T | 1 \rangle$, given in
Table~\ref{tab:BareME}, are poorly determined.  
\item
The data for the scalar charge $g_S^{u-d}$ have larger uncertainty so
we choose Fit B to include all the data except that with $\tsep=8$ and
$18$ on the $a094m280$ ensemble.  As shown in Fig.~\ref{fig:gS6}, the
2-state fits again converge from below, the three estimates from the
$a094m280$ ensemble are consistent within $1\sigma$ as shown in
Fig.~\ref{fig:3smearingCOMP}, and the estimates from the $a091m170$
and $a091m170L$ ensembles overlap.
\item
The data for the tensor charge $g_T^{u-d}$ show small excited-state
contamination and converge to the $\tsep \to \infty$ estimate from
above.  Fits to the the data for $g_T^{u-d}$ are the most stable and
all the fits give consistent estimates. We choose Fit C for the final
estimates. These estimates agree with those from Fit B, and the
$\chi^2/{\rm d.o.f.}$ of both fits are also consistent.
\item
The data for the vector charge $g_V^{u-d}$ show little variation with
$\tau$ or $\tsep$.  The excited-state contamination is highly
suppressed because $g_V^{u-d}$ is associated with a conserved charge
at $O(a)$.  As a result, statistical fluctuations in the data stand
out.  Note that the error estimates in $g_V^{u-d}$ from the
fits are comparable to the 1--2\% variance in individual data points
at the largest $\tsep$. Only the $S_7 S_7$ data with $\tsep=16$ on the
$a094m280$ ensemble deviate by about $2\sigma$ from the result of the
fit. 

\end{itemize}

In the next section, we extend the analysis to include up to 4-states
in fits to the two-point function data and 3-states in fits to the three-point
functions to evaluate the stability of estimates obtained using
2-state fits. 

\subsection{Analysis using 3-state fits}
\label{sec:3-state}

In this section, we investigate the stability of the $\tsepi$
estimates from 2-state fits by increasing the number of states kept in
the fits to the two- and three-point function data.  The additional features
introduced into the analysis, over and above those discussed
in Sec~\ref{sec:2-state} for the 2-state fits, are:
\begin{itemize}
\item
The two-point function data were analyzed using 3- and 4-state fits.
In fits with more than two states, the excited state masses and
amplitudes are, in many cases, ill-determined.  The fits were
stabilized by carrying out an empirical Bayesian analysis with
Gaussian priors for both the mass gaps and the amplitudes of the
excited states~\cite{Lepage:2001ym,Chen:2004gp}.
\item
With more fit parameters, the values of $\tmin$ and $\tskip$ were
reduced to increase the number of data points included in the fits to
both two- and three-point functions.  The results with the
final choices of these parameters for various fits are listed in
Tables~\ref{tab:2ptmulti} and~\ref{tab:BC3state}.
\item
Data with $\tsep=8,\ 10,\ 12$ and 14 were used for all four charges in
fits to the three-point functions on the $a127m285$ ensemble and with 
$\tsep=10,\ 12,\ 14$ and 16 for the three $a \approx 0.09$~fm ensembles.
\end{itemize}

The priors for the 3- and 4-state fits to the two-point function data
were determined empirically as follows:
\begin{itemize}
\item
The ground state mass and amplitude are very well constrained by the
plateau in the effective mass for $t \gtrsim 10$.  Thus, no non-trivial
priors were needed or used for determining ${\cal A}_0$ and $M_0$.
\item
Results for ${\cal A}_1$ and $M_1$ obtained from 2-state fits without
priors were used to guide the selection of their priors in 3-state
fits. The width was chosen to be large but consistent with the
requirement that the $1\sigma$ bands for $r_i \equiv ({\cal A}_i^2
/{\cal A}_0^2)$ and the $\Delta M_i \equiv (M_i - M_{i-1})$ are
positive.  These priors did not need any subsequent changes. 
\item
Results for $r_2$ and $\Delta M_2$ from the 3-state fit were used as
priors in the 4-state fits along with ${\cal A}_3^2/{\cal A}_0^2 =1
\pm 0.6$ and $\Delta M_3 \equiv M_3 -M_2 = 0.4 \pm 0.3$.  The output
estimates were used as the new central values of these two sets of
priors without decreasing their width and the fits were carried out a
second time to get the final estimates.
\item
In all cases except for the $a091m170L$ data, the final results are
close to the central value chosen for the priors for both the 3- and
4-state fits.
\item
The quoted errors are obtained using a single elimination jackknife procedure with 
the full covariance matrix and constant priors. 
\item
The augmented $\chi^2/{\rm d.o.f.}$ is given by the standard correlated $\chi^2$
plus the square of the deviation of the parameter from the prior
normalized by the width.  This is then divided by the number of
degrees of freedom calculated ignoring the priors.
\end{itemize}

\begin{table*}[!htb]
\centering
\begin{ruledtabular}
  \begin{tabular}{c|c.{5}.{4}.{4}.{4}.{4}.{4}.{4}.{3}}
   & 
  \multicolumn{1}{c}{$\mathcal{A}_0^2$} & 
  \multicolumn{1}{c}{$aM_0$} &
  \multicolumn{1}{c}{$r_1$} & 
  \multicolumn{1}{c}{$a\Delta M_1$} &
  \multicolumn{1}{c}{$r_2$} &
  \multicolumn{1}{c}{$a\Delta M_2$} &
  \multicolumn{1}{c}{$r_3$} & 
  \multicolumn{1}{c}{$a\Delta M_3$} &
  \multicolumn{1}{c}{$\chi^2/\text{d.o.f}$}
  \\\hline
  & \multicolumn{9}{c} {\underline{$a127m285$ Smearing $\sigma=5$}}     \\  
Priors      &          &          & 0.6(3)   & 0.35(20)& 0.65(35) & 0.8(4)   & 1.0(5)   & 0.4(3)     &       \\
\{2,4--20\} & 3.46(11)$\times 10^{-8}$ & 0.619(3)  & 0.57(6)  & 0.43(7) &          &          &          &            & 1.31 \\
\{3,2--20\} & 3.42(10)$\times 10^{-8}$ & 0.619(2)  & 0.49(4)  & 0.38(5) & 0.71(13) & 0.91(10) &          &            & 1.02 \\
\{4,2--20\} & 3.43(9)$\times 10^{-8}$  & 0.619(2)  & 0.54(4)  & 0.40(5) & 0.45(6)  & 1.06(9)  & 0.77(13) & 0.49(2)    & 1.07 \\
\hline
& \multicolumn{9}{c} {\underline{$a094m280$ Smearing $\sigma=5$}}     \\  
Priors      &          &          & 1.0(5)   & 0.23(12)& 1.3(6)   & 0.6(3)   & 0.8(5)   & 0.4(3)     &       \\
\{2,5--20\} & 2.67(15)$\times 10^{-8}$ & 0.468(4)  & 1.10(6)  & 0.31(4) &          &          &          &            & 0.96 \\
\{3,3--20\} & 2.49(18)$\times 10^{-8}$ & 0.464(4)  & 0.90(10) & 0.24(4) & 1.61(24) & 0.62(7)  &          &            & 0.71 \\
\{4,3--20\} & 2.45(24)$\times 10^{-8}$ & 0.463(5)  & 0.92(11) & 0.23(5) & 1.22(17) & 0.58(10) & 0.73(12) & 0.41(4)    & 0.69 \\
\hline
& \multicolumn{9}{c} {\underline{$a094m280$ Smearing $\sigma=7$}}     \\  
Priors      &          &          & 0.6(3)   & 0.23(12)& 0.8(6)   & 0.6(3)   & 0.6(4)   & 0.4(3)     &       \\
\{2,5--20\} & 5.19(40)$\times 10^{-10}$ & 0.467(5) & 0.58(7)  & 0.27(7) &          &          &          &            & 0.90 \\
\{3,3--20\} & 5.03(39)$\times 10^{-10}$ & 0.466(4) & 0.48(9)  & 0.22(5) & 1.05(21) & 0.64(8)  &          &            & 0.74 \\
\{4,3--20\} & 4.97(45)$\times 10^{-10}$ & 0.465(5) & 0.50(9)  & 0.21(6) & 0.77(16) & 0.62(8)  & 0.58(9)  & 0.40(3)    & 0.73 \\
\hline
& \multicolumn{9}{c} {\underline{$a094m280$ Smearing $\sigma=9$}}     \\  
Priors      &          &          & 0.6(3)   & 0.23(12)& 0.8(6)   & 0.6(3)   & 0.6(4)   & 0.4(3)     &       \\
\{2,4--20\} & 4.29(21)$\times 10^{-12}$ & 0.464(4) & 0.71(6)  & 0.35(6) &          &          &          &            & 0.73 \\
\{3,2--20\} & 4.16(20)$\times 10^{-12}$ & 0.462(3) & 0.55(8)  & 0.27(4) & 1.08(16) & 0.71(10) &          &            & 0.67 \\
\{4,2--20\} & 4.11(22)$\times 10^{-12}$ & 0.461(4) & 0.56(8)  & 0.26(5) & 0.67(10) & 0.64(11) & 0.57(9)  & 0.41(3)    & 0.65 \\
\hline
& \multicolumn{9}{c} {\underline{$a091m170$ Smearing $\sigma=7$}}     \\  
Priors      &          &          & 0.8(5)   & 0.30(15)& 1.0(6)   & 0.7(5)   & 1.0(5)   & 0.4(3)     &       \\
\{2,4--22\} & 4.49(16)$\times 10^{-10}$ & 0.418(3) & 0.98(4)  & 0.38(3) &          &          &          &            & 1.25 \\
\{3,2--22\} & 4.42(16)$\times 10^{-10}$ & 0.417(3) & 0.88(6)  & 0.35(3) & 1.22(20) & 0.91(13) &          &            & 1.05 \\
\{4,2--22\} & 4.44(17)$\times 10^{-10}$ & 0.417(3) & 0.91(6)  & 0.36(4) & 0.72(18) & 0.91(20) & 0.87(16) & 0.45(2)    & 1.05 \\
\hline
& \multicolumn{9}{c} {\underline{$a091m170L$ Smearing $\sigma=7$}}     \\  
Priors      &          &          & 0.5(3)   & 0.21(5) & 1.0(9)   & 0.6(4)   & 0.8(6)   & 0.30(25)   &       \\
\{2,5--22\} & 4.59(24)$\times 10^{-10}$ & 0.421(4) & 0.73(6)  & 0.33(7) &          &          &          &            & 1.46 \\
\{3,2--22\} & 4.22(23)$\times 10^{-10}$ & 0.417(3) & 0.45(10) & 0.19(2) & 1.30(8)  & 0.54(4)  &          &            & 1.00 \\
\{4,2--22\} & 4.20(25)$\times 10^{-10}$ & 0.417(3) & 0.44(10) & 0.18(2) & 0.74(5)  & 0.47(4)  & 0.60(4)  & 0.19(5)    & 1.03 \\
\end{tabular}
\end{ruledtabular}
  \caption{Results of 2-, 3- and 4-state fits to the two-point
    correlation data for the six calculations. The first column
    specifies the parameters, \{$N_\text{2pt}$,
    $t_\text{min}$--\,$t_\text{max}$\}, where $N_\text{2pt}$ number of
    states used in the fits to the two-point correlators, and
    $[t_\text{min},t_\text{max}]$ is the fit interval in lattice units. The following
    columns give the nucleon ground state amplitude $\mathcal{A}_0^2$
    and mass $aM_0$, followed by the ratio of the excited state
    amplitudes $r_i = (\mathcal{A}_i/\mathcal{A}_0)^2$, and a mass gaps 
    $a\Delta M_i = a(M_i-M_{i-1})$. For each ensemble, the first row
    gives the values of the priors used in the final fits.}
  \label{tab:2ptmulti}
\end{table*}

\begin{table*}
\centering
\begin{ruledtabular}
\begin{tabular}{c|c|c|c|.{8}.{6}.{8}.{8}}
\multicolumn{1}{c|}{ID}       & 
\multicolumn{1}{c|}{Type}     &    
\multicolumn{1}{c|}{Fit}      &  
\multicolumn{1}{c|}{$\tsep$}  &  
\multicolumn{1}{c}{$g_A^{u-d}$} & 
\multicolumn{1}{c}{$g_S^{u-d}$} & 
\multicolumn{1}{c}{$g_T^{u-d}$} & 
\multicolumn{1}{c}{$g_V^{u-d}$}    \\
\hline                                                                    
\multirow{5}{*}{a127m285}  & \multirow{5}{*}{$S_5 S_5$} &  \{2,2,3,4--20\}  &  \{10,12,14\}       &  1.423(14)    &               &    1.166(13)  &                 \\
& &  \{2,2,3,4--20\}  &  \{8,10,12,14\}     &               &  1.07(4)      &               &    1.264(5)     \\
& &  \{4,2,3,2--20\}  &  \{8,10,12,14\}     &  1.413(12)    &  1.08(3)      &    1.153(11)  &    1.265(5)     \\
& &  \{4,3,3,2--20\}  &  \{8,10,12,14\}     &  1.431(29)    &  1.09(4)      &    1.160(12)  &    1.264(7)     \\
& &  \{4,$3^\ast$,3,2--20\}  &  \{8,10,12,14\}     &  1.431(15)    &  1.09(4)      &    1.160(10)  &    1.264(5)     \\
\hline                                                         
\multirow{5}{*}{a094m280}  & \multirow{5}{*}{$S_5 S_5$} &  \{2,2,4,5--20\}  &  \{12,14,16\}       &  1.349(19)    &               &    1.071(17)  &                 \\
& &  \{2,2,4,5--20\}  &  \{10,12,14,16\}    &               &  1.18(6)      &               &    1.194(8)     \\
& &  \{4,2,4,3--20\}  &  \{10,12,14,16\}    &  1.365(34)    &  1.30(13)     &    1.025(40)  &    1.197(9)     \\
& &  \{4,3,2,3--20\}  &  \{10,12,14,16\}    &  1.369(36)    &  1.31(15)     &    1.066(23)  &    1.208(17)    \\
& &  \{4,$3^\ast$,2,3--20\}  &  \{10,12,14,16\}    &  1.369(34)    &  1.25(9)     &    1.057(19)  &    1.194(10)    \\
\hline                                                         
\multirow{5}{*}{a094m280}  & \multirow{5}{*}{$S_7 S_7$} &  \{2,2,4,5--20\}  &  \{12,14,16\}       &  1.384(28)    &               &    1.085(30)  &                 \\
& &  \{2,2,4,5--20\}  &  \{10,12,14,16\}    &               &  1.23(12)     &               &    1.199(10)    \\
& &  \{4,2,4,3--20\}  &  \{10,12,14,16\}    &  1.398(38)    &  1.34(17)     &    1.015(61)  &    1.201(11)    \\
& &  \{4,3,2,3--20\}  &  \{10,12,14,16\}    &  1.390(40)    &  1.35(21)     &    1.077(34)  &    1.243(28)    \\
& &  \{4,$3^\ast$,2,3--20\}  &  \{10,12,14,16\}    &  1.381(32)    &  1.20(12)     &    1.051(36)  &    1.202(12)    \\
\hline                                                         
\multirow{5}{*}{a094m280}  & \multirow{5}{*}{$S_9 S_9$} &  \{2,2,4,4--20\}  &  \{12,14,16\}       &  1.372(25)    &               &    1.067(25)  &                 \\
& &  \{2,2,4,4--20\}  &  \{10,12,14,16\}    &               &  1.28(9)      &               &    1.200(10)    \\
& &  \{4,2,4,2--20\}  &  \{10,12,14,16\}    &  1.355(20)    &  1.36(11)     &    1.028(34)  &    1.199(11)    \\
& &  \{4,3,2,2--20\}  &  \{10,12,14,16\}    &  1.345(30)    &  1.39(18)     &    1.038(28)  &    1.195(38)    \\
& &  \{4,$3^\ast$,2,2--20\}  &  \{10,12,14,16\}    &  1.345(23)    &  1.25(11)     &    1.038(28)  &    1.195(12)    \\
\hline                                                         
\multirow{5}{*}{a091m170}  & \multirow{5}{*}{$S_7 S_7$} &  \{2,2,4,4--22\}  &  \{12,14,16\}       &  1.388(23)    &               &    1.091(20)  &                 \\
& &  \{2,2,4,4--22\}  &  \{10,12,14,16\}    &               &  1.17(10)     &               &    1.211(11)    \\
& &  \{4,2,4,2--22\}  &  \{10,12,14,16\}    &  1.370(16)    &  1.19(11)     &    1.068(16)  &    1.211(11)    \\
& &  \{4,3,2,2--22\}  &  \{10,12,14,16\}    &  1.363(17)    &  1.09(11)     &    1.070(15)  &    1.210(13)    \\
& &  \{4,$3^\ast$,2,2--22\}  &  \{10,12,14,16\}    &  1.363(17)    &  1.09(11)     &    1.070(15)  &    1.210(13)    \\
\hline                                                         
\multirow{5}{*}{a091m170L} & \multirow{5}{*}{$S_7 S_7$} &  \{2,2,4,5--22\}  &  \{12,14,16\}       &  1.401(20)    &               &    1.067(25)  &                 \\
& &  \{2,2,4,5--22\}  &  \{10,12,14,16\}    &               &  1.15(8)      &               &    1.202(10)    \\
& &  \{4,2,4,2--22\}  &  \{10,12,14,16\}    &  1.464(31)    &  1.37(13)     &    0.962(38)  &    1.202(15)    \\
& &  \{4,3,2,2--22\}  &  \{10,12,14,16\}    &  1.480(46)    &  0.78(22)     &    1.032(30)  &    1.196(23)    \\
& &  \{4,$3^\ast$,2,2--22\}  &  \{10,12,14,16\}    &  1.477(38)    &  1.18(14)     &    1.043(26)  &    1.196(16)    \\
\end{tabular}
\end{ruledtabular}
\caption{Estimates of the bare isovector charges $g_{A,S,T,V}$ for
  four choices of $\{N_\text{2pt}, N_\text{3pt}, \tskip,
  \tmin-\tmax\}$, where $\{N_\text{2pt}, N_\text{3pt}\}$ are the
  number of states kept in fits to the two- and three-point functions,
  respectively. The $N_\text{3pt}=3^\ast$ fit is a 3-states fit
  with $\matrixe{2}{\mathcal{O}_\Gamma}{2}=0$.  Note that the choice
  of $t_\text{min}=3$ is the same as in
  Table~\protect\ref{tab:2ptmulti} except for the $a127m285$ data. Our final estimates 
  are given in the last row corresponding to the \{4,$3^\ast$\} fit.}
\label{tab:BC3state}
\end{table*}

\begin{table}
\centering
\begin{ruledtabular}
\begin{tabular}{c|c|ccc}
ID        & Type      &  $g_A/g_V$   &    $g_S/g_V$  &    $g_T/g_V$ \\
\hline
a127m285  & $S_5 S_5$ &  1.132(11)   &  0.858(31)    &  0.918(8)   \\
\hline                   
a094m280  & $S_5 S_5$ &  1.147(31)   &  1.046(77)    &  0.885(17)  \\ 
a094m280  & $S_7 S_7$ &  1.149(26)   &  0.994(99)    &  0.875(32)  \\
a094m280  & $S_9 S_9$ &  1.125(19)   &  1.048(89)    &  0.869(23)  \\
\hline                   
a091m170  & $S_7 S_7$ &  1.127(16)   &  0.898(88)    &  0.884(13)  \\
\hline                   
a091m170L & $S_7 S_7$ &  1.235(35)   &  0.983(118)   &  0.872(23)  \\
\end{tabular}
\end{ruledtabular}
\caption{Estimates of the ratios of the unrenormalized isovector
  charges $g_{A,S,T}/g_V$ from the \{4,$3^\ast$\} fits with our
  ``best'' choices of $\tmin$, $\tsep$ and $\tskip$. }
\label{tab:BC3Ratios}
\end{table}

\begin{table*}
\centering
\begin{ruledtabular}
  \begin{tabular}{c|c|.{7}.{6}.{7}.{6}|.{6}.{6}.{6}.{6}}
    \multicolumn{1}{c|}{ID}       & 
    \multicolumn{1}{c|}{Type}     &    
    \multicolumn{1}{c}{$\matrixe{0}{\mathcal{O}_A}{1}$} & 
    \multicolumn{1}{c}{$\matrixe{0}{\mathcal{O}_S}{1}$} & 
    \multicolumn{1}{c}{$\matrixe{0}{\mathcal{O}_T}{1}$} & 
    \multicolumn{1}{c|}{$\matrixe{0}{\mathcal{O}_V}{1}$} &
    \multicolumn{1}{c}{$\matrixe{1}{\mathcal{O}_A}{1}$} & 
    \multicolumn{1}{c}{$\matrixe{1}{\mathcal{O}_S}{1}$} & 
    \multicolumn{1}{c}{$\matrixe{1}{\mathcal{O}_T}{1}$} & 
    \multicolumn{1}{c}{$\matrixe{1}{\mathcal{O}_V}{1}$}    \\
\hline                                                                    
\multirow{5}{*}{a127m285}  & \multirow{5}{*}{$S_5 S_5$} &  
-0.179(21)    &               &    0.182(16)  &   & 
-0.9(2.4)     &               &   -0.2(1.2)   &              \\
& &  
&  -0.35(4)      &               &    -0.014(2) & 
&   0.6(1.1)     &               &     0.80(34)              \\
& &  
-0.172(18)    &  -0.37(4)      &    0.210(15)  &    -0.015(2)  & 
0.75(48)     &   0.8(9)      &    0.42(27)   &     0.87(28)          \\
& &  
-0.295(58)    &  -0.45(15)      &    0.167(40)  &    -0.014(6)  & 
1.5(3.0)      &   1.8(1.4)      &    0.54(86)   &     0.86(55)         \\
& &  
-0.295(57)    &  -0.45(15)      &    0.166(47)  &    -0.014(6)  & 
1.46(54)      &   1.8(1.4)      &    0.54(41)   &     0.86(28)         \\
\hline                                                         
\multirow{5}{*}{a094m280}  & \multirow{5}{*}{$S_5 S_5$} &  
-0.130(20)    &               &    0.157(15)  &   & 
 0.62(69)     &               &    0.70(40)   &              \\
& &  
&  -0.42(5)      &               &    -0.006(1)  & 
&   1.0(8)     &               &     1.35(14)              \\
& &  
-0.139(34)    &  -0.57(13)     &    0.224(46)  &    -0.008(2)  & 
 1.10(22)     &   1.4(6)     &    0.71(12)   &     1.29(9)           \\
& &  
-0.136(36)    &  -0.42(8)     &    0.141(25)  &    -0.005(2)  & 
 1.00(41)     &  -0.2(2.0)    &    0.64(34)   &     0.99(32)           \\
& &  
-0.136(36)    &  -0.42(8)     &    0.140(25)  &    -0.005(2)  & 
 1.01(25)     &   1.1(6)    &    0.84(12)   &     1.29(10)          \\
\hline                                                         
\multirow{5}{*}{a094m280}  & \multirow{5}{*}{$S_7 S_7$} &  
-0.111(36)    &               &    0.221(36)  &   & 
 0.3(1.3)     &               &    0.03(80)   &              \\
& &  
&  -0.52(12)     &               &    -0.003(3)  & 
&  1.4(1.7)      &               &     1.25(26)              \\
& &  
-0.150(48)    &  -0.69(20)     &    0.347(81)  &    -0.004(3)  & 
 1.09(42)     &   1.8(1.4)     &    0.37(26)   &     1.21(19)          \\
& &  
-0.111(47)    &  -0.39(15)     &    0.236(44)  &    -0.004(3)  & 
 0.72(92)     &  -3.2(4.4)     &   -0.24(80)   &    -0.10(92)          \\
& &  
-0.113(47)    &  -0.41(15)     &    0.231(43)  &    -0.004(3)  & 
 1.03(42)     &   1.8(1.3)     &    0.63(25)   &     1.18(20)          \\
\hline                                                         
\multirow{5}{*}{a094m280}  & \multirow{5}{*}{$S_9 S_9$} &  
-0.026(39)    &               &    0.276(28)  &   & 
-0.4(2.5)     &               &    0.6(1.6)   &              \\
& &  
&  -0.42(9)      &               &    -0.004(3)  & 
&  -0.6(3.3)     &               &     1.51(49)              \\
& &  
-0.003(34)    &  -0.53(13)     &    0.346(52)  &    -0.005(3)  & 
 1.04(45)     &   0.4(2.3)     &    0.50(39)   &     1.41(33)          \\
& &  
0.048(49)    &  -0.20(15)     &    0.287(36)  &    -0.003(3)  & 
0.7(1.3)     &  -7.9(7.5)     &    0.70(41)   &     1.4(2.0)           \\
& &  
0.048(49)    &  -0.21(15)     &    0.287(36)  &    -0.003(3)  & 
0.71(50)     &  -0.1(2.1)     &    0.70(37)   &     1.45(34)           \\
\hline                                                         
\multirow{5}{*}{a091m170}  & \multirow{5}{*}{$S_7 S_7$} &  
-0.133(33)   &               &    0.154(22)  &   & 
-2.1(2.6)    &               &   -0.2(1.7)   &              \\
& &  
&  -0.48(7)     &               &    -0.006(2) & 
&   0.1(3.5)    &               &     1.11(46)              \\
& &  
-0.142(25)    &  -0.51(8)     &    0.187(18)  &    -0.006(2) & 
 0.65(63)     &   0.3(3.4)    &    0.73(37)   &     1.13(39)           \\
& &  
-0.117(25)    &  -0.35(7)     &    0.180(16)  &    -0.006(2) & 
 0.60(62)     &   1.3(3.5)    &    0.73(37)   &     1.16(44)           \\
& &  
-0.117(25)    &  -0.35(7)     &    0.180(16)  &    -0.006(2) & 
 0.60(62)     &   1.3(3.5)    &    0.73(37)   &     1.16(44)           \\
\hline                                                         
\multirow{5}{*}{a091m170L} & \multirow{5}{*}{$S_7 S_7$} &  
-0.118(26)    &               &    0.235(23)  &   & 
-1.0(2.4)     &               &    0.54(84)   &              \\
& &  
&  -0.44(8)      &               &    -0.002(2)  & 
&   1.4(2.2)     &               &     1.26(34)              \\
& &  
-0.222(44)    &  -0.87(16)     &    0.474(54)  &    -0.003(3)  & 
 0.98(31)     &   2.7(1.4)     &   -0.17(33)    &     1.25(20)         \\
& &  
-0.258(71)    &  -0.38(21)    &    0.202(48)  &    0.004(4)  & 
 0.96(83)     &  12.7(6.8)    &    0.86(53)   &    1.25(46)            \\
& &  
-0.259(71)    &  -0.37(21)    &    0.208(48)  &    0.004(4)  & 
 1.02(35)     &   2.5(1.5)    &    0.51(24)   &    1.26(20)            \\
\end{tabular}
\end{ruledtabular}
\caption{Estimates of the matrix elements
  $\matrixe{0}{\mathcal{O}_\Gamma}{1}$ and
  $\matrixe{0}{\mathcal{O}_\Gamma}{1}$ for the three isovector
  operators. Results for the four choices of the fit parameters
  $\{N_{2pt}, N_{3pt}, \tskip, \tmin-\tmax\}$ are arranged as defined
  in Table~\protect\ref{tab:BC3state} with the the final estimates
  from the $\{N_{2pt}, N_{3pt}\} = \{4,3^\ast\}$ fit given in the last
  row.}
\label{tab:Mate3state}
\end{table*}

\begin{table*}
\centering
\begin{ruledtabular}
  \begin{tabular}{c|c|.{7}.{6}.{7}.{6}|.{6}.{6}.{6}.{6}}
    \multicolumn{1}{c|}{ID}       & 
    \multicolumn{1}{c|}{Type}     &    
    \multicolumn{1}{c}{$\matrixe{0}{\mathcal{O}_A}{2}$} & 
    \multicolumn{1}{c}{$\matrixe{0}{\mathcal{O}_S}{2}$} & 
    \multicolumn{1}{c}{$\matrixe{0}{\mathcal{O}_T}{2}$} & 
    \multicolumn{1}{c|}{$\matrixe{0}{\mathcal{O}_V}{2}$} &
    \multicolumn{1}{c}{$\matrixe{1}{\mathcal{O}_A}{2}$} & 
    \multicolumn{1}{c}{$\matrixe{1}{\mathcal{O}_S}{2}$} & 
    \multicolumn{1}{c}{$\matrixe{1}{\mathcal{O}_T}{2}$} & 
    \multicolumn{1}{c}{$\matrixe{1}{\mathcal{O}_V}{2}$}    \\
\hline                                                                    
\multirow{2}{*}{a127m285}  & \multirow{2}{*}{$S_5 S_5$} &  
 0.97(66)     &     1.2(1.4)    &    0.03(37)   &    -0.012(75)  & 
-3.7(4.7)     &   -11.3(11.3)   &    3.7(3.4)   &     0.01(62)       \\
& &  
0.97(66)      &     1.2(1.4)    &    0.03(37)   &    -0.012(90)  & 
-3.7(4.7)     &   -11.3(11.3)   &    3.7(3.4)   &     0.02(79)       \\
\hline                                                         
\multirow{2}{*}{a094m280}  & \multirow{2}{*}{$S_5 S_5$} &  
-0.02(2)      &   -0.18(7)      &    0.09(2)    &    -0.007(3)  & 
 0.21(23)     &   -0.38(44)     &    0.28(14)   &    -0.01(2)        \\
& &  
-0.02(2)      &  -0.18(7)       &    0.09(2)    &    -0.007(3)  & 
 0.21(23)     &  -0.36(44)      &    0.29(14)   &    -0.01(2)        \\
\hline                                                         
\multirow{2}{*}{a094m280}  & \multirow{2}{*}{$S_7 S_7$} &  
-0.05(4)      &  -0.36(10)     &    0.13(3)    &    -0.006(5)  & 
 0.34(42)     &   0.40(97)     &    0.18(26)   &    -0.01(3)         \\
& &  
-0.05(4)      &  -0.37(10)     &    0.14(3)    &    -0.006(5)  & 
 0.35(43)     &   0.55(99)     &    0.19(25)   &    -0.01(3)         \\
\hline                                                         
\multirow{2}{*}{a094m280}  & \multirow{2}{*}{$S_9 S_9$} &  
-0.13(5)      &  -0.39(15)     &    0.07(4)    &    -0.008(6)  & 
 0.48(63)     &  -0.7(1.8)     &    0.51(45)   &     0.06(6)         \\
& &  
-0.13(5)      &  -0.38(15)     &    0.07(4)    &    -0.008(6)  & 
 0.48(63)     &  -0.7(1.8)     &    0.51(45)   &     0.06(6)         \\
\hline                                                         
\multirow{2}{*}{a091m170}  & \multirow{2}{*}{$S_7 S_7$} &  
 0.11(6)      &  -0.09(16)     &    0.12(4)    &     0.006(8) & 
-0.4(1.2)     &  -3.8(4.0)     &    1.0(1.1)   &    -0.18(12)        \\
& &  
 0.11(6)      &  -0.09(16)     &    0.12(4)    &     0.006(8) & 
-0.4(1.2)     &  -3.8(4.0)     &    1.0(1.1)   &    -0.18(12)        \\
\hline                                                         
\multirow{2}{*}{a091m170L} & \multirow{2}{*}{$S_7 S_7$} &  
 0.01(4)      &  -0.20(12)    &    0.18(3)    &    -0.013(4)  & 
 0.03(25)     &  -1.55(87)    &    0.12(18)   &     0.03(2)          \\
& &  
 0.01(4)      &  -0.18(12)    &    0.18(3)    &    -0.013(4)  & 
 0.03(24)     &  -1.74(89)    &    0.11(18)   &     0.03(2)          \\
\end{tabular}
\end{ruledtabular}
\caption{Estimates of the matrix elements
    $\matrixe{0}{\mathcal{O}_\Gamma}{2}$ and
    $\matrixe{1}{\mathcal{O}_\Gamma}{2}$ for the three isovector
    operators obtained using 3-state fits. The two rows gives results
    for the $N_\text{3pt}=3$ and $N_\text{3pt}=3^\ast$ fits. Estimates of the
    matrix element $\matrixe{2}{\mathcal{O}_\Gamma}{2}$ from the
    $N_\text{3pt}=3$ fits are not presented as they are ill-determined.}
\label{tab:Mate3state-2}
\end{table*}

The results of fits to the two-point function data are shown in
Table~\ref{tab:2ptmulti} for the three cases, 2-, 3- and 4-state fits
using our best choices of $\tmin$.  The results of the 2-state fits
are reproduced from Table~\ref{tab:2ptfits}. 
Overall, the results presented in Tables~\ref{tab:2ptmulti} 
exhibit the following behavior:
\begin{itemize}
\item
The 2-, 3- and 4-state fits to the two-point data on the $a127m285$
ensemble data are very stable and the central values show little
variation with changes in $\tmin$ and/or the number of states.
Similarly, the estimates of all four charges are stable within
$1\sigma$.  Fits to the data from the $a094m280$ and $a091m170$
ensembles were also stable but the variation in the results was
larger. 
\item
Estimates of the ground-state mass $M_0$ and the mass gap $\Delta M_1$
obtained from the 3-state and 4-state fits are essentially
identical. Even estimates of $\Delta M_2$ are consistent.
\item
All three ratios of amplitudes, $r_i$, decrease with the smearing size
$\sigma$ between $\sigma=5$ and $7$ and then are essentially flat
between $\sigma=7$ and $9$ on the $a094m280$ ensemble. Note also that
the amplitudes for $S_9 S_9$ are essentially the same for $\tmin=2$
and 3.
\item
The 3- and 4-state fits to the two-point data on the $a091m170L$
ensemble are sensitive to the choice of the priors, their widths
and $\tmin$. Furthermore, for any choice of fit parameters for the
two-point functions, the results for the four charges are sensitive to
the choice of $\tskip$. As remarked in Sec.~\ref{sec:2-state}, we
attribute this sensitivity to low statistics in the $a091m170L$
calculation and reiterate that the quoted errors are underestimates
since the auto-correlations between configurations, that are
significant, have not been taken into account.
\end{itemize}

With 3- and 4-state fits to the two-point data in hand we carried out
three analyses to estimate the isovector charges: $\{\Ntwo,\Nthree\}=
\{4,2\} $, $\{4,3\}$ and $\{4,3^\ast\}$.  The
$\{\Ntwo,\Nthree\}=\{4,3^\ast\}$ is a 3-state fit with $\langle 2 |
\mathcal{O}_\Gamma | 2 \rangle$ set to zero. The reason for this
additional analysis is that $\langle 2 | \mathcal{O}_\Gamma | 2
\rangle$ is essentially undetermined in the 
$\{\Ntwo,\Nthree\}= \{4,3\}$ fits. This is because (i) the
contribution of $\langle 2 | \mathcal{O}_\Gamma | 2 \rangle$ for any
of the four charges is suppressed by at least $e^{-6}$ relative to
$\langle 1 | \mathcal{O}_\Gamma | 1 \rangle$ as can be deduced from
Eq.~\eqref{eq:3pt} and the data in Table~\ref{tab:2ptmulti}; (ii) the
three matrix elements, $\langle 0 | \mathcal{O}_\Gamma | 0 \rangle$,
$\langle 1 | \mathcal{O}_\Gamma | 1 \rangle$, and $\langle 2 |
\mathcal{O}_\Gamma | 2 \rangle$ are only sensitive to $\tsep$, and the
data at the four values of $\tsep$ overlap within $1\sigma$. Thus,
three matrix elements cannot be determined reliably from overlapping data at four
values of $\tsep$.  (iii) Even $\langle 1 | \mathcal{O}_\Gamma | 1 \rangle$
is poorly determined as shown by the data in
Table~\ref{tab:Mate3state}.

We find that setting $\langle 2 | \mathcal{O}_\Gamma | 2 \rangle = 0$ leads
to a significant improvement over the unconstrained
$\{\Ntwo,\Nthree\}=\{4,3\}$ fit. Thus, our final unrenormalized
estimates for the four charges are taken from the $\{\Ntwo,\Nthree\}=\{4,3^\ast\}$
fits and given in Tab.~\ref{tab:BC3state}. These fits are shown in the
right panels of Figs.~\ref{fig:gA6},~\ref{fig:gS6},~\ref{fig:gT6}
and~\ref{fig:gV6}.  Estimates for the matrix elements $\langle 0 |
\mathcal{O}_\Gamma | 1 \rangle$, $\langle 1 | \mathcal{O}_\Gamma | 1 \rangle$,
$\langle 0 | \mathcal{O}_\Gamma | 2 \rangle$ and $\langle 1 | \mathcal{O}_\Gamma
| 2 \rangle$ are given in Tabs.~\ref{tab:Mate3state}
and~\ref{tab:Mate3state-2}.

The data in Table~\ref{tab:BC3state} show that estimates for the four
charges from the four analyses, $\{\Ntwo,\Nthree\}=\{2,2\}, \{4,2\} $,
$\{4,3\}$ and $\{4,3^\ast\}$, are consistent.  Based on this stability
and the small size of the variation in estimates under changes in the
values of $\tmin$ and $\tskip$ that have been investigated, we
conclude that estimates for $g_{A,T,V}$ can be obtained with $O(3\%)$
uncertainty from $\{\Ntwo,\Nthree\}=\{4,3^\ast\}$ fits to data
comprising $O(10^5)$ measurements. Our statistical tests also indicate
that this estimate of the number of measurements required will
increase as the lattice spacing and the pion mass are decreased. The
data in Figs.~\ref{fig:gA6},~\ref{fig:gS6},~\ref{fig:gT6}
and~\ref{fig:gV6} further indicate that increasing the statistics to
$O(10^6)$ measurements on each ensemble will lead to results for
$g_{A,T,V}$ with $O(1\%)$ uncertainty. This factor of ten increase in
statistics will have to come primarily from increasing the number of
independent gauge configurations analyzed since the $O(100)$
measurements per configuration that we have made in this study were
shown to be optimal in Ref.~\cite{Yoon:2016dij}.

Our final results for the isovector charges, using bare estimates from the
$\{\Ntwo,\Nthree\}=\{4,3^\ast\}$ fits given in Tables~\ref{tab:BC3state} and~\ref{tab:BC3Ratios}
and renormalized using the factors given in Table~\ref{tab:Zall}, are
given in Tables~\ref{tab:gren-1} and~\ref{tab:gren-2}.

\section{Renormalization Constants}
\label{sec:Zfac}

We calculated the renormalization constants $Z_\Gamma$ for the
isovector quark bilinear operators $\overline{u} \Gamma d$ on the
lattice using the non-perturbative RI-sMOM
scheme~\cite{Martinelli:1994ty,Sturm:2009kb}. Details of the procedure
for calculating the three-point functions and the renormalization
conditions used are given in Ref.~\cite{Bhattacharya:2013ehc}.  In
short, in the RI-sMOM scheme we require the projected
amputated three-point function $\Lambda^\text{R}$, renormalized at the scale $q^2$, to satisfy the condition
\begin{multline}
\left.\Lambda^\text{R}_\Gamma(p_a, p_b) \right|_{p_a^2 = p_b^2 = q^2} = 1 = \\
\left.\left(Z_\psi^{-1}Z_\Gamma~\Lambda^\text{PA}_\Gamma(p_i, p_f) \right)\right|_{p_a^2 = p_b^2 = q^2} ,
\end{multline}
where $p_a^\mu$ and $p_b^\mu$ are the 4-momenta in the two fermion
legs, $q^\mu = p_b^\mu -p_a^\mu$ and they satisfy the kinematic
constraint $p_a^2 = p_b^2 = q^2$.  Here $\Lambda^\text{PA}$ is the 
projected amputated three-point function discussed below and 
$Z_\psi$ is the wavefunction
renormalization constant defined by
\begin{equation}
\left.(Z_\psi)^{-1} \frac{i}{12} \Tr \left(\frac{\slashed{p}  \langle S(p) \rangle^{-1}  }{p^2}\right)\right|_{p^2 = q^2} = 1 \,.
\end{equation}
It is obtained from the momentum space quark propagator $S(p)$ 
calculated on lattices fixed to the Landau gauge defined as the
maximum of the sum of the trace of the gauge links.  The notation $\langle
\cdots \rangle$ denotes ensemble average. The projected amputated
three-point function is
\begin{equation}
\label{eq:pa-correlator}
\Lambda^\text{PA}_\Gamma(p_a, p_b) = \frac{1}{12} \Tr \Big(P_\Gamma \Lambda^\text{A}_\Gamma(p_a, p_b) \Big), 
\end{equation}
where the amputated vertex $\Lambda^\text{A}_\Gamma(p_a, p_b)$ is defined as
\begin{multline}
\Lambda^\text{A}_\Gamma(p_a, p_b) =\\
\langle \langle S(p_a) \rangle^{-1} S(p_a) \, \Gamma \, (\gamma_5 S^\dagger(p_b) \gamma_5) \langle(\gamma_5 S^\dagger(p_b) \gamma_5)\rangle^{-1} \rangle.
\end{multline}
The projector $P_\Gamma$ for the RI-sMOM scheme is $I$ (scalar),
$(q_\mu/{q^2}) \slashed{q}$ (vector), $(q_\mu/{q^2}) \gamma_5
\slashed{q}$ (axial-vector) or $(i/12)\gamma_{[\mu} \gamma_{\nu]}$
(tensor). In $\Lambda^\text{PA}_\Gamma(p_a, p_b) $, lattice artifacts
due to the breaking of the rotational symmetry to $O(4)$ can induce
dependence on the momenta $p_a$ and $p_b$ in addition to that on
$q^2$.  This systematic is significant in our data as discussed below.

We analyzed 132, 100 and 100 configurations on the three ensembles,
$a127m285$, $a094m280$ and $a091m170$, respectively, to get estimates
at the three distinct values of $a$ and $M_\pi$ simulated. With this
sample size, we find that the statistical errors in the
data are much smaller than the systematics discussed below.

Operationally, we first translate the lattice data, $Z_\Gamma^{\rm
  RI-sMOM}(p_a,p_b,q^2)$, to the $\overline{\text{MS}}$ scheme at
$2$~GeV. This is done by matching estimates at a given squared
momentum transfer $q^2$ to the $\overline{\text{MS}}$ scheme in the
continuum at the same $q^2$ (horizontal matching) using 2-loop
perturbative relations expressed in terms of the coupling constant
$\alpha_{\overline{\text{MS}}}(q^2)$~\cite{Gracey:2011fb}. These
results in the $\overline{\text{MS}}$ scheme are then run to $2$~GeV
using the 3-loop anomalous dimension relations for the scalar and
tensor bilinears~\cite{Gracey:2000am,Agashe:2014kda} and labeled 
$Z_\Gamma(p_a,p_b,q^2)$. 

The calculation of $\alpha_{\overline{\text{MS}}}(q^2)$ was carried out
as follows. Starting with the 5-flavor
$\alpha_{\overline{\text{MS}}}(M_Z=91.1876\ {\rm GeV})=0.1185$, we
used the 4-loop expression in ${\overline{\text{MS}}}$
scheme~\cite{Chetyrkin:2000yt} to run to the bottom quark threshold at
$m_b=4.18$~GeV, and then to $m_c=1.275$~GeV using the 4-flavor
evolution.  This 4-flavor result was converted to 3-flavor at this
scale and then run to the final desired $q^2$ using the 3-flavor
evolution.

Ideally, after removing the dependence on $p_a$ and $p_b$
from $Z_\Gamma(p_a,p_b,q^2)$, one expects a window, $\eta \Lambda_{\rm
  QCD} \ll q \ll \xi \pi/a$, in which the data for the renormalized
$Z(\overline{\text{MS}},2\ {\rm GeV})$ are independent of $q$; that
is, at sufficiently small values of the lattice spacing $a$, the data
should show a plateau versus $q$. The lower cutoff $\eta \Lambda_{\rm
  QCD}$ is dictated by nonperturbative effects and the upper cutoff
$\xi \pi/a$ by discretization effects. Here $\eta $ and $\xi$ are,
{\it a priori}, unknown dimensionless numbers of $O(1)$ that depend on
the lattice action and the gauge-link smearing procedure.

\begin{figure*}
\centering
\subfigure{
 \includegraphics[width=0.46\linewidth]{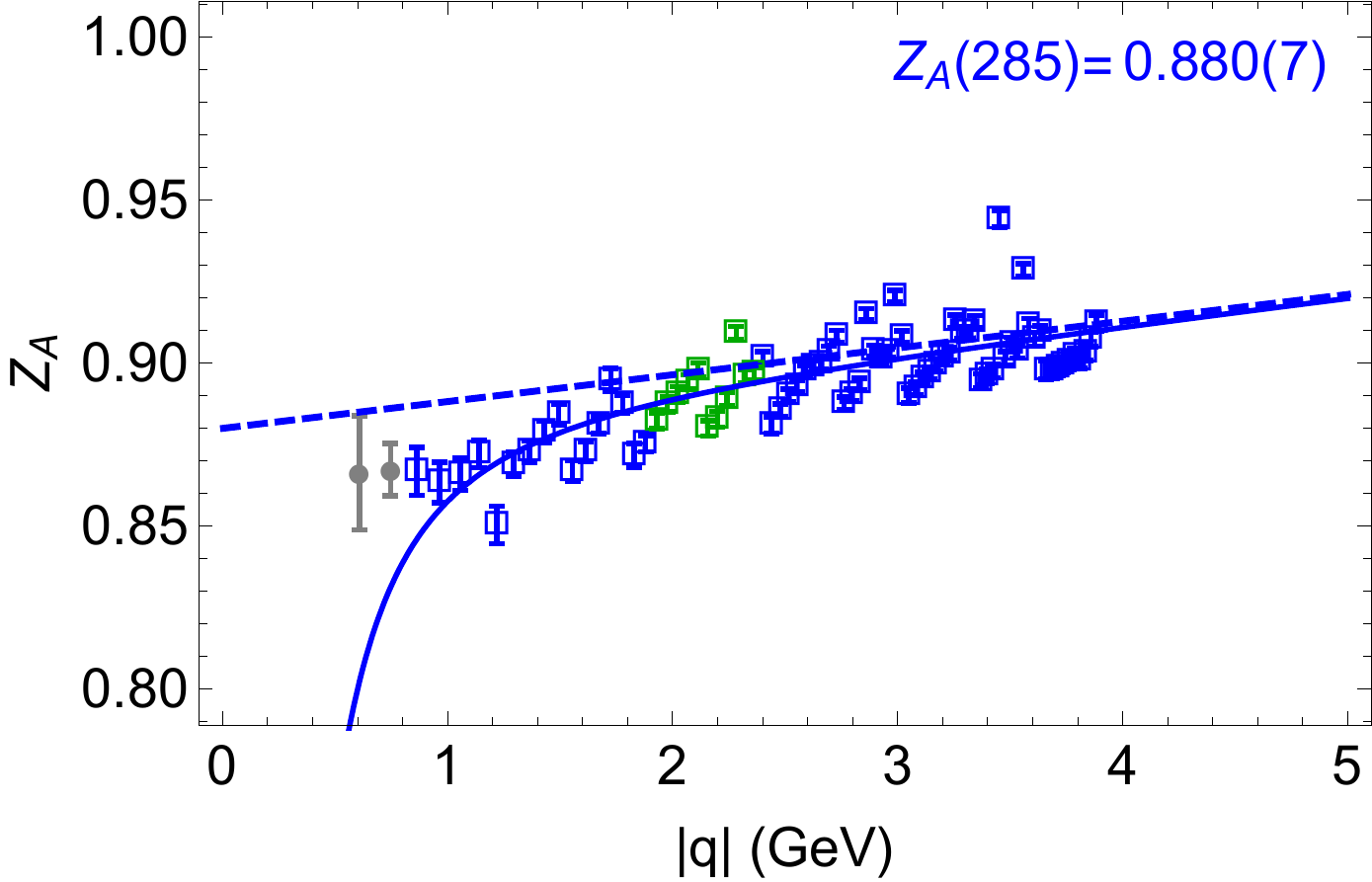}
 \includegraphics[width=0.46\linewidth]{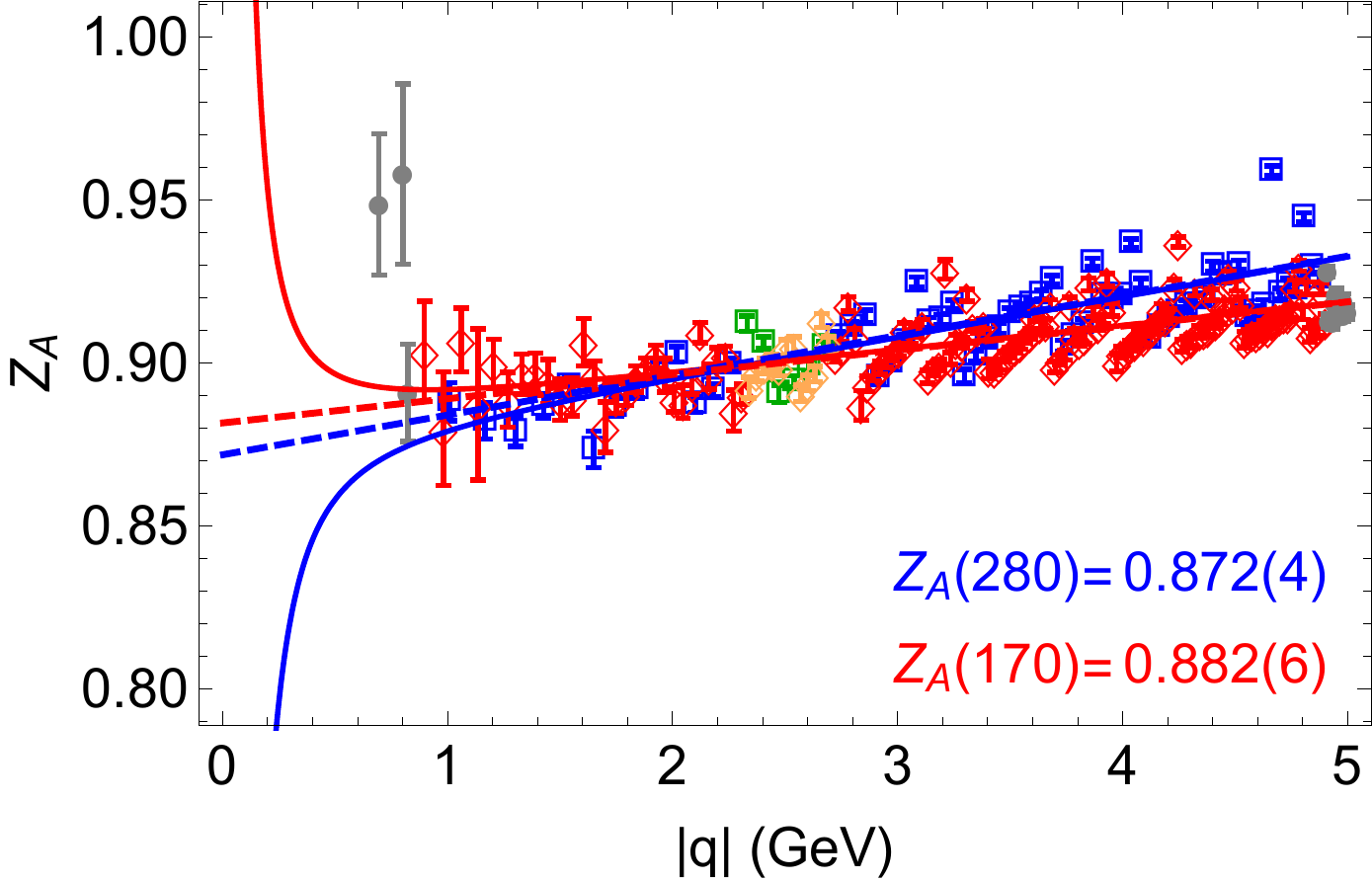}
}
\subfigure{
 \includegraphics[width=0.46\linewidth]{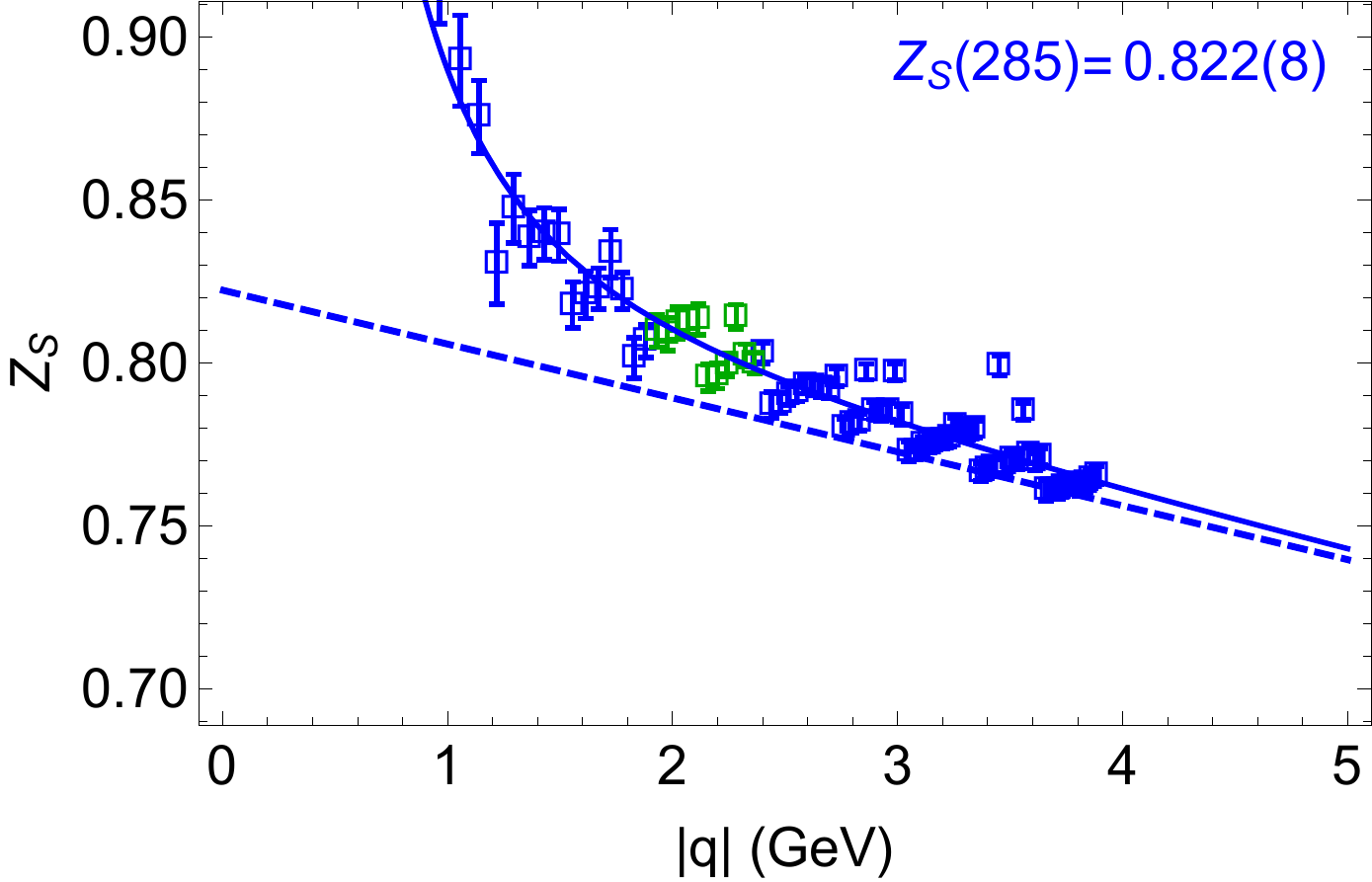}
 \includegraphics[width=0.46\linewidth]{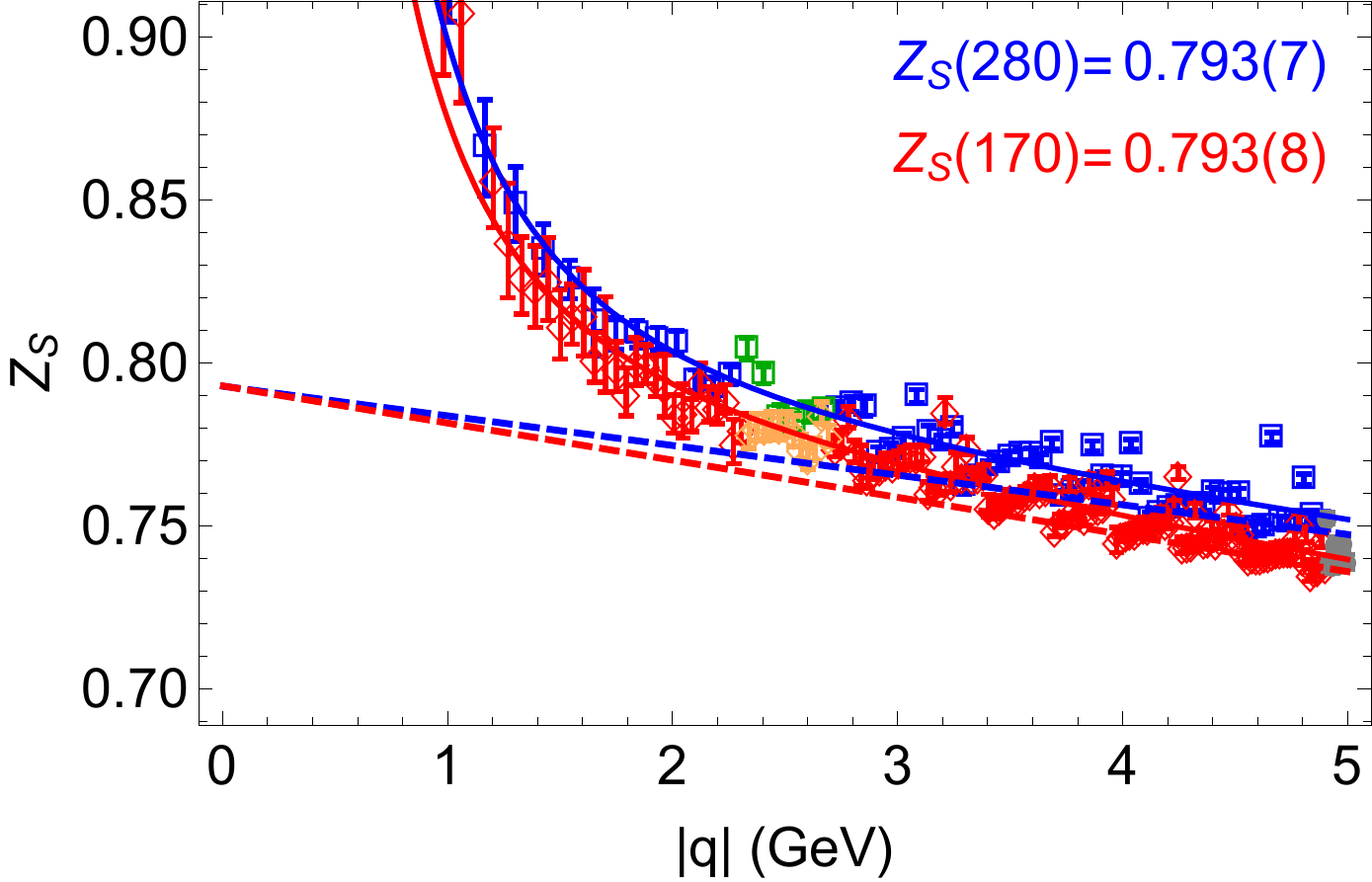}
}
\subfigure{
 \includegraphics[width=0.46\linewidth]{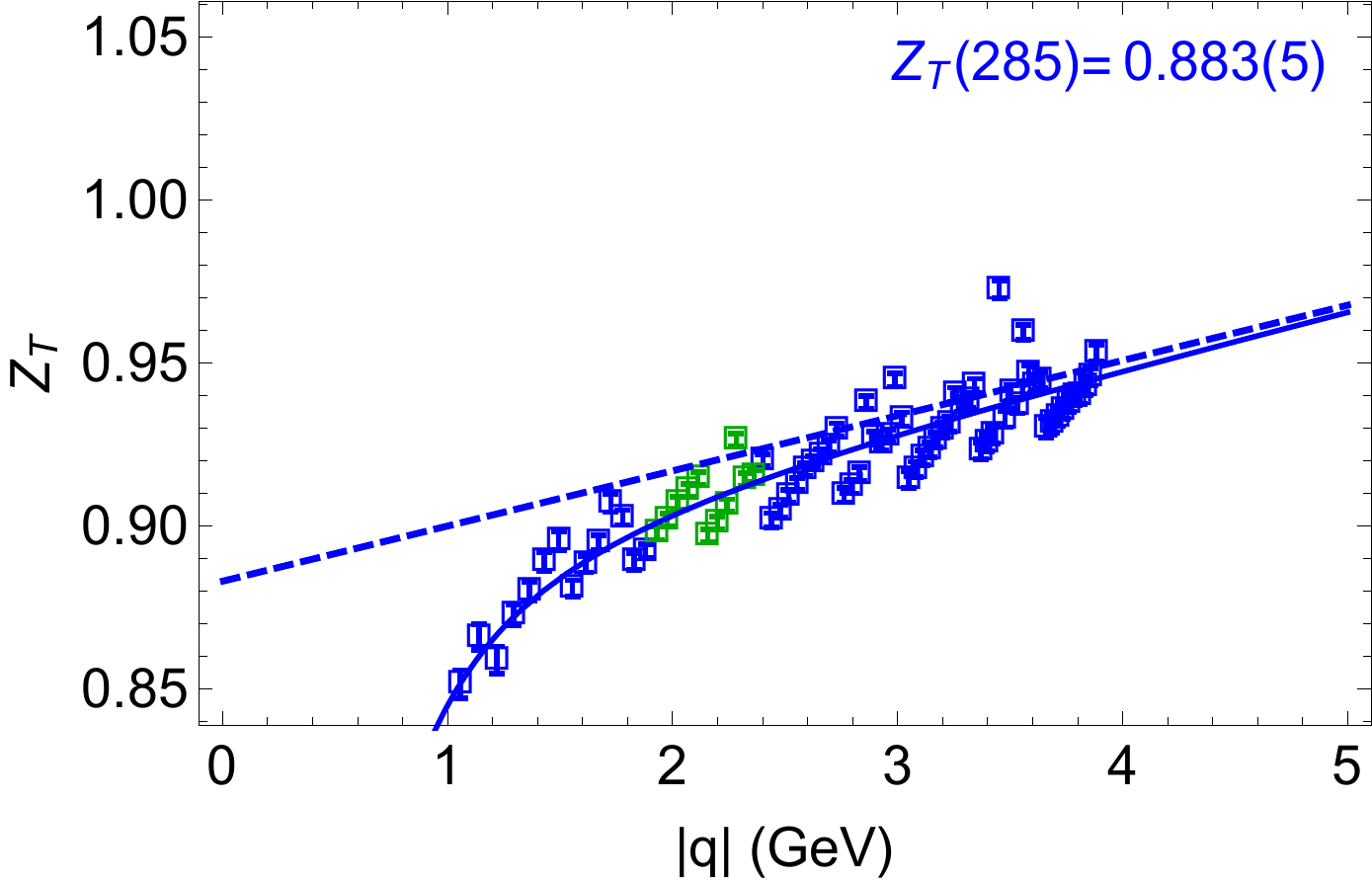}
 \includegraphics[width=0.46\linewidth]{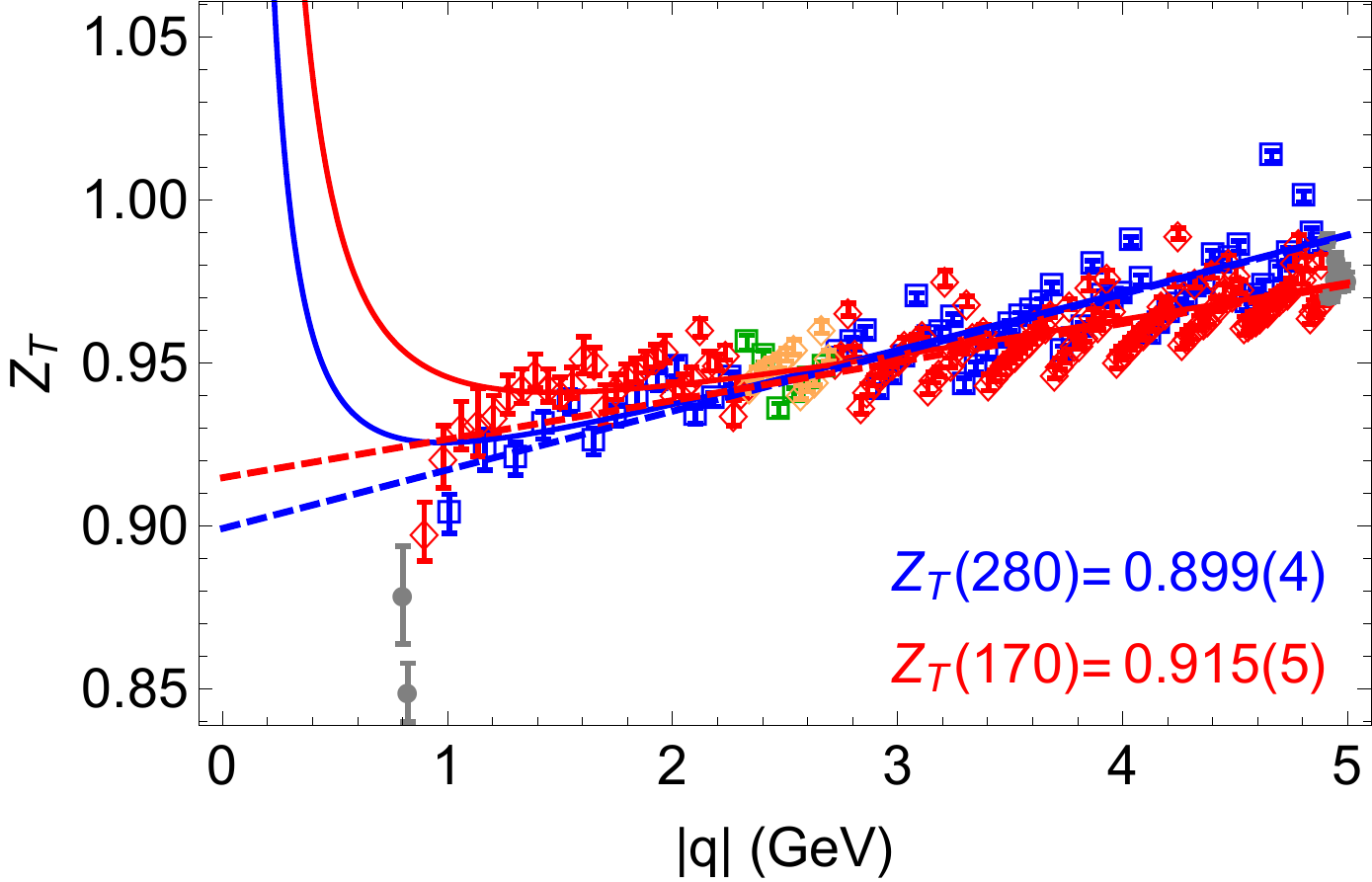}
}
\subfigure{
 \includegraphics[width=0.46\linewidth]{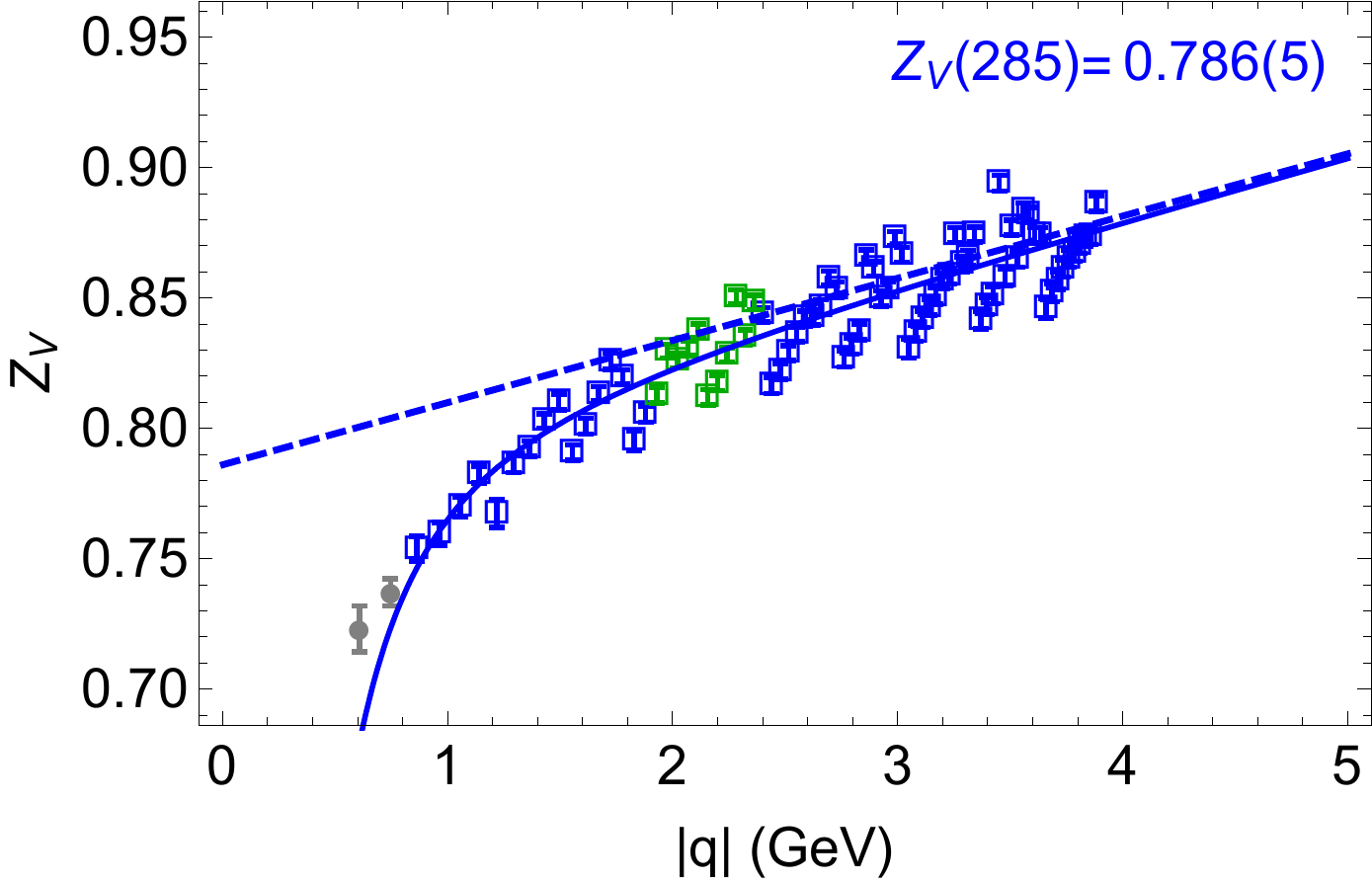}
 \includegraphics[width=0.46\linewidth]{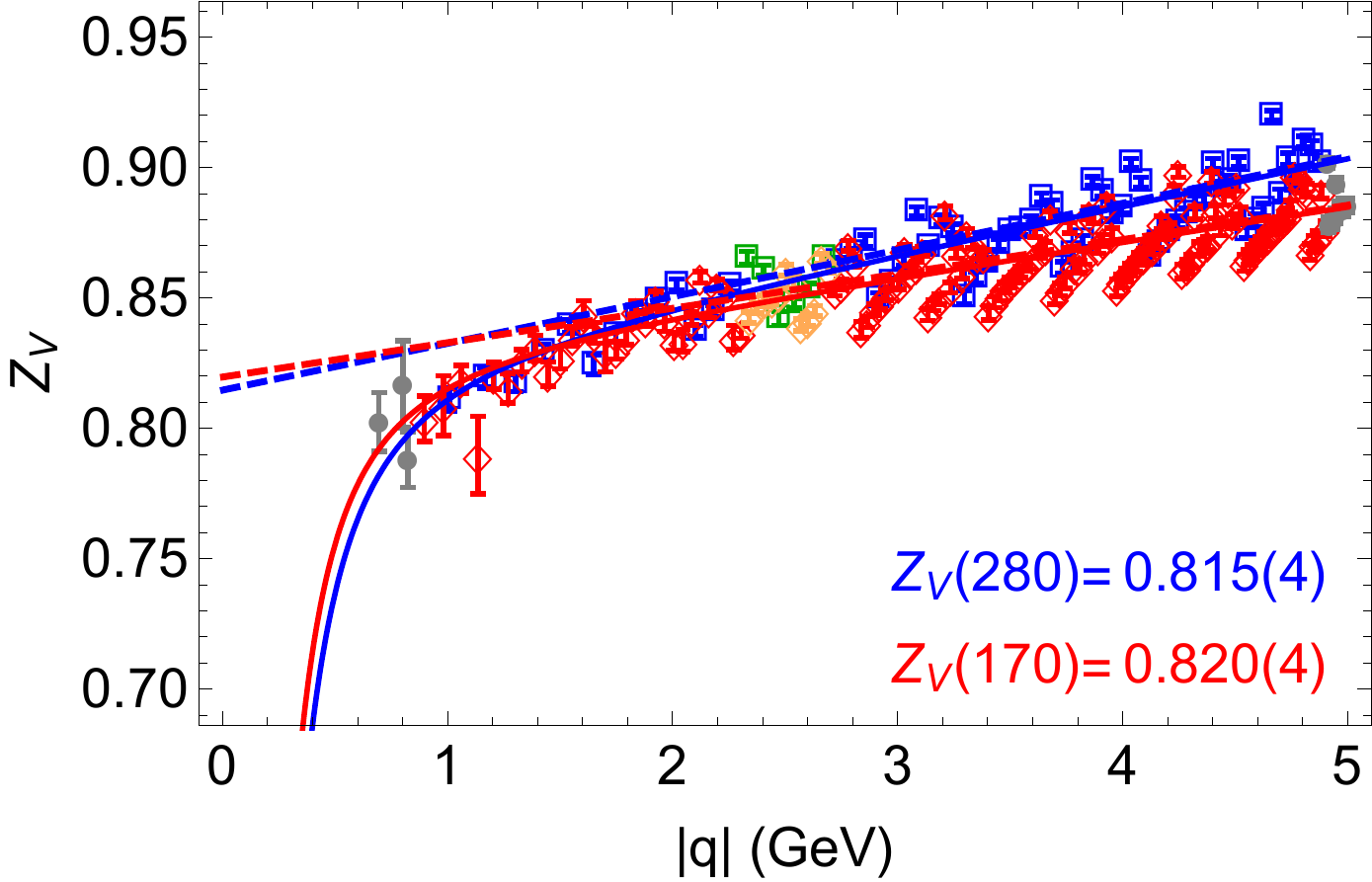}
}
\caption{Data for the renormalization constants $Z_A$, $Z_S$, $Z_T$
  and $ Z_V$ in the $\overline{\text{MS}}$ scheme at $2\GeV$, keeping
  only points that minimizes $\sum_\mu (p^a_\mu)^4 + \sum_\mu
  (p^b_\mu)^4$ for each $q^2$ as described in the text.  Estimates
  from the $a127m285$ ensemble are shown as blue squares in the left
  panels.  The data points used in estimating the $Z$'s in method B
  are shown as green squares.  The right panel shows estimates for the
  $a094m280$ (blue squares) and the $a091m170$ (red diamonds)
  ensembles.  The data points included in the estimate of $Z$ using
  method B are shown as green squares ($a094m280$) and yellow diamonds
  ($a091m170$).  The results given in the labels are from method A
  using the fit ansatz $c/q^2 + Z + d_1 q $ as described in the
  text. The fit is shown by the solid blue (red) line and the result
  $Z + d_1 q $ by the dashed blue (red) line. }
\label{fig:Z}
\end{figure*}

\begin{figure*}
\centering
\subfigure{
 \includegraphics[width=0.46\linewidth]{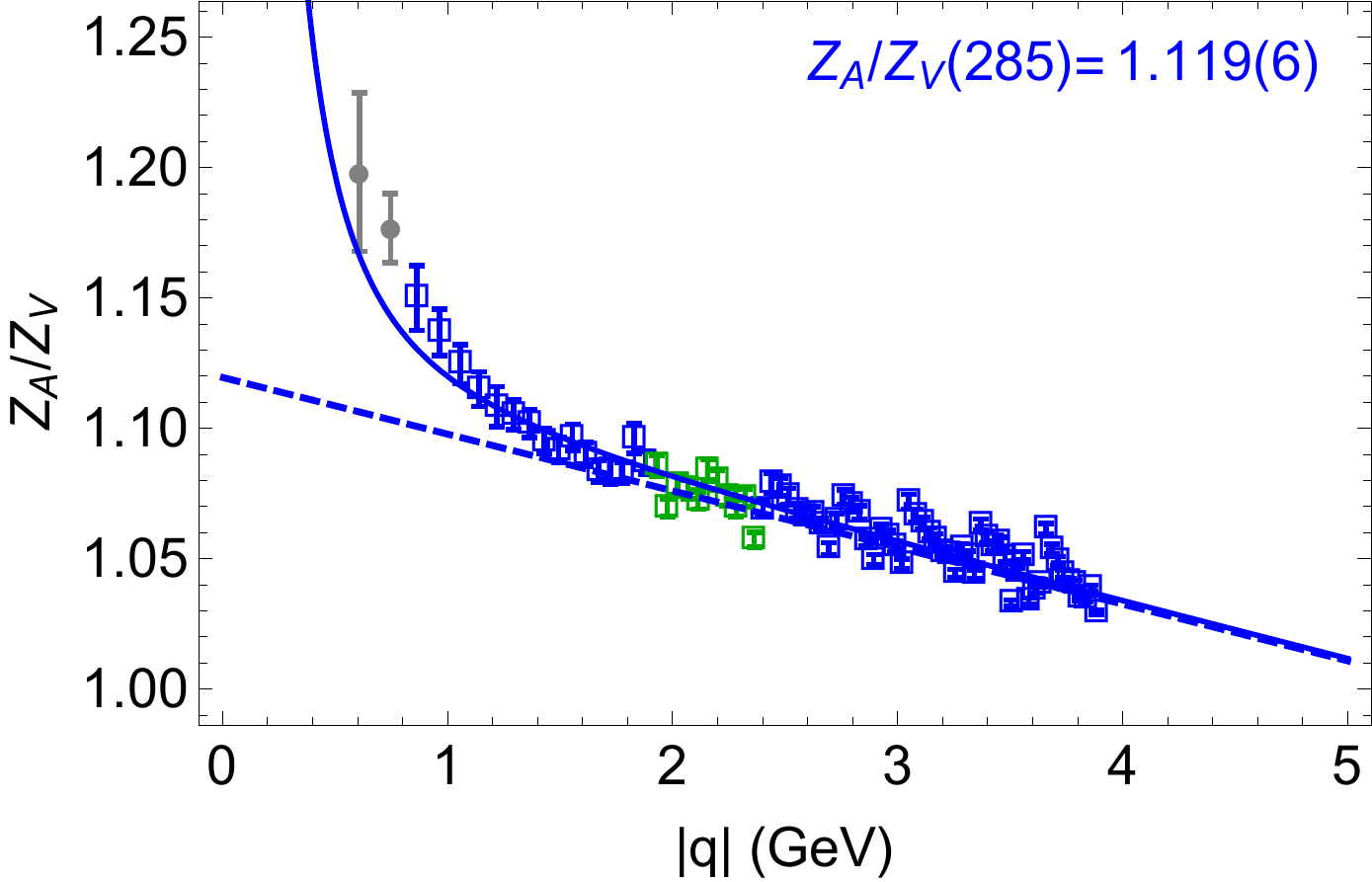}
 \includegraphics[width=0.46\linewidth]{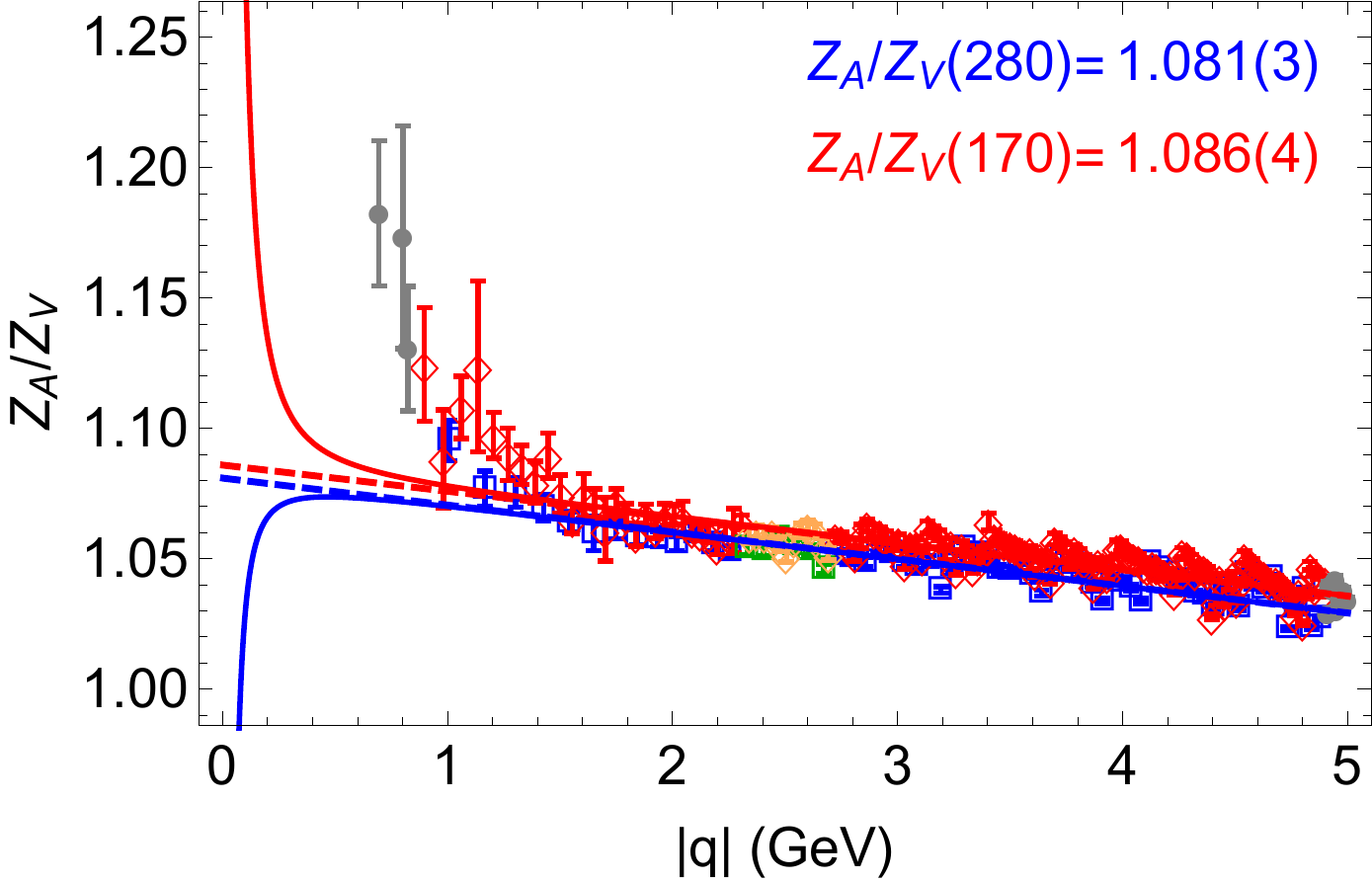}
}
\subfigure{
 \includegraphics[width=0.46\linewidth]{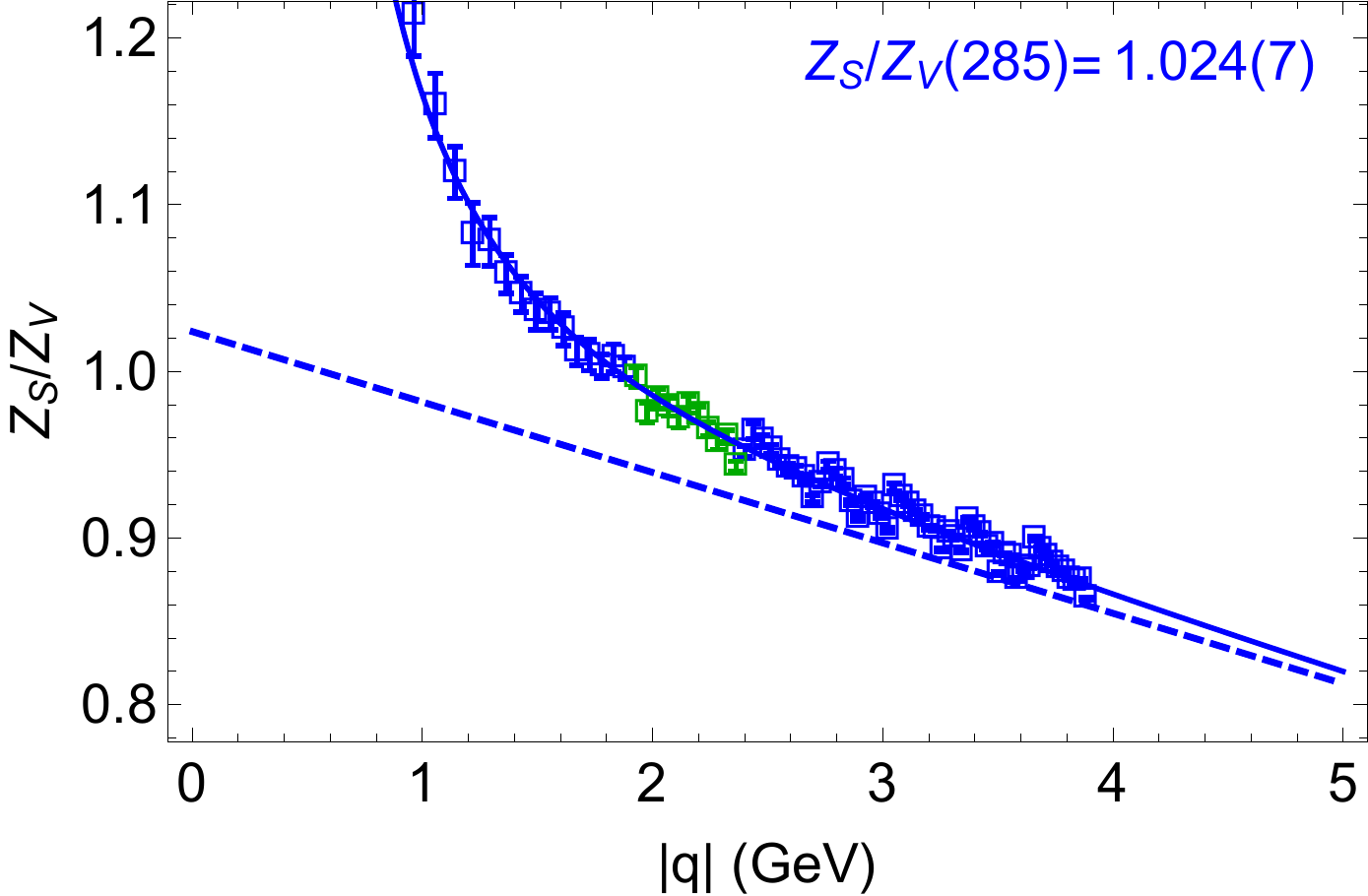}
 \includegraphics[width=0.46\linewidth]{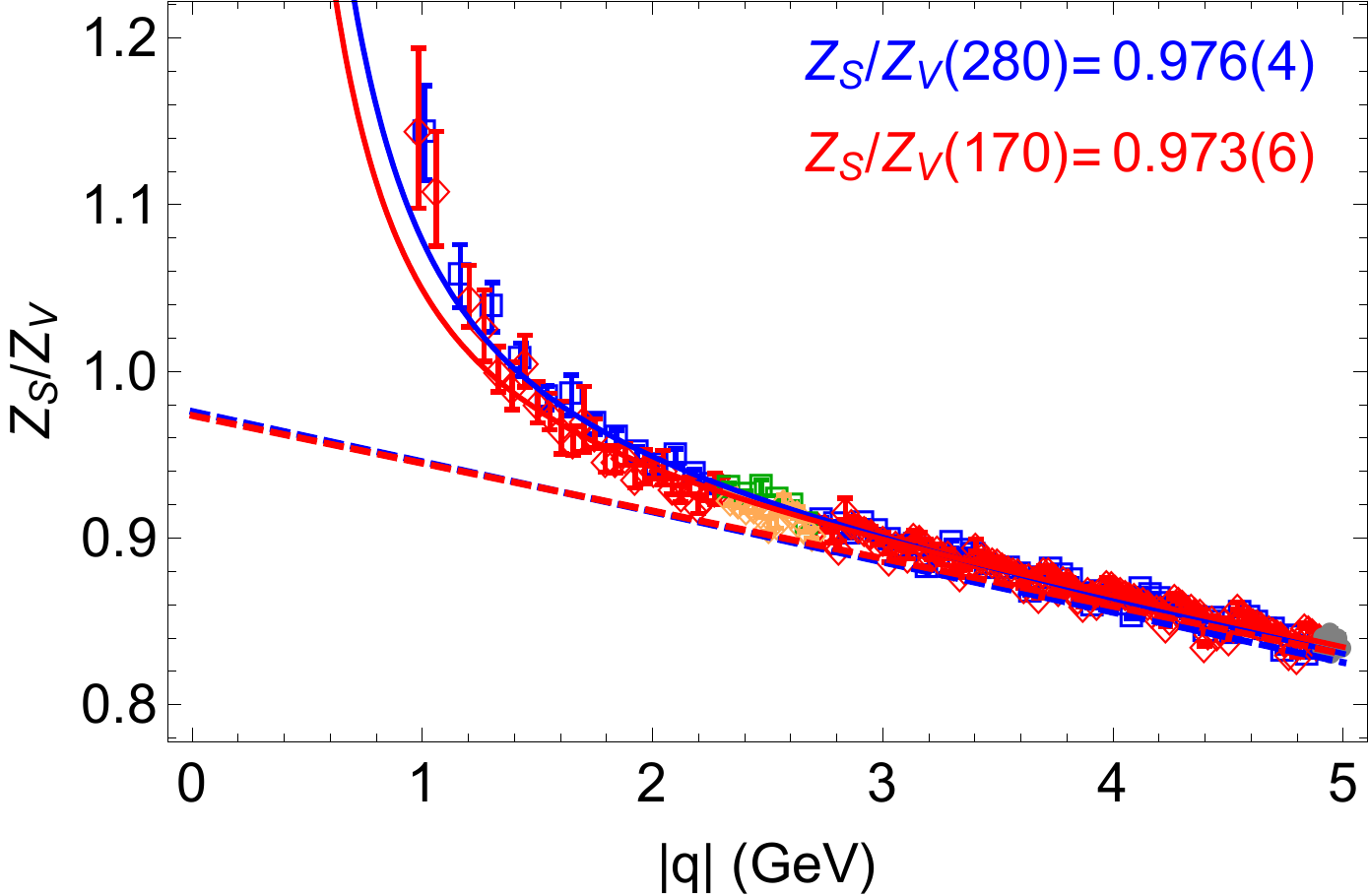}
}
\subfigure{
 \includegraphics[width=0.46\linewidth]{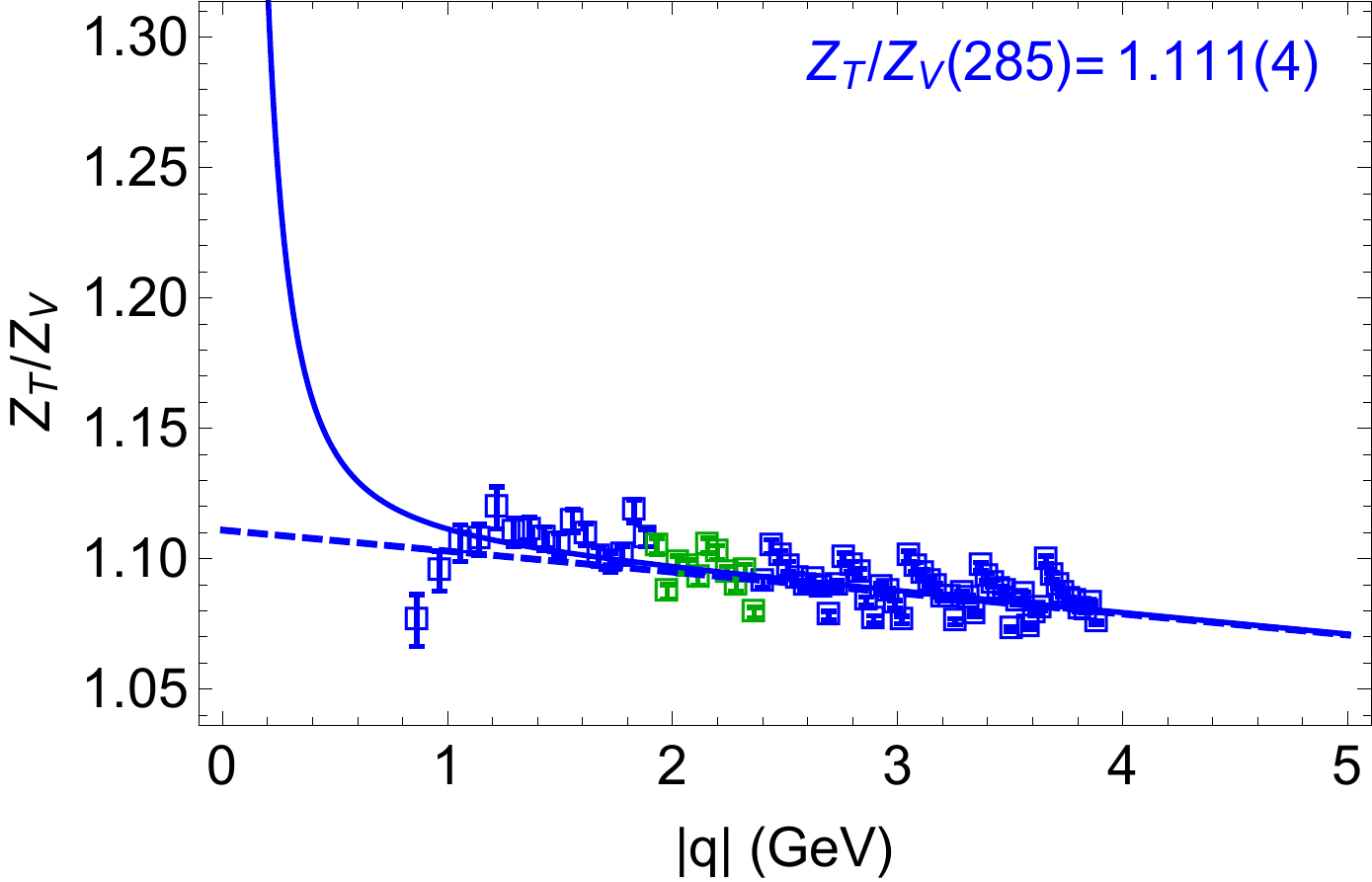}
 \includegraphics[width=0.46\linewidth]{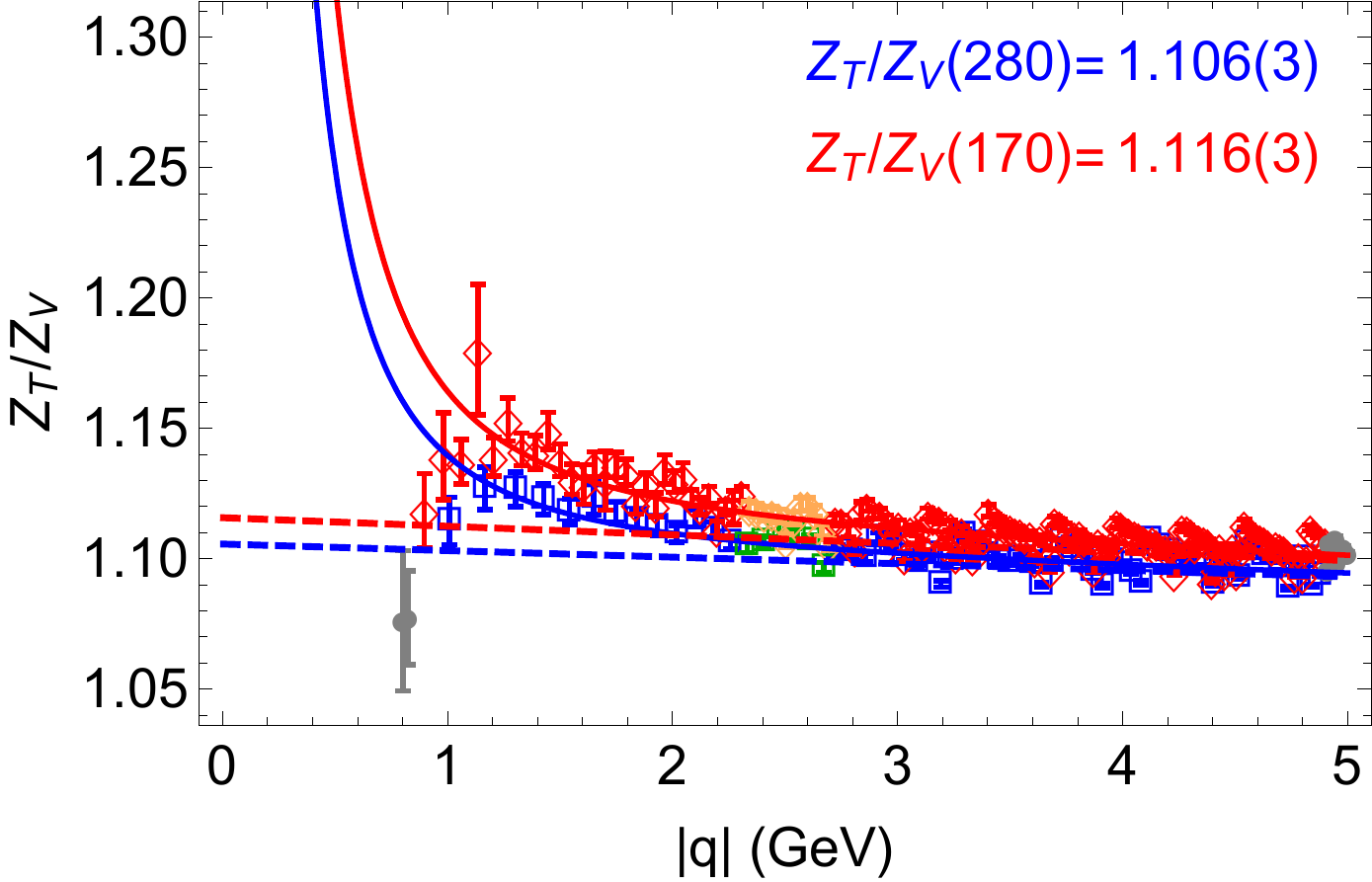}
}
\caption{Data for the ratios of renormalization constants $Z_A/Z_V$, $Z_S/Z_V$ and
  $Z_T/Z_V$ in the $\overline{\text{MS}}$ scheme at $2\GeV$ on the
  $a127m285$ (left) and the $a094m280$ and $a091m170$ (right)
  ensembles. The rest is the same as in Fig.~\protect\ref{fig:Z}.}
\label{fig:Zrat}
\end{figure*}

\begin{figure*}
\centering
\subfigure{
 \includegraphics[width=0.33\linewidth]{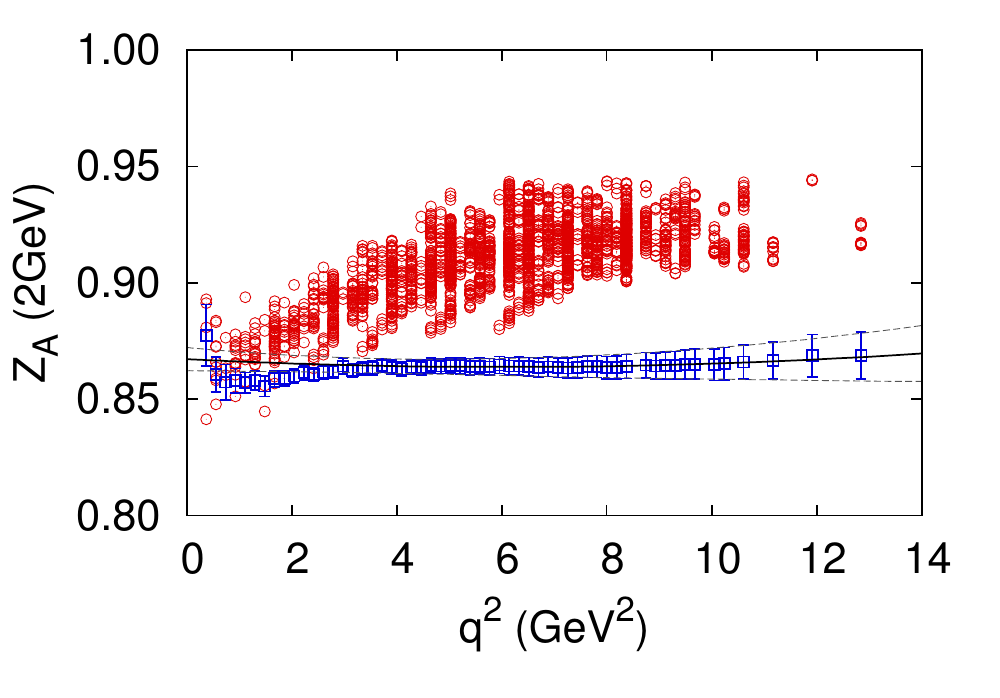}
 \includegraphics[width=0.33\linewidth]{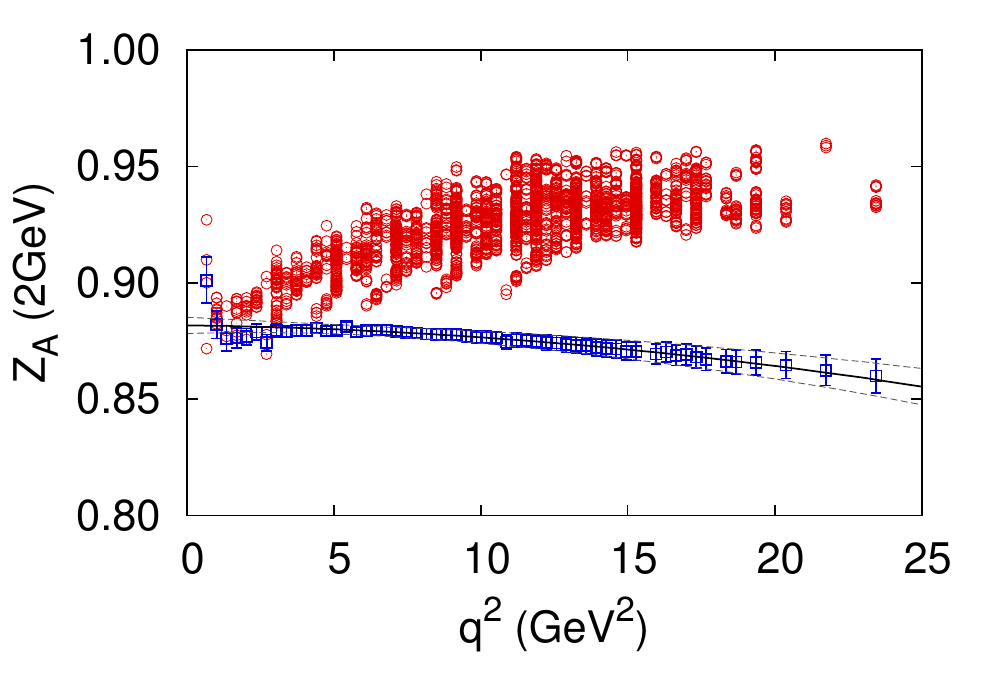}
 \includegraphics[width=0.33\linewidth]{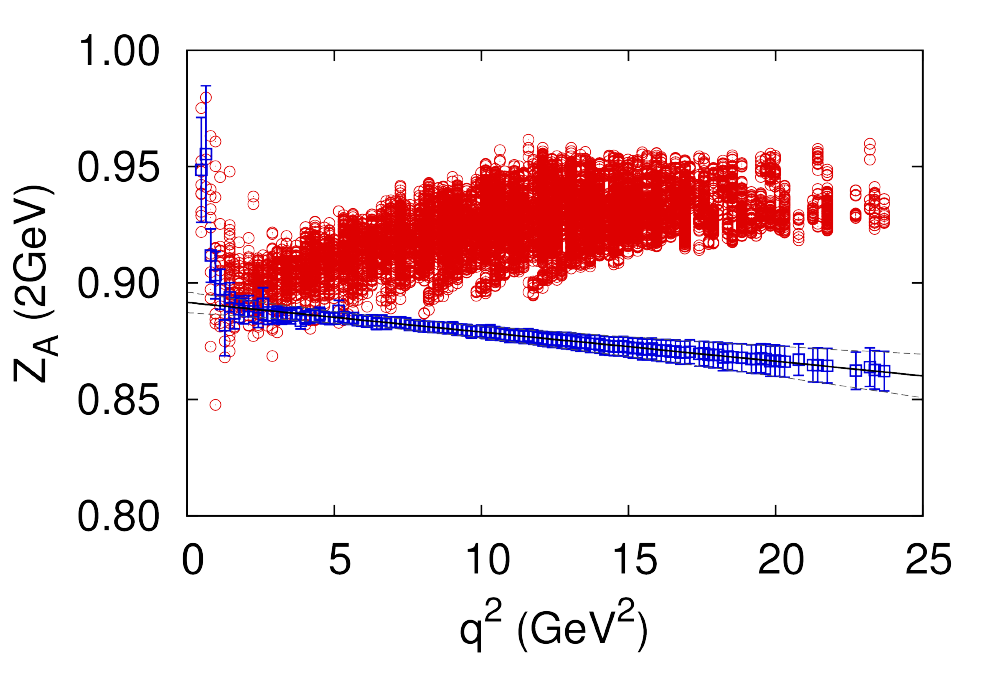}
}
\subfigure{
 \includegraphics[width=0.33\linewidth]{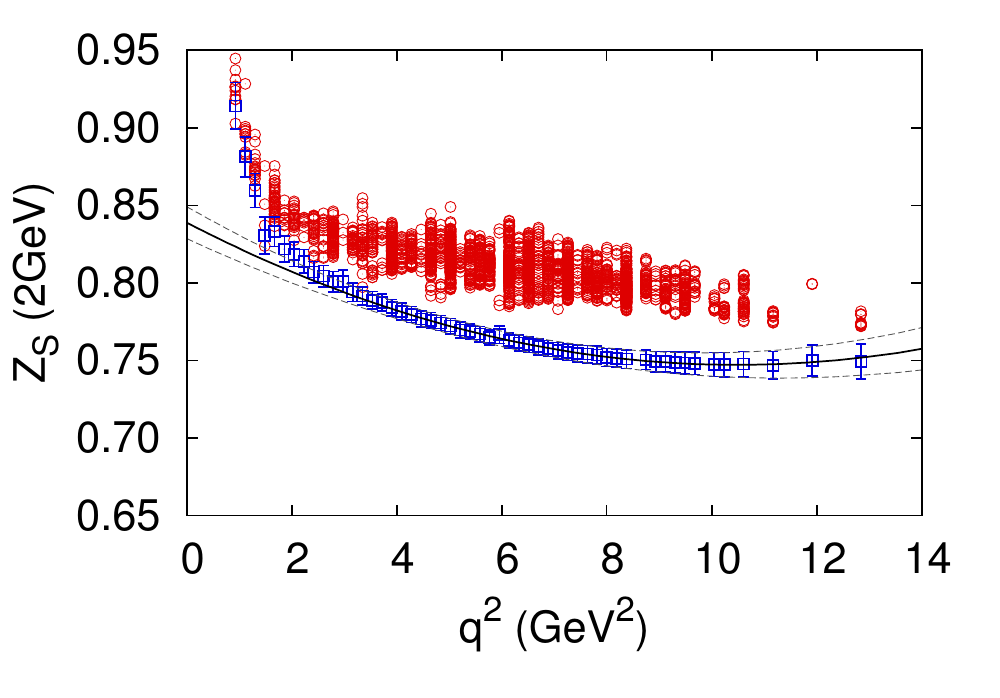}
 \includegraphics[width=0.33\linewidth]{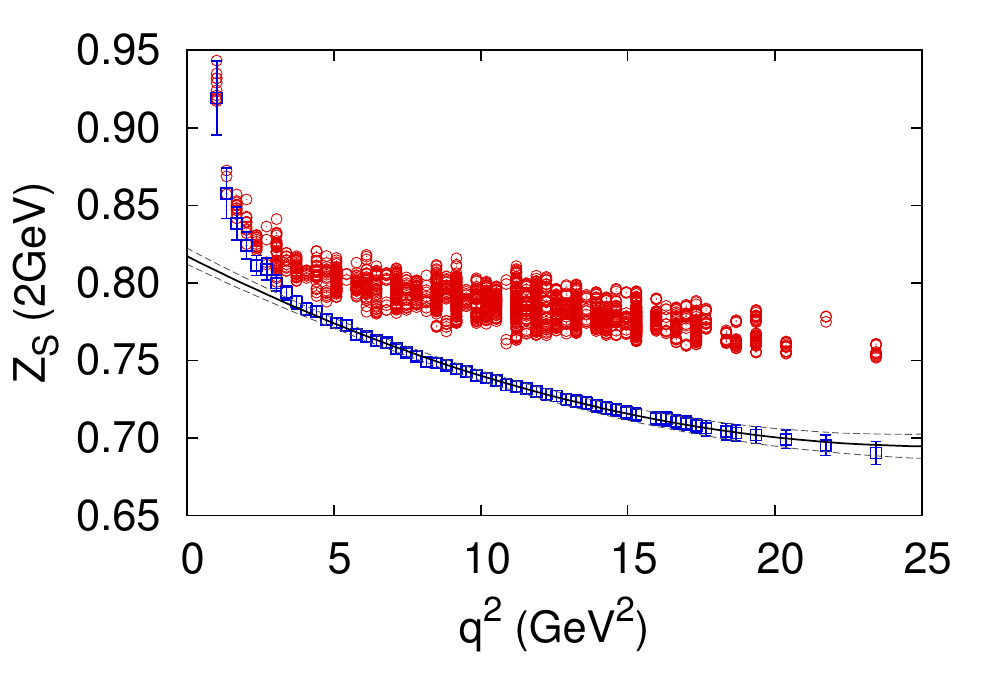}
 \includegraphics[width=0.33\linewidth]{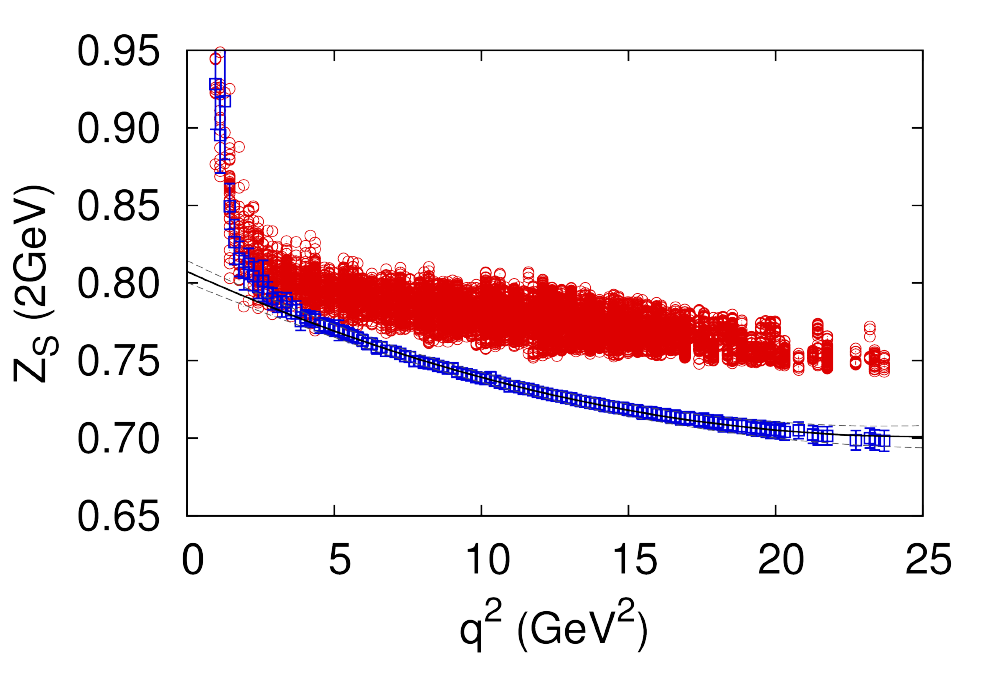}
}
\subfigure{
 \includegraphics[width=0.33\linewidth]{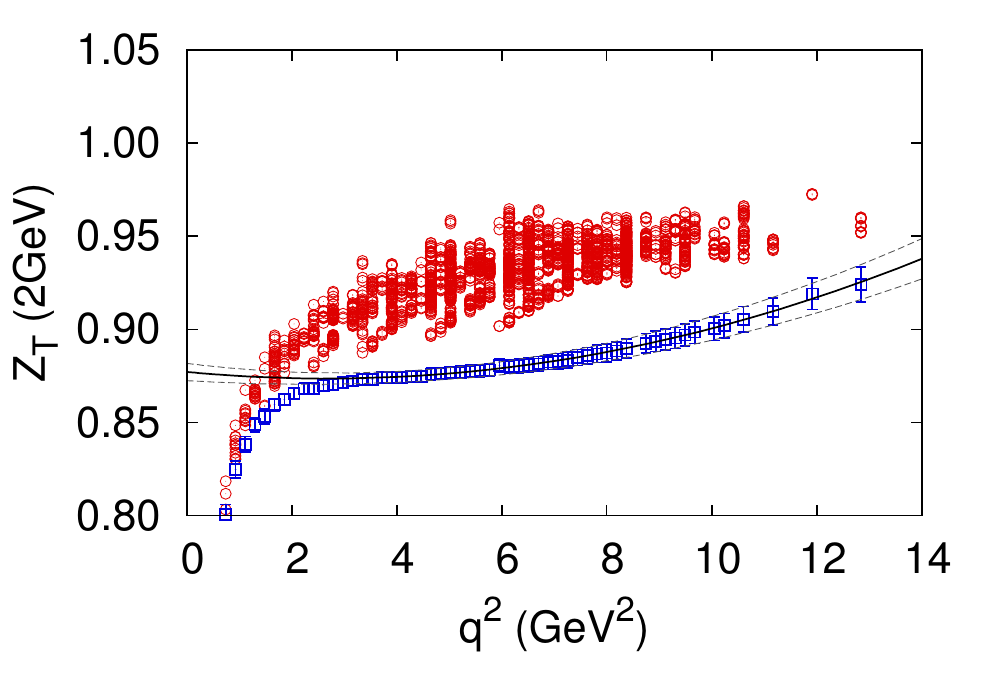}
 \includegraphics[width=0.33\linewidth]{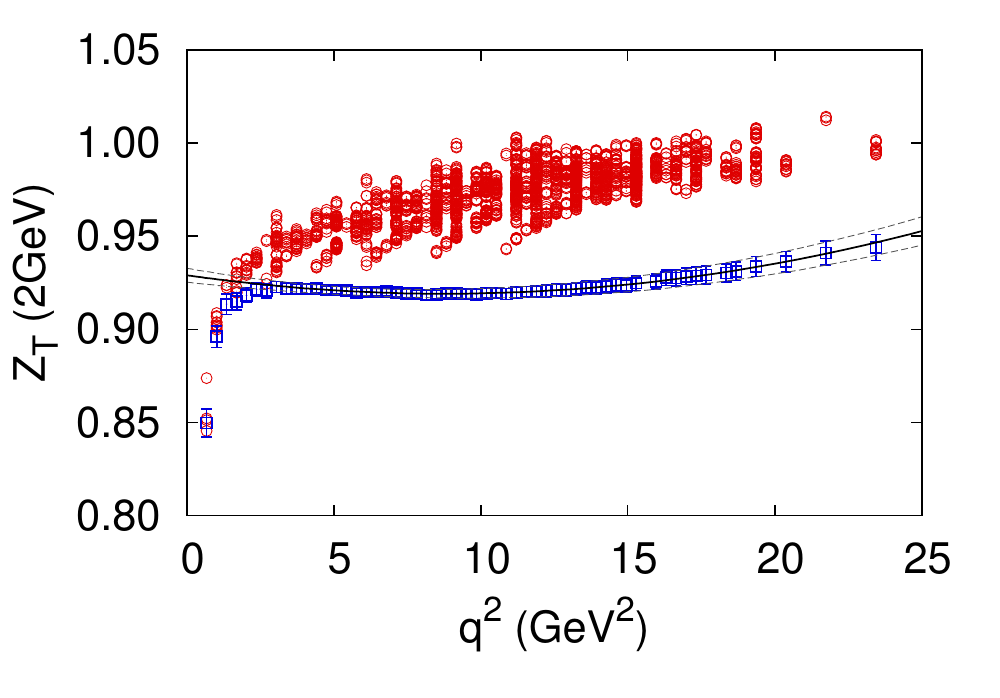}
 \includegraphics[width=0.33\linewidth]{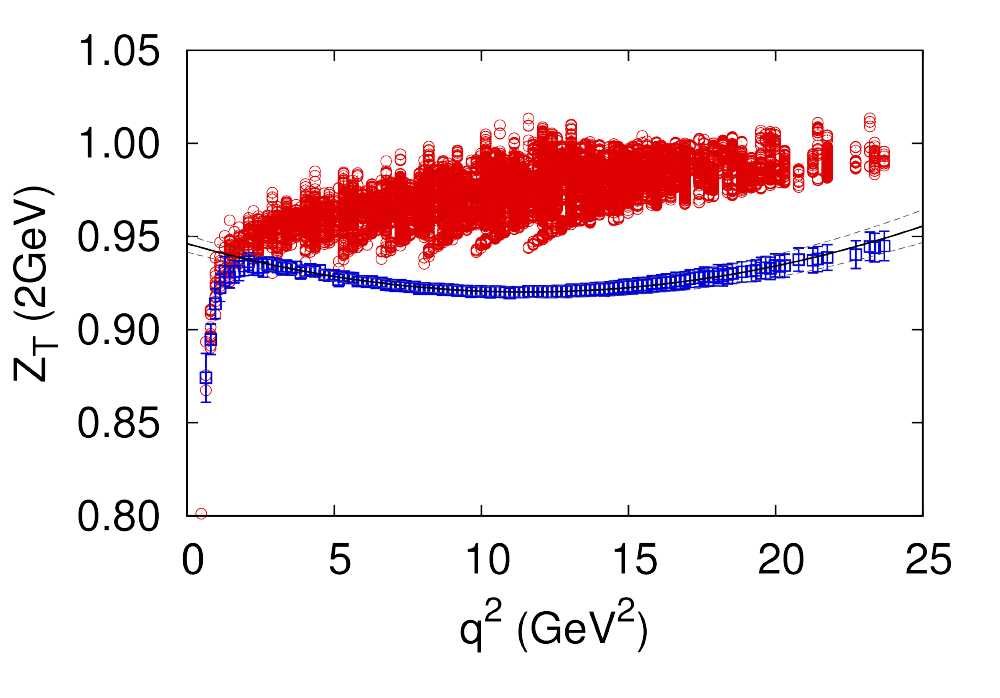}
}
\subfigure{
 \includegraphics[width=0.33\linewidth]{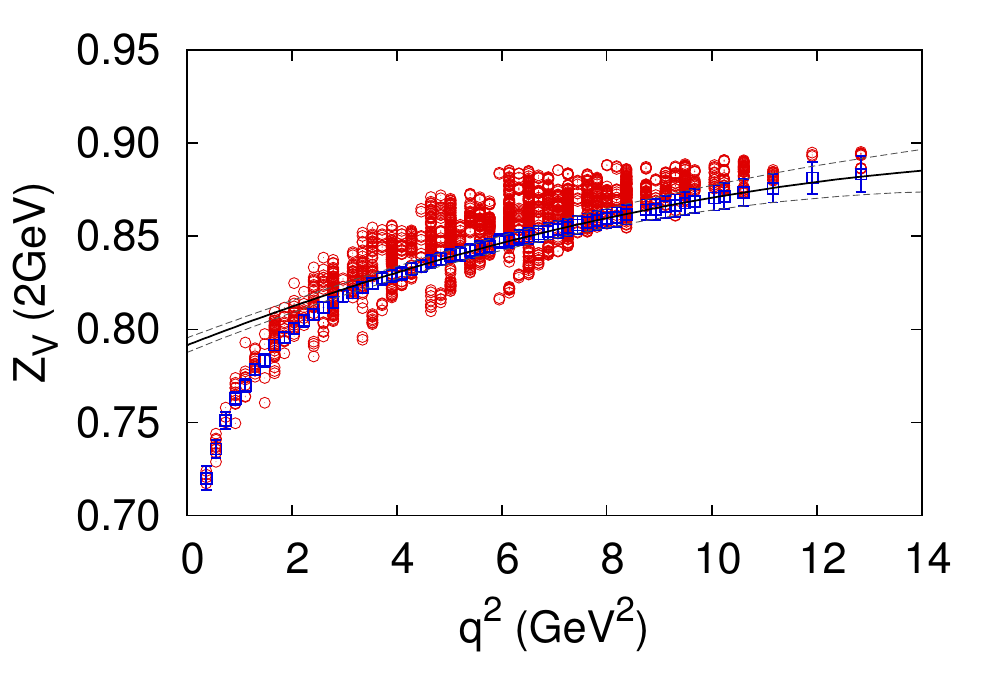}
 \includegraphics[width=0.33\linewidth]{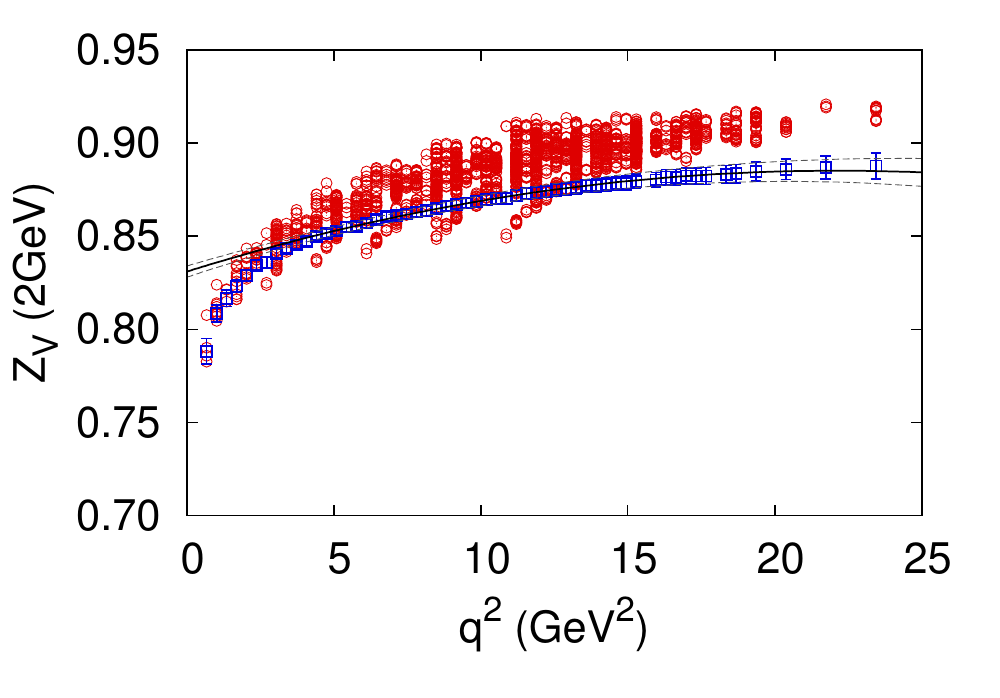}
 \includegraphics[width=0.33\linewidth]{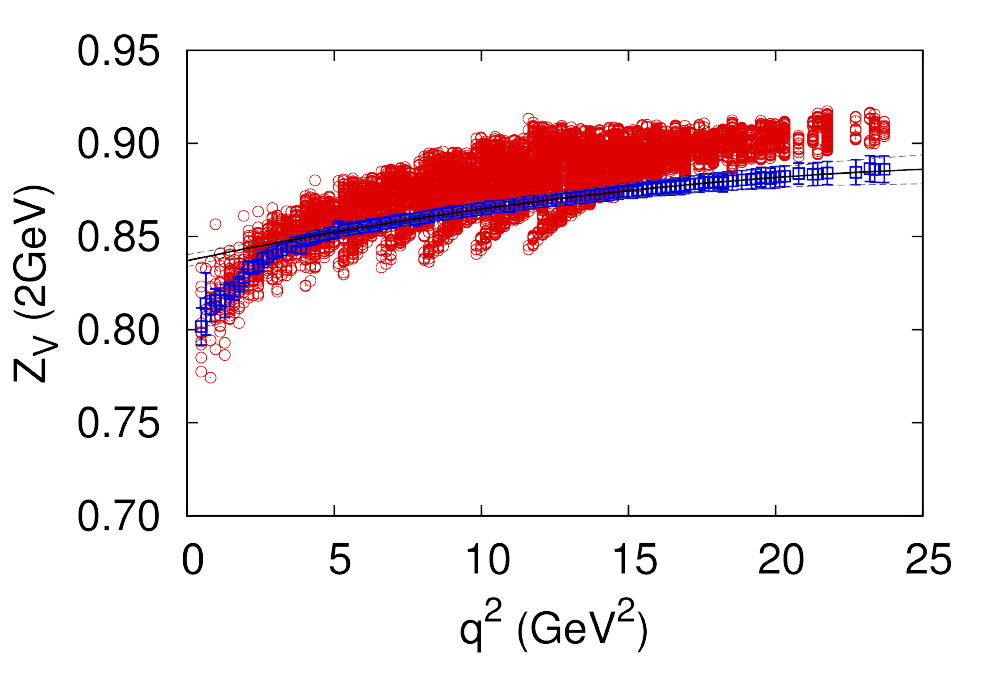}
}
\caption{Data for the renormalization constants $Z_A$, $Z_S$, $Z_T$
  and $ Z_V$ in the $\overline{\text{MS}}$ scheme at $2\GeV$ using
  method C described in the text. All the data from the $a127m285$
  (left) $a094m280$ (middle) and $a091m170$ (right) ensembles are
  shown as red circles.  The data after correcting for the rotational
  symmetry breaking are shown as blue squares. The corrected data
  (blue squares) are then fit using the ansatz $Z + e_2 q^2 + e_4 q^4
  $ to estimate the $Z$. The fit is shown by the solid black line. }
\label{fig:Z3}
\end{figure*}

\begin{figure*}
\centering
\subfigure{
 \includegraphics[width=0.33\linewidth]{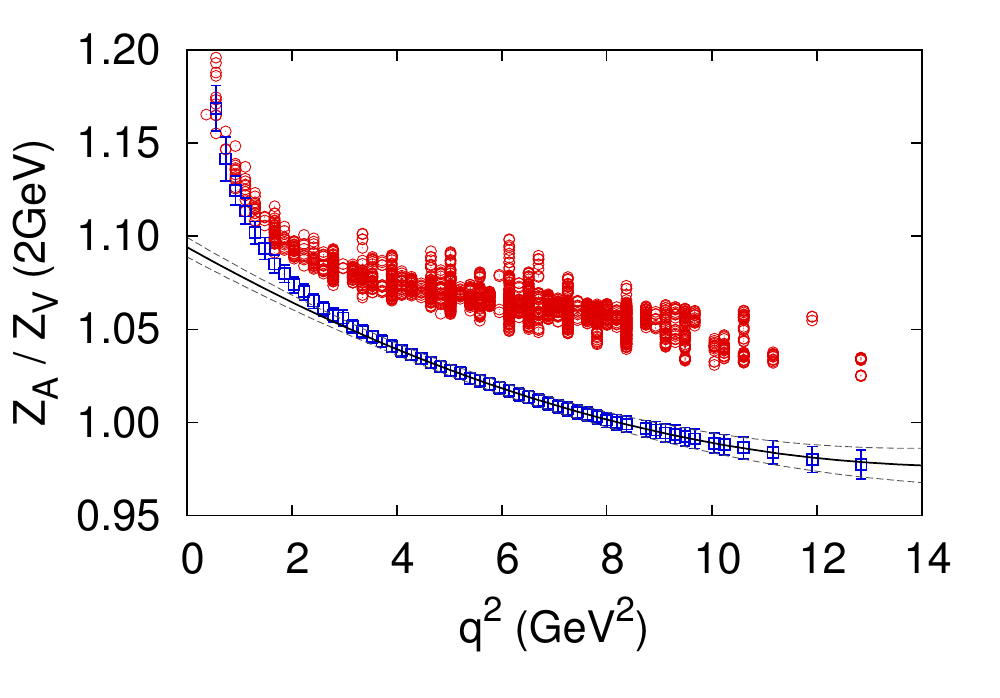}
 \includegraphics[width=0.33\linewidth]{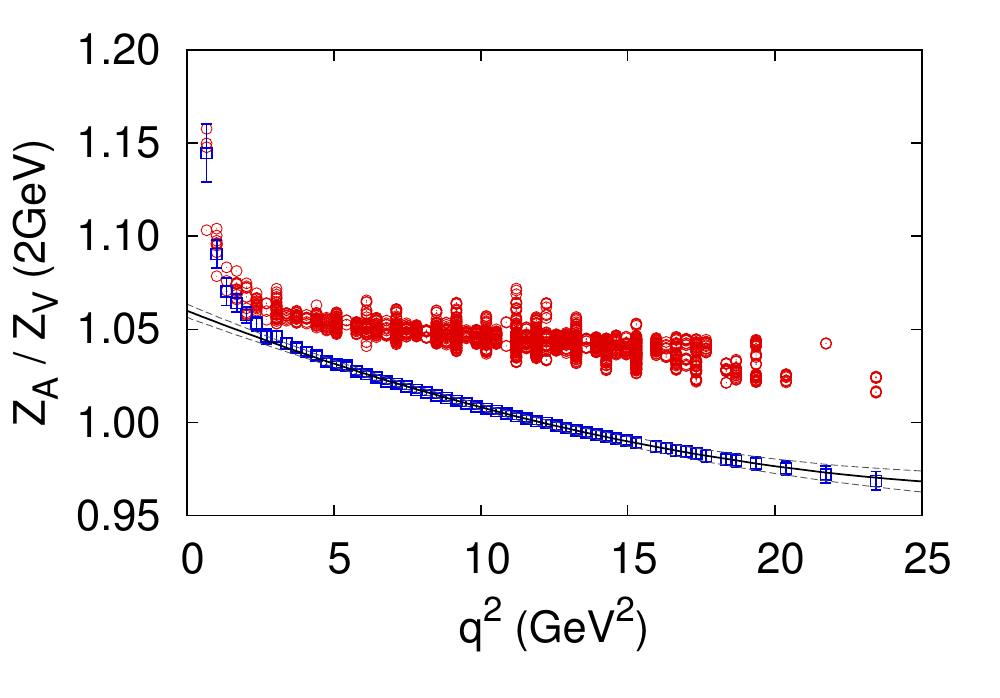}
 \includegraphics[width=0.33\linewidth]{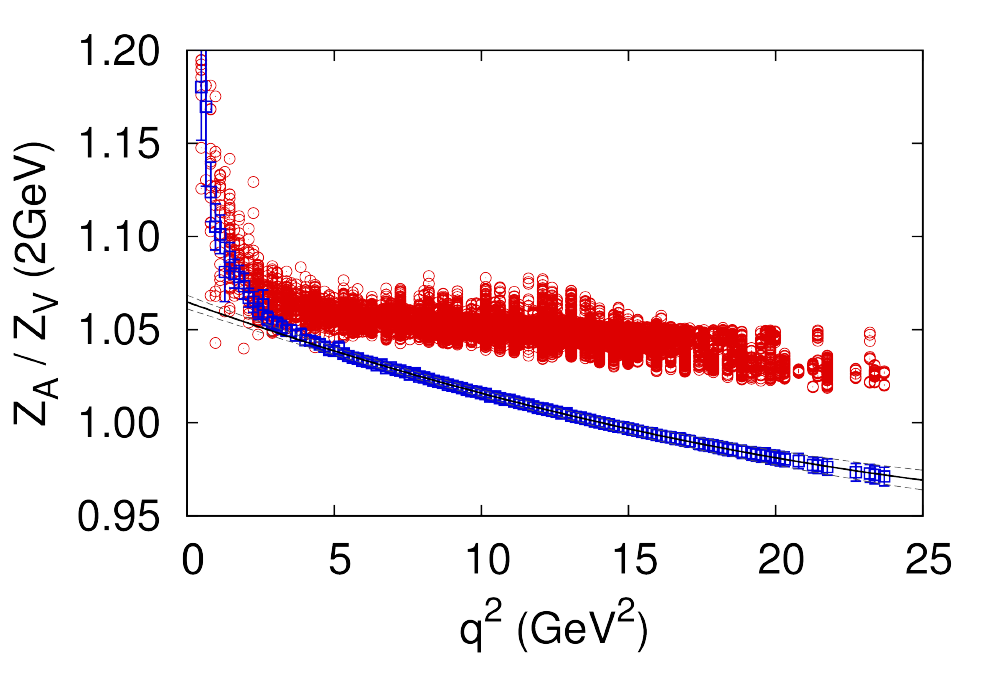}
}
\subfigure{
 \includegraphics[width=0.33\linewidth]{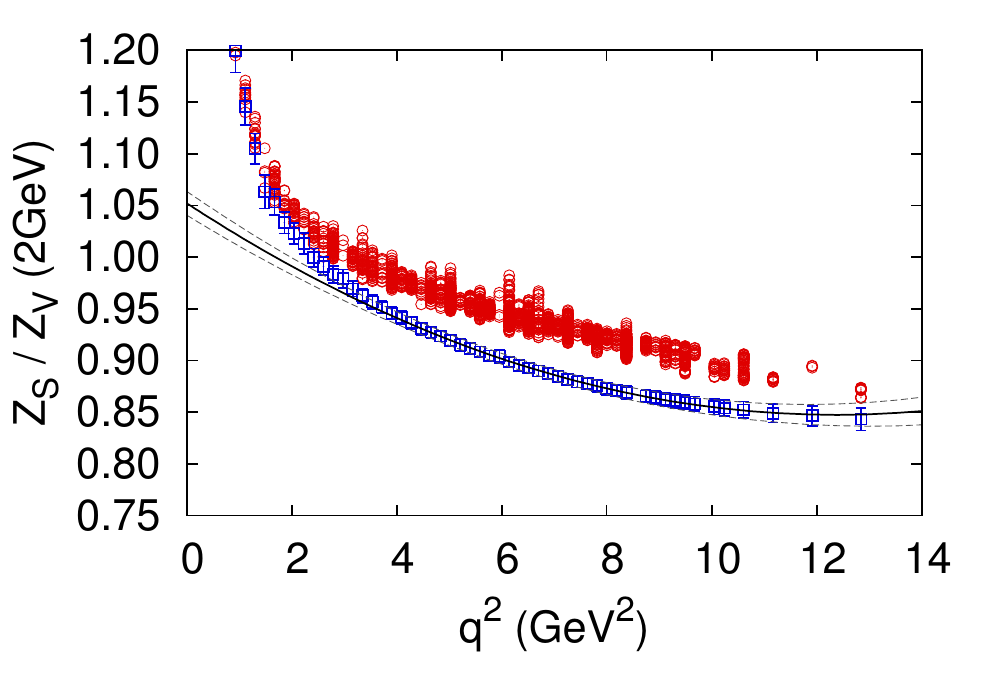}
 \includegraphics[width=0.33\linewidth]{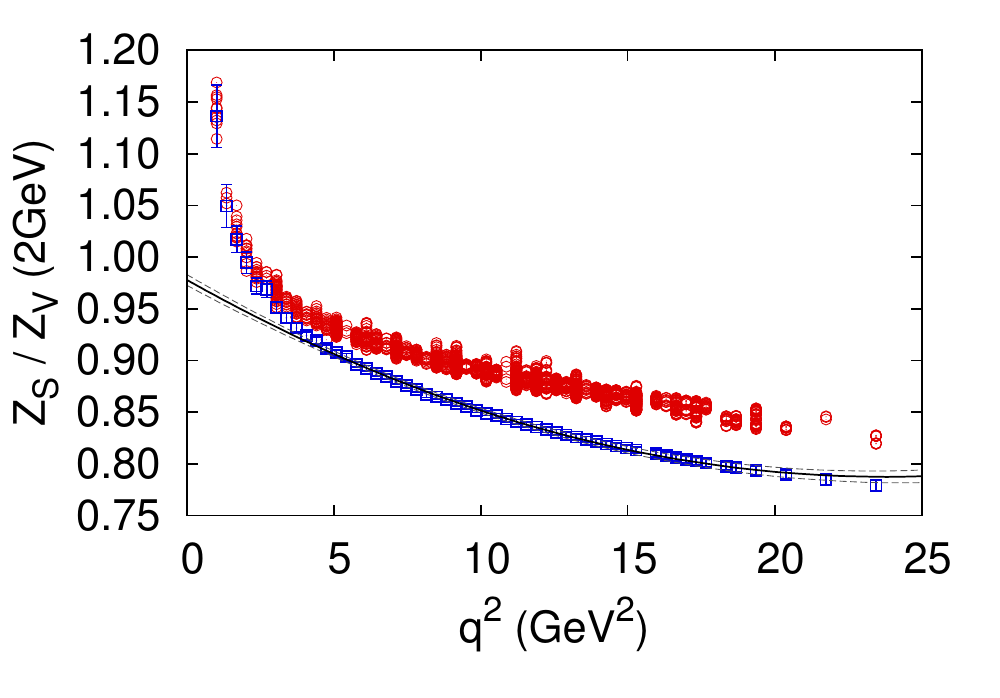}
 \includegraphics[width=0.33\linewidth]{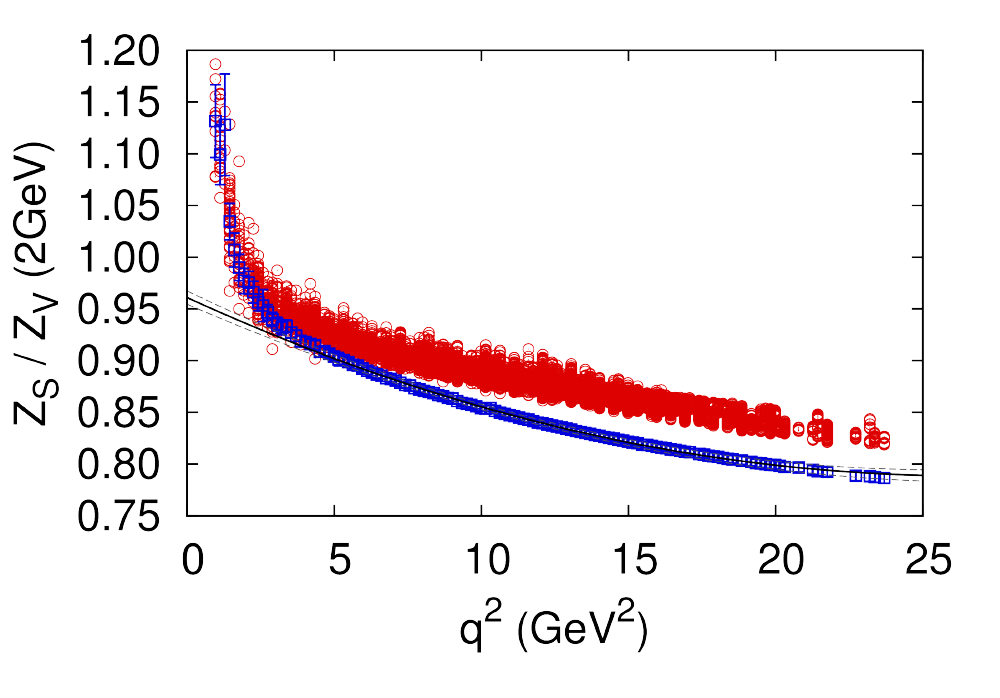}
}
\subfigure{
 \includegraphics[width=0.33\linewidth]{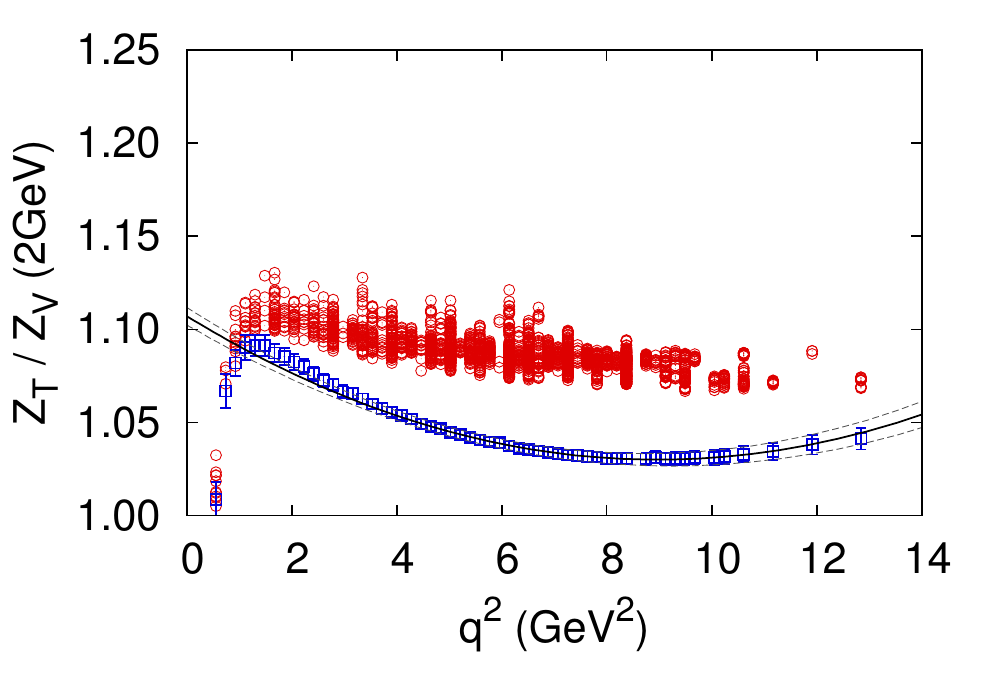}
 \includegraphics[width=0.33\linewidth]{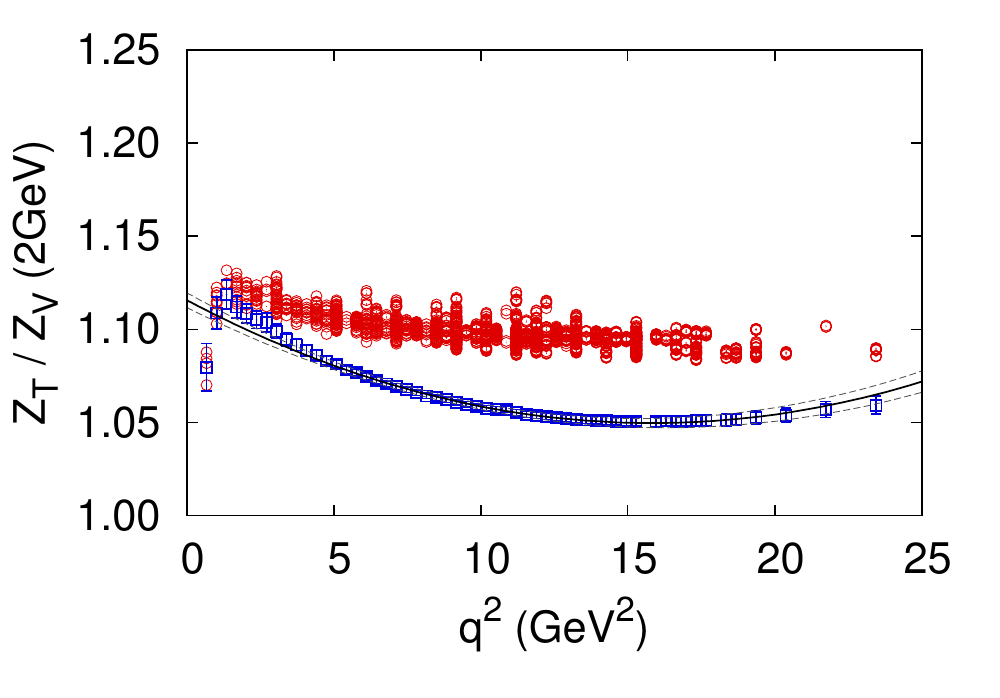}
 \includegraphics[width=0.33\linewidth]{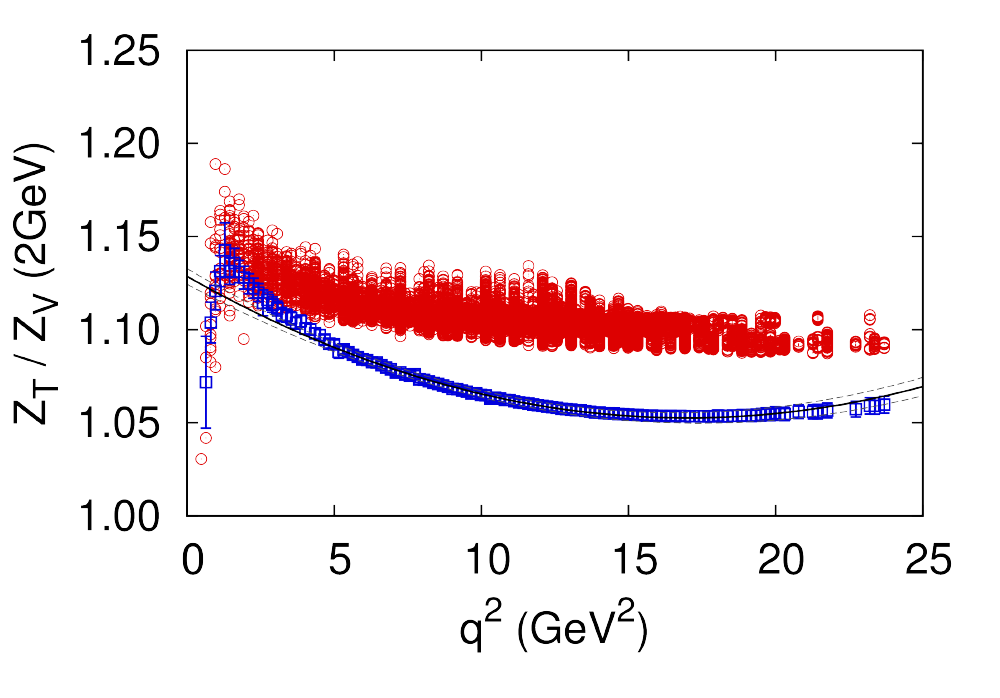}
}
\caption{Ratios of renormalization constants $Z_A/Z_V$, $Z_S/Z_V$ and
  $Z_T/Z_V$ in the $\overline{\text{MS}}$ scheme at $2\GeV$ using
  method C.  The rest is the same as in Fig.~\protect\ref{fig:Z3}.}
\label{fig:Z3_rat}
\end{figure*}

The data, shown in Figs.~\ref{fig:Z},~\ref{fig:Zrat},~\ref{fig:Z3}
and~\ref{fig:Z3_rat}, do not exhibit such a window in which they are
independent of $q^2$, as needed for a unique determination of the
$Z_\Gamma$ and the ratios $Z_\Gamma/Z_V$.  The lattice artifacts are
much larger than the statistical errors. The four main systematics
contributing to the lack of such a window and the resulting
uncertainty in the extraction of the renormalization constants are (i)
breaking of the Euclidean $O(4)$ rotational symmetry to the hypercubic
group, because of which different combinations of momenta with the
same $q^2$ give different results in the RI-sMOM scheme; (ii)
discretization errors at large $q^2$ other than these $O(4)$ breaking
effects; (iii) nonperturbative effects at small $q^2$;
and (iv) truncation errors in the perturbative matching to the
$\overline{\text{MS}}$ scheme and the running to $2\GeV$.

To reduce these systematics, we estimate $Z_\Gamma$ using
three methods. In methods A and B, to reduce artifacts
due to the breaking of rotational symmetry on the lattice, we only
keep points that minimize $\sum_\mu [(p_a^\mu)^4 + (p_b^\mu)^4]$ when
there are multiple combinations of momenta $p_a^\mu$ and $p_b^\mu$
that have the same $q^2$.  These points, after conversion to the
$\overline{\text{MS}}$ scheme at $\mu = 2\GeV$ are shown in
Figs.~\ref{fig:Z} and~\ref{fig:Zrat} as a function of $\sqrt{q^2}$,
the momentum flowing in all three legs in the RI-sMOM scheme. Using
this subset of the data, the first two estimates are obtained as
follows:

Method A: We fit the data in the $\overline{\text{MS}}$ scheme at $\mu
= 2\GeV$ for $q^2 > 0.85$ GeV${}^2$ using the ansatz $c/q^2 + Z_\Gamma + d_1 q
$. The first term, $c/q^2$, is introduced to account for
non-perturbative artifacts and the third, $d_1 q$, for discretization
errors. These fits are shown in Figs.~\ref{fig:Z}
and~\ref{fig:Zrat}. In these figures, the data from the ensembles
$a094m280$ and $a091m170$ are plotted together to show that possible
dependence on the pion mass is much smaller than the statistical
errors or the lattice artifacts.

Method B: We choose the estimate for $Z_\Gamma$ by taking an average
over data points about $q^2 = \Lambda /a $, where $\Lambda=3$~GeV is a
scale chosen to be small enough to avoid discretization effects,
large enough to avoid non-perturbative effects, and above which
perturbation theory is expected to be reasonably well-behaved. With
this choice, both $q a \to 0$ and $\Lambda/q \to 0$ in the continuum
limit as desired. In our simulations, the values of $q^2$ are $4.7$
and $6.4$~GeV${}^2$ for the $a =0.127$ and $0.09$~fm ensembles,
respectively.\footnote{ For these choices of $q^2$, a given momentum
  component $k$, evaluated as $a^2 q^2 = 4k^2$, satisfies the
  condition $k - \sin(k) < 0.05$, which provides a bound on some of
  the tree-level discretization effects.}  Thus, the value from method
B and the error in it is taken to be the mean and the standard
deviation of the data over the ranges 3.7--5.7 and 5.4--7.4 GeV${}^2$
for the ensembles at $a=0.127$ and $0.09$~fm, respectively.

Method C: We first isolate the $O(4)$ breaking artifacts in the data,
$Z(p_a,p_b,q^2)$, by using a fit. This was done in
Refs.~\cite{Boucaud:2003dx,Boucaud:2005rm} for different kinematics
($p_a=p_b$) with terms up to $O(p^4)$. We generalize that ansatz to
our kinematics:
\break
\begin{equation}
\begin{split}
 Z_\Gamma(p_a,p_b,q^2) &= Z_\Gamma^0(q^2)
\\ &\quad
+ a^4\sum_{i=0}^2 c^{(1)}_i p_S^{[4,i]}
+ a^2\sum_{i=0}^2 c^{(2)}_i \frac{p_S^{[4,i]}}{q^2}
\\ &\quad
+ a^4\sum_{i=0}^3 c^{(3)}_i \frac{p_S^{[6,i]}}{q^2}
+ a^4\sum_{\substack{i,j=0 \\ i\leq j}}^2 c^{(4)}_{ij} \frac{p_S^{[4,i]}p_S^{[4,j]}}{q^4}
\\ &\quad
+ a^4\sum_{\substack{i,j=0 \\ i\leq j}}^1 c^{(5)}_{ij} \frac{p_A^{[4,i]}p_A^{[4,j]}}{q^4},
\end{split}
\end{equation}
where
\begin{equation}
 p_{S,A}^{[n,i]} = \sum_\mu\left[ (p_a^\mu)^{n-i}(p_b^\mu)^i
   \pm (p_a^\mu)^i(p_b^\mu)^{n-i} \right] \, .
\end{equation}
Here, for each tensor structure $\Gamma$, the $Z_\Gamma^0(q^2)$ are independent parameters for each $q^2$, and
the nineteen $c_i$, whose $q^2$ dependence is ignored, parameterize
terms that break the $O(4)$ symmetry. Also, only
$Z_\Gamma(p_a,p_b,q^2)$ with momenta satisfying
$\{|p_a^\mu|,|p_b^\mu|,|q^\mu|\} \leq \pi/(2a)$ are included in the
fit. The $Z_0(q^2)$, after conversion to the $\overline{\text{MS}}$
scheme at $\mu = 2\GeV$, are then fit over the ranges $q^2= $4--16
($a127m285$) and $q^2= $4--25~GeV${}^2$ ($a094m280$ and $a091m170$)
using the ansatz $Z + e_2 q^2 + e_4 q^4 $ to extract the desired
$Z_\Gamma$. We show all the data, $Z_\Gamma(p_a,p_b,q^2)$, as red
circles in Figs~\ref{fig:Z3} and~\ref{fig:Z3_rat} and the
$Z_\Gamma^0(q^2)$ as blue squares.  The final fit, along with the
error band, is shown by the black lines. The data have been analyzed
to obtain both $Z_\Gamma$ and the ratios $Z_\Gamma/Z_V$, and their
final values are collected in Table~\ref{tab:Zall}.

On comparing the raw data presented in Figs~\ref{fig:Z3} and~\ref{fig:Z3_rat} 
(red circles), we find the data for the ratios $Z_\Gamma/Z_V$ show a smaller
spread, presumably because some of the systematics cancel. As a
result, the errors in estimates from all three methods, shown in
Table~\ref{tab:Zall}, are smaller with the ratio method. Also, in all
three methods, the region of $q^2$ that contributes to the fits is
consistent with the general requirement that $\eta \Lambda_{\rm QCD}
\ll q \ll \zeta \pi/a$ with $\eta$ and $\zeta$ of $O(1)$ to avoid both
non-perturbative and discretization artifacts.

\begin{table*}
\centering
\begin{ruledtabular}
\begin{tabular}{c|c|cccc|ccc}
ID          &  Method &   $Z_A$       & $Z_S$        &  $Z_T$       & $Z_V$        & $Z_A/Z_V$    & $Z_S/Z_V$    & $Z_T/Z_V$    \\
\hline                  
$a127m285$  &   A     &   0.880(7)    &   0.822(8)   &   0.883(5)   &   0.786(5)   &   1.119(6)   &   1.024(7)   &   1.111(4)   \\
$a127m285$  &   B     &   0.891(9)    &   0.807(7)   &   0.908(9)   &   0.829(13)  &   1.075(8)   &   0.974(13)  &   1.096(8)   \\
$a127m285$  &   C     &   0.867(5)    &   0.839(10)  &   0.877(5)   &   0.791(4)   &   1.094(5)   &   1.052(12)  &   1.107(5)   \\
\hline                  
$a127m285$  &   AV    &  0.879(12)    &   0.823(16)  &   0.889(16)  &   0.802(22)  &   1.096(22)  &   1.017(39)  &   1.105(7)   \\
\hline                  
$a094m280$  &   A     &   0.872(4)    &   0.793(7)   &   0.899(4)   &   0.815(4)   &   1.081(3)   &   0.976(4)   &   1.106(3)   \\
$a094m280$  &   B     &   0.901(9)    &   0.790(9)   &   0.947(8)   &   0.855(10)  &   1.054(4)   &   0.924(8)   &   1.106(5)   \\
$a094m280$  &   C     &   0.889(4)    &   0.817(5)   &   0.929(4)   &   0.831(3)   &   1.060(4)   &   0.978(5)   &   1.116(4)   \\
\hline                  
$a094m280$  &   AV    &  0.887(15)    &   0.800(14)  &  0.925(24)   &  0.834(20)   &   1.065(14)  &  0.959(27)   &   1.109(5)   \\
\hline                  
$a091m170$  &   A     &   0.882(6)    &   0.793(8)   &   0.915(5)   &   0.820(4)   &   1.086(4)   &   0.973(6)   &   1.116(3)   \\
$a091m170$  &   B     &   0.899(6)    &   0.779(4)   &   0.949(6)   &   0.850(8)   &   1.058(4)   &   0.916(7)   &   1.116(5)   \\
$a091m170$  &   C     &   0.892(4)    &   0.807(7)   &   0.946(5)   &   0.837(3)   &   1.065(4)   &   0.961(7)   &   1.129(4)   \\
\hline                  
$a091m170$  &   AV    &   0.891(9)    &  0.793(14)   &  0.937(17)   &  0.836(15)   &  1.070(14)   &  0.950(29)   &   1.120(6)   \\
\end{tabular}
\end{ruledtabular}
\caption{The renormalization constants $Z_A$, $Z_S$, $Z_T$, $Z_V$ and
  the ratios $Z_A/Z_V$, $Z_S/Z_V$ and $Z_T/Z_V$ in the
  $\overline{\text{MS}}$ scheme at $2\GeV$ at the two values of the
  lattice spacings. For each ensemble, the three rows give estimates
  for the three methods (A, B and C) described in the text. The fourth
  row gives the mean (AV) with the error given by the larger of the
  two---half the spread or the largest statistical error. This average
  value is taken to be our final estimate of the renormalization
  factor.  }
\label{tab:Zall}
\end{table*}

The estimates from the three methods, given in Tab~\ref{tab:Zall},
have different systematics.  For example, as shown in Figs~~\ref{fig:Z}, 
\ref{fig:Zrat}, \ref{fig:Z3} and~\ref{fig:Z3_rat}, the variation
with $q^2$, in many cases, is large.  Nevertheless, the estimates from the three
methods agree to within about 2\%. We, therefore, take the average of
the three as our final estimate. To assign a conservative error, we
use half the spread between the three estimates since it is larger
than the statistical errors.

We also point out that the 2-loop perturbative expression for the
matching of $Z_T$ between the RI-sMOM scheme and the
$\overline{\text{MS}}$ scheme is badly behaved over the range of $q^2$
investigated. For example, the successive terms in the loop expansion
are $1 + 0.0052 + 0.0159$ at $q^2=4$~GeV${}^2$
($\alpha^{\overline{MS}}=0.2979$) and $1 + 0.0037 + 0.0078$ at
$q^2=25$~GeV${}^2$ ($\alpha^{\overline{MS}}=0.2041$) using the
matching expressions given in Ref.~\cite{Gracey:2011fb}.  We,
therefore, take the average of the two 2-loop correction at $q^2=4$
and $25$~GeV${}^2$, 0.012, as a systematic error in the estimates of
$Z_T$ due to truncation errors.  The series for $Z_S$ at
$q^2=4$~GeV${}^2$, $1 - 0.0157 -0.0039$, is much better
behaved. Again, we take the average of the 2-loop value at $q^2=4$ and
$25$~GeV${}^2$, 0.003, as the additional systematic uncertainty.  Note
that these estimates of systematics are smaller than the final errors
estimates given in Table~\ref{tab:Zall}.

\section{Renormalized Charges}
\label{sec:results}

%
\begin{table}
\centering
\begin{ruledtabular}
\begin{tabular}{c|c|cccc}
ID          &  Analysis   & $g_A^{u-d}$ & $g_S^{u-d}$& $g_T^{u-d}$& $g_V^{u-d}$     \\
\hline                                                                           
   $a127m285$  &  $S_5 S_5$  &  1.258(22)  &  0.90(4)   &  1.031(21)  &  1.014(28)   \\
\hline                          							
   $a094m280$  &  $S_5 S_5$  &  1.214(36)  &  1.00(7)   &  0.978(31)  &  0.996(25)   \\
\hline                          							
   $a094m280$  &  $S_7 S_7$  &  1.225(35)  &  0.96(10)  &  0.972(42)  &  1.002(26)   \\
\hline                          							
   $a094m280$  &  $S_9 S_9$  &  1.193(29)  &  1.00(9)   &  0.960(36)  &  0.997(26)   \\
\hline                          							
   $a094m280$  &  Average    &  1.206(33)  &  0.99(9)   &  0.972(36)  &  0.998(26)   \\
\hline                          							
   $a094m280$  &  VAR579     &  1.221(26)  &  0.97(7)   &  1.034(32)  &  1.012(27)   \\
\hline                          							
   $a091m170$  &  $S_7 S_7$  &  1.214(19)  &  0.86(9)   &  1.003(23)  &  1.012(21)   \\
\hline                          							
   $a091m170L$ &  $S_7 S_7$  &  1.316(36)  &  0.94(11)  &  0.977(30)  &  1.000(22)   \\
\end{tabular}
\end{ruledtabular}
\caption{Estimates of the renormalized isovector charges using the
  product $Z_\Gamma \times g_\Gamma^{\rm bare}$ that is labeled Method
  (i) in the text. We also give the weighted average of the three
  measurements on the $a081m210$ ensemble and the estimate from the
  variational calculation, VAR579 using $3 \times 3$ smearing $\sigma
  = 5, 7, 9$ at a single value of $\tsep=12$, reported in
  Ref.~\cite{Yoon:2016dij}. }
\label{tab:gren-1}
\end{table}

\begin{table}
\centering
\begin{ruledtabular}
\begin{tabular}{c|c|ccc}
ID             &  Analysis   &  $g_A^{u-d}$& $g_S^{u-d}$ & $g_T^{u-d}$     \\
\hline                                                              
   $a127m285$  &  $S_5 S_5$  &  1.241(28)  &  0.873(46)  &  1.014(11)   \\
\hline                          					  
   $a094m280$  &  $S_5 S_5$  &  1.222(37)  &  1.003(79)  &  0.981(19)   \\
\hline                          					  
   $a094m280$  &  $S_7 S_7$  &  1.224(32)  &  0.953(99)  &  0.970(39)   \\
\hline                          					  
   $a094m280$  &  $S_9 S_9$  &  1.198(26)  &  1.005(90)  &  0.964(26)   \\
\hline                          					  
   $a094m280$  &  Average    &  1.210(31)  &  0.991(89)  &  0.975(28)   \\
\hline                          					  
   $a094m280$  &  VAR579     &  1.208(22)  &  0.953(68)  &  1.021(15)   \\
\hline                          					  
   $a091m170$  &  $S_7 S_7$  &  1.206(23)  &  0.853(88)  &  0.990(15)   \\
\hline                          					  
   $a091m170L$ &  $S_7 S_7$  &  1.321(41)  &  0.934(116) &  0.977(26)   \\
\end{tabular}
\end{ruledtabular}
\caption{Estimates of the renormalized isovector charges using the
  product of the ratios $(Z_\Gamma/Z_V) \times (g_\Gamma^{\rm
    bare}/g_V^{\rm bare})$ and the conserved vector current relation
  $Z_V g_V^{\rm bare} = 1 $ that is labeled Method (ii) in the text.
  We also give the weighted average of the three measurements on the
  $a081m210$ ensemble and the estimate from the variational
  calculation, VAR579, reported in Ref.~\cite{Yoon:2016dij}.}
\label{tab:gren-2}
\end{table}

\begin{table}
\centering
\begin{ruledtabular}
\begin{tabular}{c|ccc}
ID           &  $g_A^{u-d}$& $g_S^{u-d}$& $g_T^{u-d}$  \\
\hline                                                 
$a127m285$   &  1.249(28)  &  0.885(46)  &  1.023(21)  \\
\hline          					  
$a094m280$   &  1.208(33)  &  0.990(89)  &  0.973(36)  \\
\hline          					  
$a091m170$   &  1.210(23)  &  0.859(89)  &  0.996(23)  \\
\hline          					  
$a091m170L$  &  1.319(41)  &  0.935(116) &  0.977(30)  \\
\end{tabular}
\end{ruledtabular}
\caption{Our final Estimates of the renormalized isovector charges obtained by averaging the 
  estimates given in Tables~\protect\ref{tab:gren-1} and~\protect\ref{tab:gren-2} 
  as explained in the text. }
\label{tab:FinalValues}
\end{table}

\begin{table*}
\centering
\begin{ruledtabular}
\begin{tabular}{c|c|cc|cc|cc}
ID        & Type      &  $g_A^u$    &   $g_A^d$     &    $g_S^u$   &    $g_S^d$   &  $g_T^u$    &   $g_T^d$     \\
\hline
a127m285  & $S_5 S_5$ &  0.919(16)  &  -0.319(08)  &  3.28(15)  &  2.39(12)  &   0.839(18)  &   -0.195(07)    \\
a127m285* & $S_5 S_5$ &  0.932(17)  &  -0.325(09)  &  3.33(11)  &  2.43(09)  &   0.831(18)  &   -0.201(08)    \\
\hline
a12m310   & clover-on-HISQ &  0.914(11)  &  -0.315(6)    &   3.07(6)    &  2.23(4)     &  0.848(29)  &  -0.209(8)    \\
\hline                   
\hline                   
a094m280  & $S_5 S_5$ &  0.909(22)  &  -0.297(10)  &  3.61(17)  &  2.65(13)  &   0.780(25)  &   -0.203(09)    \\
a094m280* & $S_5 S_5$ &  0.911(29)  &  -0.302(14)  &  3.82(27)  &  2.82(23)  &   0.770(28)  &   -0.208(12)    \\
a094m280  & $S_7 S_7$ &  0.940(30)  &  -0.294(13)  &  3.79(37)  &  2.82(28)  &   0.796(31)  &   -0.196(12)    \\
a094m280* & $S_7 S_7$ &  0.929(32)  &  -0.296(16)  &  3.99(41)  &  3.03(35)  &   0.777(34)  &   -0.195(14)    \\
a094m280  & $S_9 S_9$ &  0.907(27)  &  -0.307(13)  &  3.66(19)  &  2.66(14)  &   0.783(29)  &   -0.209(11)    \\
a094m280* & $S_9 S_9$ &  0.904(24)  &  -0.289(13)  &  3.74(20)  &  2.74(16)  &   0.777(30)  &   -0.183(11)    \\
a094m280  & Average   &  0.915(26)  &  -0.299(12)  &  3.65(24)  &  2.67(18)  &   0.784(28)  &   -0.203(11)    \\
a094m280* & Average   &  0.911(28)  &  -0.296(14)  &  3.80(29)  &  2.80(25)  &   0.774(31)  &   -0.196(12)    \\
\hline
a09m310   & clover-on-HISQ &  0.926(26)  &  -0.304(15)   &   3.40(32)   &  2.56(25)    &  0.823(33)  &  -0.200(13)   \\
\hline                   
\hline                   
a091m170  & $S_7 S_7$ &  0.909(22)  &  -0.299(15)  &  4.23(20)  &  3.31(16)  &   0.814(22)  &   -0.221(14)    \\
a091m170* & $S_7 S_7$ &  0.886(16)  &  -0.329(10)  &  4.30(24)  &  3.43(20)  &   0.798(19)  &   -0.204(09)    \\
\hline                   
a09m220   & clover-on-HISQ &  0.911(26)  &  -0.337(16)   &   3.78(30)   &  2.98(23)    &  0.823(31)  &  -0.215(11)   \\
a09m130   & clover-on-HISQ &  0.891(20)  &  -0.338(15)   &   4.97(41)   &  4.08(35)    &  0.784(31)  &  -0.191(11)   \\
\hline                   
\hline                   
a091m170L & $S_7 S_7$ &  0.917(18)  &  -0.331(11)  &  4.39(33)  &  3.39(19)  &   0.804(24)  &   -0.196(10)    \\
a091m170L*& $S_7 S_7$ &  0.960(30)  &  -0.356(22)  &  4.86(33)  &  3.93(29)  &   0.808(28)  &   -0.170(18)    \\
\end{tabular}
\end{ruledtabular}
\caption{Estimates of the connected part of the flavor diagonal axial,
  scalar and tensor charges $g_{A,S,T}^{u,d}$ with our ``best''
  choices of $\tmin$, $\tsep$ and $\tskip$. For each ensemble we give
  the two estimates from $\{\Ntwo,\Nthree\}=\{2,2\}$ and
  $\{4,3^\ast\}$ fits with the latter marked with an asterisk. These
  are renormalized using the same factors as for the isovector charges
  given in Table~\protect\ref{tab:Zall}. Results from a 2+1+1-flavor
  clover-on-HISQ calculation are reproduced from Table~XII in
  Ref.~\protect\cite{Bhattacharya:2016zcn} to facilitate the following
  comparisons $a127m285* \leftrightarrow a12m310$, $a094m280*
  \leftrightarrow a09m310$ and $a091m170* \leftrightarrow (a09m220,a09m130)$
  between estimates with similar lattice parameters.  }
\label{tab:FDCharges}
\end{table*}

\begin{table*}
\centering
\begin{ruledtabular}
\begin{tabular}{c|ccc|cccc}
ID                  &  Lattice Theory       & $a$~fm    & $M_\pi$(MeV)   &  $g_A^{u-d}$ & $g_S^{u-d}$ & $g_T^{u-d}$ & $g_V^{u-d}$   \\
\hline                                                     
$a127m285$          &  2+1 clover-on-clover & 0.127(2)  &  285(6)        &  1.249(28)   &  0.89(5)    &  1.023(21)  &  1.014(28)    \\  
$a12m310$           &  2+1+1 clover-on-HISQ & 0.121(1)  &  310(3)        &  1.229(14)   &  0.84(4)    &  1.055(36)  &  0.969(22)    \\  
\hline                                                                 
$a094m280$          &  2+1 clover-on-clover & 0.094(1)  &  278(3)        &  1.208(33)   &  0.99(9)    &  0.973(36)  &  0.998(26)    \\     
$a09m310$           &  2+1+1 clover-on-HISQ & 0.089(1)  &  313(3)        &  1.231(33)   &  0.84(10)   &  1.024(42)  &  0.975(33)    \\     
\hline                                                                 
$a091m170$          &  2+1 clover-on-clover & 0.091(1)  &  166(2)        &  1.210(19)   &  0.86(9)    &  0.996(23)  &  1.012(21)    \\     
$a09m220$           &  2+1+1 clover-on-HISQ & 0.087(1)  &  226(2)        &  1.249(35)   &  0.80(12)   &  1.039(36)  &  0.969(32)    \\     
$a09m130$           &  2+1+1 clover-on-HISQ & 0.087(1)  &  138(1)        &  1.230(29)   &  0.90(11)   &  0.975(38)  &  0.971(32)    \\     
\end{tabular}
\end{ruledtabular}
\caption{Comparison of the renormalized isovector charges with those
  from four 2+1+1-flavor clover-on-HISQ ensembles with similar values
  of the lattice spacing and pion mass.  The clover-on-clover data
  have been reproduced from Tables~\protect\ref{tab:gren-1}
  and~\protect\ref{tab:FinalValues} and the clover-on-HISQ data have
  been reproduced from Tables~I, XII and XIII
  in Ref.~\protect\cite{Bhattacharya:2016zcn}. }
  \label{tab:ChargesHISQ}
\end{table*}


Combining our final estimates of the unrenormalized charges on the four
ensembles given in Tab~\ref{tab:BC3state} and for the ratios in
Tab~\ref{tab:BC3Ratios} with the renormalization
factors given in Table~\ref{tab:Zall}, the renormalized charges are
extracted in two ways:
\begin{itemize}
\item
Method (i): using the product $Z_\Gamma \times g_\Gamma^{\rm
  bare}$. These results are given in Table~\ref{tab:gren-1}.
\item
Method (ii): using the product of the ratios $(Z_\Gamma/Z_V) \times
(g_\Gamma^{\rm bare}/g_V^{\rm bare})$ and the conserved vector current
relation $Z_V g_V^{\rm bare} = 1 + O(a^2)$. These results are given in
Table~\ref{tab:gren-2}.
\end{itemize} 
In both cases, the errors in the $Z$'s ($Z_\Gamma/Z_V$) are combined
in quadrature with the error in the unrenormalized charges,
$g_\Gamma^{\rm bare}$ ($g_\Gamma^{\rm bare}/g_V^{\rm bare}$), to get
the final estimates.  The results for the $a094m280$ ensemble, labeled
Average, is an average, weighted by $1/{\rm error}^2$, over the three estimates with different 
smearing parameter $\sigma$.

The two sets of estimates given in Tables~\ref{tab:gren-1}
and~\ref{tab:gren-2} are consistent: the difference is less than
$1\sigma$ and the deviation of $Z_V g_V$ from unity (column labeled
$g_V$ in Table~\ref{tab:gren-1}) is $\lesssim 1\%$ and smaller than
the errors.  The two estimates have their relative strengths but we
have no obvious reason for choosing one over the other.  We,
therefore, use the average of the two estimates and the larger of the
two errors for our final values given in Table~\ref{tab:FinalValues}.

Lastly, in Table~\ref{tab:FDCharges} we give results for the
renormalized connected parts of the flavor diagonal charges. The
renormalization is carried out using the $Z_{A,S,T} \times
g_{A,S,T}^{u,d}$ method with the $Z_{A,S,T}$ given in
Table~\ref{tab:Zall}. Technically, the flavor diagonal operators are a
combination of flavor singlet and non-singlet currents and the
renormalization factors are different for the
two~\cite{Bhattacharya:2005rb}. In this work we are ignoring the
difference.

\section{Comparison with previous results}
\label{sec:comparison}

The results presented here are on three ensembles with lattice spacing
$a \approx 0.127$ and $\approx 0.09$~fm and two values of the light
quark masses corresponding to $M_\pi \approx 280$ and 170~MeV.  Note that 
we regard estimates on the $a091m170L$ ensemble as preliminary. These 
three data points are not sufficient to reliably extrapolate
to the continuum limit or to the physical light quark mass. We,
therefore, compare these results with other similar
calculations.

A number of collaborations have performed calculations of the
isovector charges. For recent results see
Refs.~\cite{Bali:2014nma,Bhattacharya:2016zcn,Abdel-Rehim:2015owa,Alexandrou:2016tuo,Green:2012ej}.
The lattice action used, the statistics, the handling of systematic
uncertainties, and the overall strategy for the analysis is different
in each case.  In this work, the first comparison we therefore make is
with calculations done using the same methods but with a 2+1+1-flavor
clover-on-HISQ lattice
formulation~\cite{Bhattacharya:2015wna,Bhattacharya:2016zcn,Gupta:2016rli}.
Results for the renormalized isovector charges given in
Table~\ref{tab:FinalValues} are compared with the clover-on-HISQ
estimates with the closest values of the lattice spacing and the pion
mass given in Table XII of Ref.~\cite{Bhattacharya:2016zcn}. Both sets
of results are reproduced in Table~\ref{tab:ChargesHISQ} to facilitate
comparison.  We find that the estimates for $g_{A,S,T}^{u-d}$ from
simulations using two different lattice formulations and slightly
different lattice parameters agree within one combined $\sigma$.  Note
that the systematics at a given value of the lattice spacing, the
lattice volume, or the pion mass can be different in any two
calculations with different the lattice formulations. Thus, our
conclusions are mostly qualitative.


Comparing the results for the unrenormalized transition matrix elements $\langle 0 |
\mathcal{O}_{A,S,T} | 1 \rangle$ given in Table~\ref{tab:BareME} to
those in Tables 6--8 in Ref.~\cite{Bhattacharya:2016zcn}, we find that
they have the same sign and are similar in magnitude. Our 
rough estimates for these matrix elements are:
$\langle 0 | \mathcal{O}_{A} | 1 \rangle \approx -0.1$, 
$\langle 0 | \mathcal{O}_{S} | 1 \rangle \approx -0.4$ and 
$\langle 0 | \mathcal{O}_{T} | 1 \rangle \approx 0.2$. 
Since these matrix elements account for most of the observed
excited-state contamination in these two calculations, the size and
pattern of the excited-state contamination is similar.  In both calculations, the errors in
the estimates for $\langle 1 | \mathcal{O}_{A,S,T} | 1 \rangle$ are
too large to warrant a comparison.

The renormalized connected parts of the flavor diagonal charges given
in Table~\ref{tab:FDCharges} are also in very good agreement with
those from the 2+1+1-flavor clover-on-HISQ calculation. To facilitate
comparison, we have reproduced the relevant results from Table~XII of
Ref.~\cite{Bhattacharya:2016zcn} in Table~\ref{tab:FDCharges}.

%


\begin{figure}[tb]
\centering
    \includegraphics[width=0.99\linewidth]{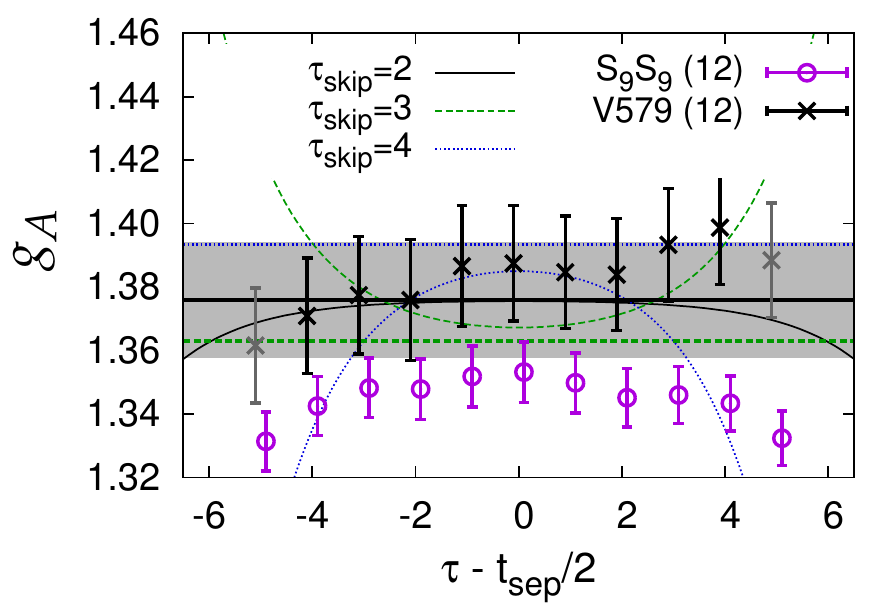} \\
    \includegraphics[width=0.99\linewidth]{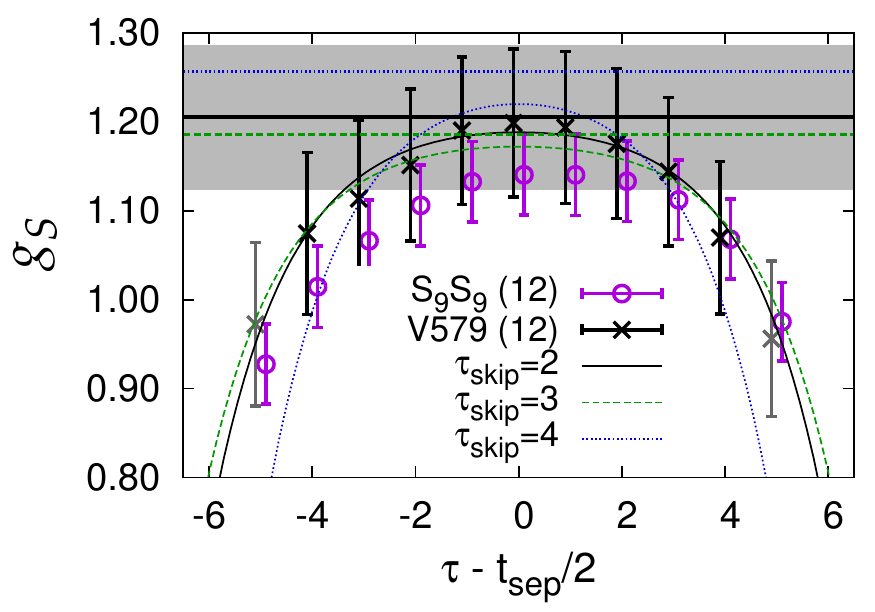} \\
    \includegraphics[width=0.99\linewidth]{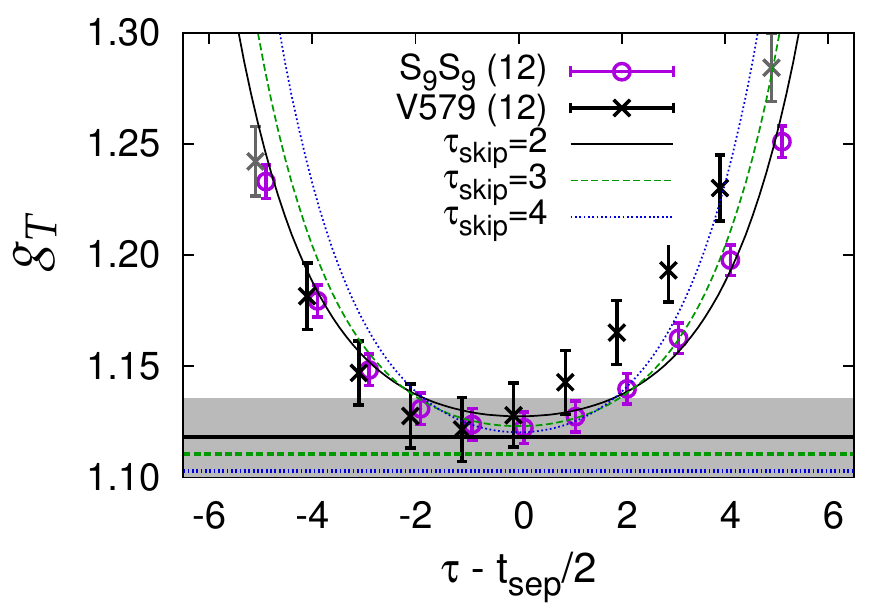}
 \caption{The projected variational three-point correlation function
   (black crosses) for the three isovector charges obtained from
   simulations with $\tsep=12$ on the $a094m280$ ensemble. The raw
   data is the same as that presented in
   Ref.~\protect\cite{Yoon:2016dij} and labeled V579.  The new fits
   with the full covariance matrix are shown for three values of
   $\tskip=2,3,4$. The variational result (solid black line) and the
   grey error band is the result of the fit with $\tskip =2$. For
   comparison, we also show the data points for the $S_9 S_9$
   correlation function (purple circles) with $\tsep=12$.  }
\label{fig:VAR}
\end{figure}

The second comparison we make is with results given in
Ref.~\cite{Yoon:2016dij} obtained using the variational method on the
$a094m280$ ensemble. The results from the variational analysis of
the $3 \times 3$ matrix of two- and three-point correlations functions
constructed using the smearing parameter values $\sigma = 5, 7 $ $9$
at a single value of $\tsep=12$ are also given in
Tables~\ref{tab:gren-1} and ~\ref{tab:gren-2}. These numbers, labeled
VAR579, are different from those presented in Ref.~\cite{Yoon:2016dij}
as the fits have now been done using the full covariance matrix and
the renormalization factors have been included.  The results from the
2-state fits presented in this work and those from the variational
method are in very good agreement.

In Fig.~\ref{fig:VAR}, we show the projected variational three-point
correlation function for the three isovector charges on the $a094m280$
ensemble, taken from the data presented in
Ref.~\protect\cite{Yoon:2016dij}.  The curved lines show the 2-state fit to
these data for three values of $\tskip=2,3,4$.  The corresponding three
$\tsepi$ estimates agree and lie within the shaded $1\sigma$ band for
the $\tskip=2$ fit.  The panels also show the $S_9 S_9$ correlation
function data with $\tsep=12$ (purple circles). Note that the variational
analysis has been done on only 443 configurations versus the full set
of 1005 for the $S_9 S_9$ study. The errors in the two sets of data
points are comparable once this difference in statistics is taken into
account.

The data for $g_A^{u-d}$ in the top panel of Fig.~\ref{fig:VAR} are
almost flat in $\tau$ for both methods, suggesting that the
contribution of the $\langle 0 | \mathcal{O}_{A} | 1 \rangle$ term to
both the variational and the tuned $S_9 S_9$ correlation functions is
small.  The variational estimates lie about $ 3\%$ higher. A
difference of this size can easily be explained by possible
differences in the $\langle 1 | \mathcal{O}_{A} | 1 \rangle$ or higher
terms that cannot be isolated from fits to data with a single value of
$\tsep$.\footnote{In an n-state generalization of
  Eqs.~\protect\eqref{eq:2pt} and~\protect\eqref{eq:3pt}, one can
  divide terms into those that depend on $\tau$ and those that do
  not. The $\tau$ dependent terms are proportional to ${\rm
    cosh}(\Delta M (\tau - \tsep/2))$, where $\Delta M$ is the mass
  gap. The contribution of each of these terms has the same form as
  the observed curvature, but the amplitude for each higher state
  decreases due to the exponential suppression with the associated
  mass gap. On the other hand, each of the $\tau$ independent terms
  gives an overall shift---up or down depending on the sign of the matrix element,  
  and the magnitude is again suppressed exponentially, i.e., by $\exp{(-\Delta M \tsep})$. }

The data for the scalar channel exhibits significant curvature in both
correlation functions and this $\tau$-dependence is again almost entirely
accounted for by the $\langle 0 | \mathcal{O}_{S} | 1 \rangle$ term.  The
$\approx 1 \sigma$ difference between the two correlation functions is
most likely again due to differences in the contributions of the
$\langle 1 | \mathcal{O}_{S} | 1 \rangle$ and higher terms. In the
tensor channel, the data show very little change with the smearing
parameter $\sigma$, and the variational and the $S_9 S_9$ estimates
essentially overlap.

This comparison indicates that the most significant gain on using the
variational method with $\tsep=12$ is in $g_A^{u-d}$. Further
calculations are needed to understand why the excited-state behavior
is so different in the three charges. Based on the current data, 
to confirm that estimates in the $\tsepi$ limit have been obtained, the
variational analysis needs to be repeated at values of $\tsep > 12$
and the 2-state fit with multiple values of $\tsep$ requires
high-precision data at $\tsep > 1.2$~fm.

\section{Conclusions}
\label{sec:conclusions}

We have presented a high statistics study of isovector charges of the
nucleon using four ensembles of (2+1)-flavor clover lattices generated
using the RHMC algorithm~\cite{Duane:1987de}. The all-mode-averaging
method~\cite{Bali:2009hu,Blum:2012uh} and the coherent source
sequential propagator technique~\cite{Bratt:2010jn,Yoon:2016dij} are
shown to be cost-effective ways to increase the statistics. We
demonstrate control over excited-state contamination by performing
simulations at multiple values of the source-sink separation $\tsep$,
and by showing the stability of the 2-, 3- and 4-state fits.

The first highlight of the analysis is that $O(10^5)$ measurements allowed
us to carry out 2-, 3- and 4-state fits to the two-point functions and
2- and 3-state fits to the three-point correlation functions using the
full covariance matrix. In all cases, except for the $a091m170L$
ensemble that is statistics limited, the results for the nucleon mass,
the mass-gaps and the charges show stability with respect to variations in the fit
parameters and the number of states included in the fits. Based on
this analysis, we estimate that it will take $O(10^6)$ measurements to
obtain results for $g_{A,T,V}$ with $O(1\%)$ ($g_{S}$ with $O(3\%)$) error on
each ensemble.

The second highlight is that our clover-on-clover results are in good
agreement with calculations done using the clover-on-HISQ lattice
formulation with similar values of the lattice
parameters~\cite{Bhattacharya:2015wna,Bhattacharya:2016zcn,Gupta:2016rli}.
Estimates of $g_A$, considered a litmus test of lattice QCD's promise
to provide precise estimates of the nucleon structure, are within
$5\%$ of the experimental value even with light quark masses
corresponding to $M_\pi \approx 280$ and $170$~MeV.  These
calculations are being extending to lighter quarks to study the chiral
behavior and to finer lattice spacings to carry out the continuum
extrapolation.

\begin{acknowledgments}
We thank Stefan Meinel for discussions and for sharing his unpublished
results on lattice scale setting.  This research used resources of the
Oak Ridge Leadership Computing Facility at the Oak Ridge National
Laboratory, which is supported by the Office of Science of the
U.S. Department of Energy under Contract No. DE-AC05-00OR22725. The
calculations used the Chroma software suite~\cite{Edwards:2004sx} and
Mathematica~\cite{ram2010}. The work of T.B., R.G. and B.Y. is
supported by the U.S. Department of Energy, Office of Science, Office
of High Energy Physics under contract number~DE-KA-1401020 and the
LANL LDRD program. The work of J.G. was supported by the PRISMA
Cluster of Excellence at the University of Mainz.  The work of
H-W.L. is supported in part by the M. Hildred Blewett Fellowship of
the American Physical Society.  B.J., K.O., D.G.R., S.S. and F.W. are
supported by the U.S. Department of Energy, Office of Science, Office
of Nuclear Physics under contract DE-AC05-06OR23177.
\end{acknowledgments}

\clearpage
%
\bibliography{ref} 

\end{document}